\documentclass{aastex62}
\usepackage{graphicx}
\usepackage{longtable}
\usepackage{subfigure}
\usepackage{natbib}
\usepackage{amsmath}
\usepackage{mathtools}

\newcommand{\mum}{\ensuremath{\mu \rm{m}}}

\submitjournal{ApJ}

\shorttitle{Unraveling SEDs}
\shortauthors{Mart\'{\i}nez-Galarza et al.}

\begin{document}

\title{Unraveling the Spectral Energy Distributions of Clustered YSOs}

\correspondingauthor{J.~Rafael Mart\'{\i}nez-Galarza}
\email{jmartine@cfa.harvard.edu}

\author{J. Rafael Mart\'{\i}nez-Galarza}
\affil{Harvard-Smithsonian Center for Astrophysics \\
60 Garden Street \\
Cambridge, MA 02138, USA}
\nocollaboration

\author{Pavlos Protopapas}
\affil{Institute for Applied Computational Science, Harvard University \\
Northwest B162, 52 Oxford Street \\
Cambridge, MA 02138, USA}
\nocollaboration

\author{Howard A. Smith}
\affil{Harvard-Smithsonian Center for Astrophysics \\
60 Garden Street \\
Cambridge, MA 02138, USA}
\nocollaboration

\author{Esteban F. E. Morales}
\affil{Max-Planck-Institut f\"ur Astronomie \\
K\"onigstuhl 17, 69117 \\
Heidelberg, Germany }

\begin{abstract}

Despite a significant body of evidence suggesting that intermediate- and high-mass stars form in clustered environments, how  stars form  when the available resources are shared  is still not well understood. A related question is whether the IMF is in fact universal across  galactic environments, a galactic initial mass function (IGIMF), or whether it is an average of IMFs that differ, for example, in massive versus low-mass molecular clouds.  If the distribution of stellar masses depends on the birth environment, then the preferred modes of star formation must also vary, since not all models derive in self-regulated star formation. One of the long-standing problems in resolving  these questions and in the study of young clusters is observational: accurately combining multi-wavelength datasets obtained using telescopes with different spatial resolutions. The emission from multiple sources is frequently seen as blended either because the cluster complexities are unresolved, because at different wavelengths or with different telescopes the beam sizes are different, or a combination of these.  The confusion hinders our ability to fully characterize clustered star formation. Here we present a new method that uses a genetic algorithm and Bayesian inference to fit the blended SEDs and images of \emph{individual} YSOS in confused clusters. We apply this method to the infrared photometry of a sample comprising 70 \emph{Spitzer}-selected, low-mass ($M_{\rm{cl}}<100~\rm{M}_{\odot}$) young clusters in the galactic plane, and use the derived physical parameters to investigate the distributions of masses and evolutionary stages of clustered YSOs, and the implications of those distributions for studies of the IMF and the different models of star formation. We find that for low-mass clusters composed of class I and class II YSOs, there exists a non-trivial relationship between the total stellar mass of the cluster ($M_{\rm{cl}}$) and the mass of its most massive member ($m_{\rm{max}}$). The properties of the derived correlation are most compatible with the random sampling of a Kroupa IMF, with a fundamental high-mass limit of $150~\rm{M}_{\odot}$. Our results are also compatible with SPH models that predict a dynamical termination of the accretion in protostars, with massive stars undergoing this stopping at later times in their evolution.

\end{abstract}

\keywords{stars:formation --- 
stars:protostars --- methods: statistics --- galaxy:clusters}


\section{Introduction} 
\label{sec:intro}

The physical characterization of individual young stars at different stages of their formation, from an initial cold clump that undergoes gravitational collapse to the onset of the hydrogen burning phase, and through a period of gas accretion that can last for several million years, is at the base of our understanding of the process of star formation at global scales. About 25\% of all young stars in the galaxy are forming in clustered environments, close enough to each other to mutually affect the phases of this accretion process \citep{Bressert10}. Some important questions in this respect are whether and how early clustering affects the shape and cutoff of the stellar initial mass function (IMF) across the range of cluster masses, and the extent to which such effects influence massive stars and low-mass stars similarly.   Although important, the effects of clustering on star formation remain poorly understood,  in large part due to observational limitations: sensitivity and/or resolution, especially  at mid-infrared (MIR) wavelengths where most of the emission of dust-processed radiation takes place.   

Multi-wavelength surveys of the galactic plane in the past decade, using the \emph{Spitzer}, \emph{Herschel} and \emph{WISE} observatories, have greatly improved the number statistics of individual observations, allowing us to tackle these questions by comparing models of star formation with a growing number of multi-wavelength observations of clustered star formation.  The scenario that has emerged agrees with early studies that suggested that a large fraction of stars form in embedded clusters \citep[e.g.][]{Lada03}. The observations also revealed a continuous distribution of young stellar surface densities in the galactic plane, suggesting that there is no clear-cut distinction between isolated and clustered star formation but rather a broad range of environments forming stars. In addition to core clustering developing during the collapse and fragmentation of a massive primordial molecular cloud, a multiplicity of young stars is expected to develop at smaller scales when the conditions for disk fragmentation are met, as has been demonstrated by numerical simulations \citep{Stamatellos09, Lomax15} and observations \citep{Tobin16}. The ubiquity of embedded clusters and multiple systems in star-forming regions predicted by the theory and revealed by infrared surveys implies that the formation of individual stars cannot be disconnected from the properties of the clusters in which they form, and that the shape and cutoffs of the IMF are most likely related to the individual mass distributions of these young embedded clusters where the properties of the initial distribution of masses have not yet been washed out by stellar death or dynamical evolution. 
% * <hsmith@cfa.harvard.edu> 2018-02-12T23:30:55.216Z:
%
% ^.

The observed properties of clusters inform us about fundamental aspects of star formation, such as the timescales available for the formation of massive stars and the self-regulation of star formation in regions of limited resources. Both of these aspects have an effect on the distributions of masses and ages in young associations. Observations also allow for critical comparisons between existing models. In the competitive accretion scenario \citep{Bonnell01}, for example, stars in a young cluster accrete from a shared reservoir of gas, and in the cluster core the high relative velocities between stars results in Bondi-Hoyle accretion that in turn produces a fragmented IMF that is steeper at the high-mass end. This self-regulated process results in optimal sampling \citep{Kroupa13}, and populates the cluster with an optimal  number of stars starting from the most massive star in the cluster. Not all models of star formation result in optimal sampling of an IMF: if star-formation is not self-regulated, the IMF can be understood as a probability distribution, and stellar masses are randomly sampled from it \citep{Weidner13}. Whether one of these two sampling modes or others dominate in different regimes of cluster mass has deep implications for the properties of the most massive stars that can form, since the sampling mode will set the masses of the most massive stars in a given region; on the galactic scale, it  affects the conclusions about the dynamical and chemical evolution of galaxies.
% * <hsmith@cfa.harvard.edu> 2018-02-12T23:34:20.385Z:
%
% ^.

SED fitting is one of the most common tools in the analysis of young stellar objects. It allows a transformation from the observed photometry of a particular object into a set of physical properties that can be compared to the models, but its results must be interpreted with care, due to limitations in both the models themselves and the photometric data. When applied to the formation of stars in clusters, and the implications for the IMF and massive star formation, significant data analysis challenges need to be addressed. Clusters are difficult to study because they are usually highly embedded, distant, and spatially unresolved, containing tens, hundreds, or thousands of YSOs. Even for nearby clusters where we have good chances of resolving individual YSOs at the shortest infrared wavelengths, the emission from individual stars is often blended together within the beam of infrared telescopes at longer wavelengths.  The spatial confusion of multiple YSOs in infrared observations has not yet been properly addressed in the literature.  

Moreover, YSO spectral energy distribution (SED) models are often applied to photometric datasets taken across optical and infrared bands without accounting for unresolved multiplicities.  Some of the most sophisticated attempts to address the problem so far involve replacing photometric bands with spatially resolved spectrophotometric points \citep{Forbrich10} and including bands in the fitting only as upper limits if there is evidence for multiplicity \citep{Mottram11}.   The \citet{Robitaille06} (R06 hereafter) SED models are by far the most common set of models used to characterize YSOs. They have been used to study the properties of entire star-forming regions both within the Milky Way \citep[e.g.][]{Indebetouw07, Azimlu15} and in the nearby Magellanic clouds \citep[e.g.][]{Simon07, Carlson12}. The limitations of the R06 models have been described in detail in \citet{Robitaille08b}, but perhaps the most severe of these limitations is not due to the model, but to the dataset which is assumed to result from the emission of a single YSO. This has rendered the physical characterization of confused and clustered YSOs an ambiguous exercise. 

This problem has been recognized before.  \citet{Morales17} use the higher resolution images from the UKIRT InfraRed Deep Sky Surveys \citep[UKIDSS,][]{Hewett06} to identify multiple sources within a single IRAC YSO source from the Galactic Legacy Infrared Midplane Survey Extraordinaire \citep[GLIMPSE,][]{Churchwell09, Benjamin03} and MIPS Inner Galactic Plane Survey \citep[MIPSGAL,][]{Carey05} surveys. After a PSF fitting procedure, they examine the near-IR SED of the source which at the UKIDSS bands has photometric points for \textit{each} of the sub-sources. The authors then apply a optimal spline-fitting algorithm to associate the GLIMPSE SED points with each of the UKIDSS sources, and to identify the predominant UKIDSS source; they conclude that the large majority (87\%) of the GLIMPSE YSO sources have only one dominant source; they note that their conclusions apply to the mass range covered by GLIMPSE YSO candidates, between about $3-20~\rm{M}_{\odot}$. Another possible pathway to deal with the problem of SED fitting of multiple YSOs has been recently discussed in \citet{Lomax18}. They have produced hydrodynamical simulations and pathetic SEDs for multiple systems, accounting for episodic accretion, and propose to use the results from a large number of these simulations as informative priors for the Bayesian fitting of multiple YSOs. This would require, as the authors point out, a significant computational effort.

We propose here an alternative method. We use existing SED models to create informative priors for the photometry of unresolved multiple systems, and then perform Bayesian inference to obtain the most likely physical parameters of \emph{individual} YSOs given a set of confused photometry. We construct a probabilistic model for the integrated emission of the cluster across NIR and MIR bands, which includes both the SEDs and images. We optimize the resulting posterior probabilities for the parameters using a genetic algorithm and sample from these posteriors using a Monte Carlo Markov Chain (MCMC) method in order to estimate uncertainties in the derived parameters. We apply this method to the SEDs of of 70 clusters whose mass, $M_{\rm{cl}}$ is below $100~\rm{M}_{\odot}$, and that are also confused within the \emph{Spitzer} beam. We characterize them using data from the UKIRT InfraRed Deep Sky Surveys (UKIDSS) and the Galactic Legacy Infrared Midplane Survey Extraordinaire (GLIMPSE) surveys, and show how the SED method can be used not only to identify the dominant source as in the \citet{Morales17} approach, but also to obtain the most-probable SEDS and YSO properties for \textit{all} of the constituent sources, using Bayesian techniques with a grid of YSO models. We use the derived physical parameters such as mass, age and optical extinction, to investigate trends in the properties of individual stars and clusters. Finally, we interpret the results in the context of different models of star formation and discuss their implications for the IMF in the mass range of these observations. 

This paper is organized as follows. In \S~\ref{sec:observations} we describe the photometry used in the present study, the determination of distances to the sources, and the matching technique used to associate multiple UKIDSS sources to single GLIMPSE detections. \S~\ref{sec:algorithm} describes the probabilistic algorithm to simultaneously fit the SEDs and images of confused YSOs. We describe the results of applying this algorithm to the 70 clusters in \S~\ref{sec:results}, where we also describe the overall statistics of the SED parameters estimated with our method. In \S~\ref{sec:discusion} we discuss the implications of our results for the sampling of the IMF in the mass range of the studied clusters and compare the correlations found between individual YSOs and their parent clusters with theoretical and semi-empirical models of star formation. Finally, in \S~\ref{sec:conclusions} we present our conclusions.

\section{Observational datasets} 
\label{sec:observations}

We use observations from the GLIMPSE survey, which were carried out with the \emph{Spitzer Space Telescope}'s InfraRed Array Camera \citep[IRAC,][]{Fazio04} using bands IRAC 1 (3.6~\mum), IRAC 2 (4.5~\mum), IRAC 3 (5.6~\mum), and IRAC 4 (8.0~\mum). The dataset of observations studied here cover the inner Galactic plane ($|\ell| \le 65\degr$). A first attempt to isolate intrinsically red sources from the large ($> 30$ million sources) GLIMPSE catalog was made by \citet{Robitaille08}. They established several criteria to come up with a photometrically reliable set of red $\sim 2\times 10^4$ sources that was not affected by saturation, sensitivity issues, or variability. In their catalog, which is at least 65\% complete and consists of $\sim 19000$ sources, approximately 30\%-50\% are likely to be AGB stars, and approximately 50\%-70\% are likely to be YSOs. The authors point out that their catalog does not provide a complete picture of Galactic star formation as seen by \emph{Spitzer}, since it does not include blended sources, extended sources, or sources with molecular emission that blue-shifts them in the IRAC bands. 

\citet{Morales17} have isolated a sample of 8325 GLIMPSE YSO candidates which have corresponding UKIDSS coverage. UKIDSS includes images in the the NIR bands J (1.17\mum), H (1.49\mum) and K (2.03\mum), and has a better angular resolution than that of GLIMPSE by a factor $>2$. Following their approach, we use an empirical method for source matching which evaluates the smoothness of the SED transition between NIR and MIR bands for each of the UKIDSS sources with respect to the GLIMPSE/MIPSGAL fluxes by comparing the cubic spline that fits the SED of each UKIDSS source (and associated GLIMPSE/MIPSGAL source) with the simple quadratic function that fits the four middle points that define the NIR to MIR transition ($H$, $K$, 3.6~\mum, and 4.5~\mum). Similarity between these two curves indicates a higher likelihood that a given UKIDSS source contributes to the GLIMPSE flux. This is a generic method which could be robustly applied for matching SEDs across gaps at other wavelengths. 

\begin{figure}
  \centering
  \subfigure{
\includegraphics[scale=0.55,angle=0,trim=0cm 0cm 0cm 0cm,clip=true]{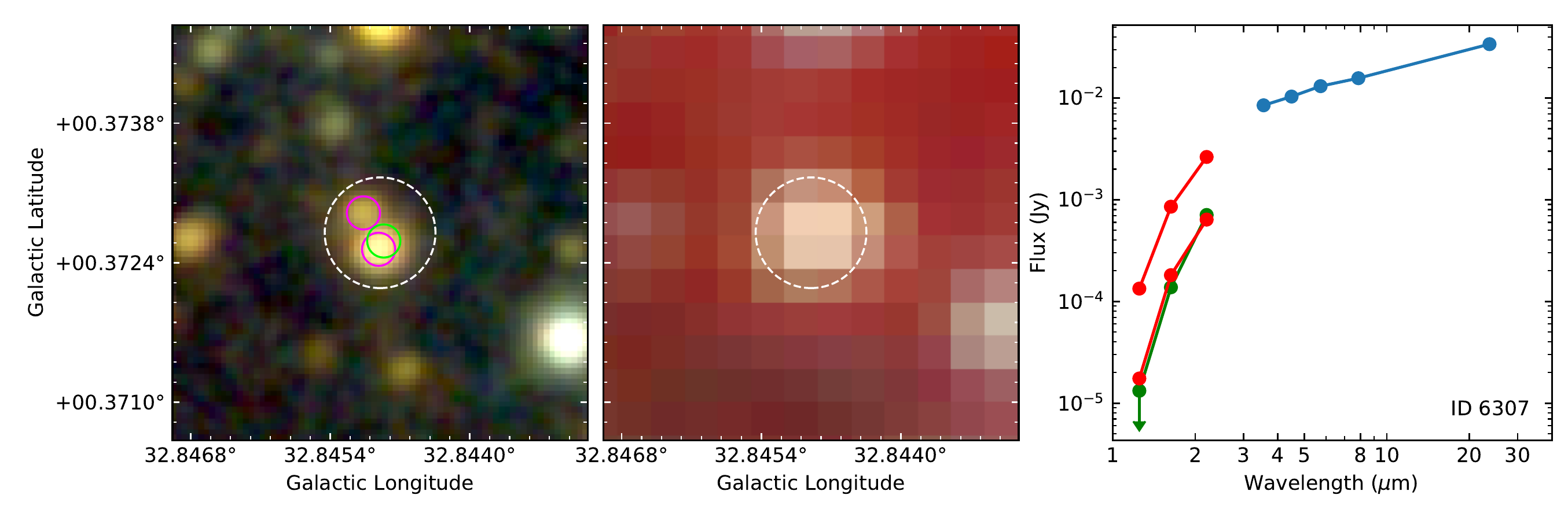}
\label{fig:s1}
}
  \subfigure{
\includegraphics[scale=0.55,angle=0,trim=0cm 0cm 0cm 0cm,clip=true]{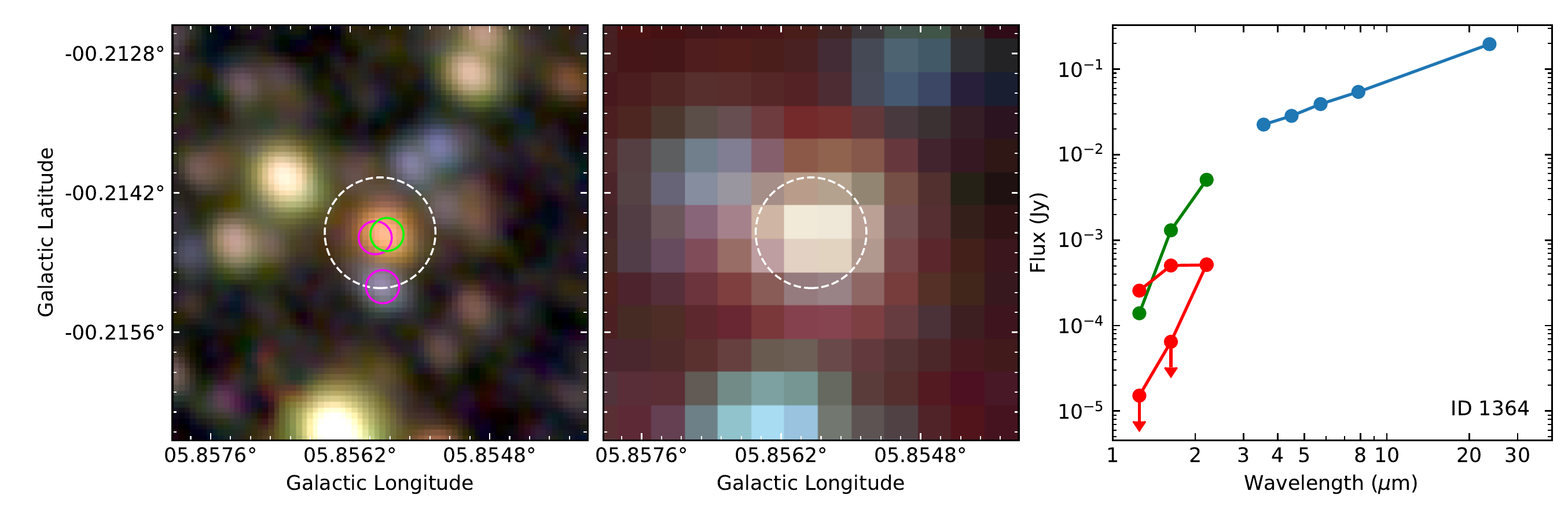}
\label{fig:s2}
}
  \subfigure{
\includegraphics[scale=0.55,angle=0,trim=0cm 0cm 0cm 0cm,clip=true]{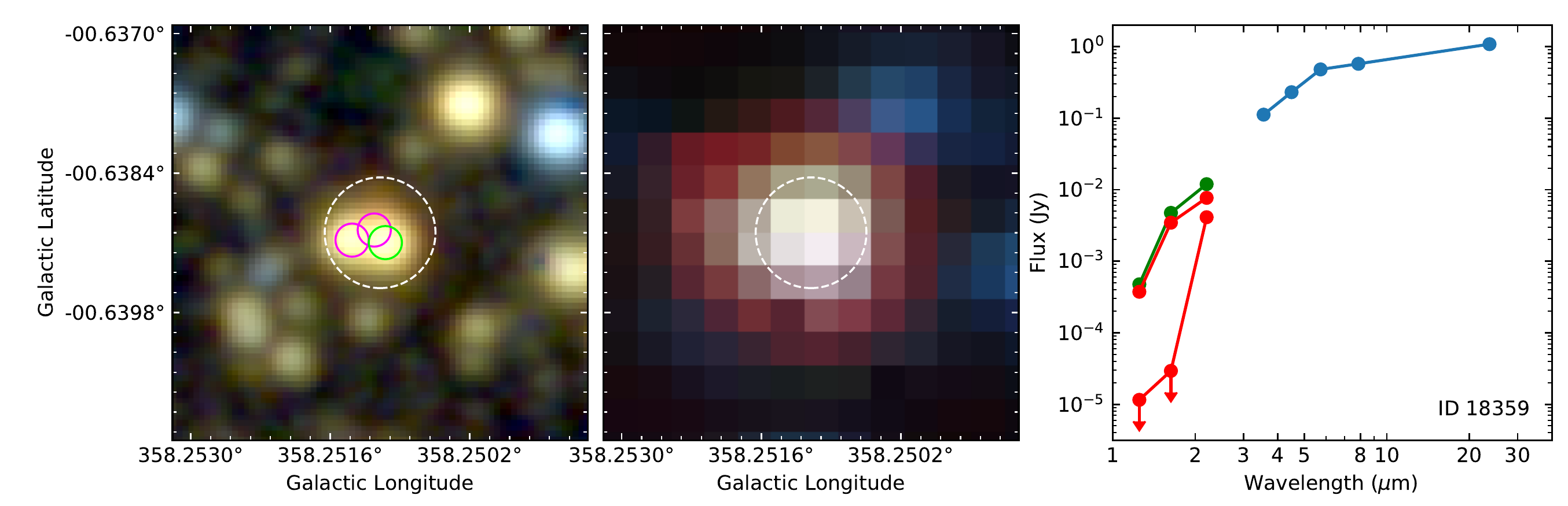}
\label{fig:s3}
}
  \caption{Three GLIMPSE sources with multiple UKDISS matches. In each figure, the left panel is a composite 3 color UKIDSS image with the J band in blue, the H band in green and the K band in red, and the circles indicating the UKIDSS sources detected within $2^{\prime\prime}$ of the corresponding GLIMPSE source. The green circle is for the nearest source in angular separation, the magenta circles for the others, and the bigger dashed-line circle indicates the $2^{\prime\prime}$ search radius. The center panel is a composite 3 color GLIMPSE image with the 3.6~\mum\ band in blue, the 5.8~\mum\ band in green, and the 8.0~\mum\ band in red. The right panel shows the SEDs of the corresponding  GLIMPSE (blue points) and UKIDSS sources (green points for the nearest source and red points for the others).}

\label{fig:sed_examples}
\end{figure}

We use the same quantitative criteria by \citet{Morales17} (their equation 2) to identify UKIDSS sources that match the SED of the corresponding GLIMPSE object, but here we increase the angular distance threshold to $2\arcsec$ in order not to miss likely multiple matching UKIDSS sources that could be farther from the GLIMPSE object than the stricter original threshold of $0.57\arcsec$. After manually discarding a few targets with extended emission for which the PSF-fitting photometry erroneously assigns multiple UKIDSS sources, we ended up with a sample of 194 GLIMPSE objects with more than one UKIDSS source contributing to the flux in the IRAC and MIPS bands, where they appear blended into a singe source. We remark that this empirical SED matching only represents a preliminary step for cluster identification, and that the detailed cluster characterization done in the present paper does not adopt any assumption on the UKIDSS sources from that step. Indeed, all UKIDSS sources within $2\arcsec$ (the width of the IRAC PSF) will be initially modeled here as potential contributors to the IRAC/MIPS fluxes. Fig.~\ref{fig:sed_examples} shows multi-wavelength views for three of the selected clusters. Shown in the figure are three color images using both UKIDSS and IRAC colors, together with the SEDs in each case derived from the images.

Our final list of 70 low-mass YSO clusters (see Table~\ref{tab:sources}) results from restricting the sample to those objects with available distance estimates, needed for the physical modeling performed in this paper. For objects that are part of a star-forming region (infrared dark cloud, sub-millimeter clump, \ion{H}{2} region) we use the distances to the regions reported in the literature.  For objects with LSR velocity measurements we use a kinematic distance solution based on a Galactic rotation model following \citet[their appendix B.4]{Morales13}. For many other objects in our sample we use the large dataset of LSR velocities and KDA resolutions provided by \citet{Wienen15}, who carried out molecular line follow-up observations towards several sources detected in the APEX Telescope Large Area Survey of the Galaxy (ATLASGAL).

\section{Bayesian Parameter Estimation} 
\label{sec:algorithm}

This section describes the Bayesian parameter estimation methods that we have designed to characterize clustered YSOs\footnote{The python code containing the detailed algorithms can be found at \url{https://github.com/juramaga/Bayesian_fitter}}. The statistical model for SED fitting is introduced first, followed by a discussion of the optimization and sampling of the resulting posterior probability distributions. The statistical model for image fitting is described at the end of the section, together with the strategy to combine results from SED and image fitting to reduce the variance of our results.

\subsection{Likelihood}
Our goal is to simultaneously fit the SEDs of $m$ sources that are observed in $n$ different bands. The sources appear resolved as individual objects in the first $n'$ bands, but are blended together within the beams of the remaining $n-n'$ bands. Thus, in the unresolved bands we have a single photometric measurement for all $m$ sources. We fit the sources using a grid of pre-calculated model SEDs $M$ that depend on parameters $\{\theta\}$. For source $i$, then, we have a set $\{\theta_i\}$ of parameters and we can denote the full set of parameters for all sources $i = 1,\ldots,m$ as $\Theta = \{\theta_1,\ldots,\theta_m\}$. Data $D$ consists of the fluxes observed in all bands, resolved and unresolved, and their respective uncertainties. For source $i$ and resolved band $j$, the measured flux is $F_{ij}$, with associated measurement error $\sigma_{ij}$, whereas for the unresolved sources the integrated measured flux is $F_j$, with associated error $\sigma_{j}$. The modeled flux for source $i$ in band $j$ is $M_{ij}$. With this nomenclature, the log-likelihood of observing a particular set of photometry for the cluster, if we assume that it was drawn from model $M$, is:

\begin{equation}
\begin{multlined}
\log P(D|\Theta) \propto \frac{1}{n'}\sum_{j=1}^{n'}  \left[-\sum_{i=1}^{m}\frac{\left(F_{ij} - M_{ij}(\Theta)\right)^2}{2\sigma_{ij}^2} \right] \\  - \frac{1}{n-n'}\sum_{j=n'+1}^{n}\frac{\left(F_j-\sum_{i=1}^{m} M_{ij}(\Theta)\right)^2}{2\sigma_j^2}
\end{multlined}
\label{eq:prob}
\end{equation}

Our problem reduces to finding the set of model parameters ${\Theta}$ that optimizes the probabilistic model above. We are therefore faced with a maximum likelihood estimation problem. But rather than just finding the best-fitting model, we can use our predictive model to estimate the probabilities over all possible solutions, i.e., we perform Bayesian inference. 

\subsection{Bayesian formulation}
Rather than being interested in knowing which observations are expected given the model parameters, we would like to infer the physics, i.e., we want to know which parameter models are more probable given the observed photometry. This means that we assume the model parameters to be random variables with associated probability distributions $P(\Theta|D)$, as opposed to the frequentist view, in which parameters have absolute, true values. Once we infer $P(\Theta|D)$, the \emph{posterior} distribution, we can obtain a point estimate for the parameter values, usually the maximum a posteriori (MAP) estimate. The marginalization of the joint posterior with respect to each model parameter also provides complete information about the uncertainties on those parameters. Formally, the chain rule from probability theory provides a relation between the likelihood and the posterior distributions:

\begin{equation}
P(\Theta|D) \times P(D) = P(\Theta, D) = P(D|\Theta) \times P(\Theta)
\end{equation}

which gives the well-known Bayes theorem:

\begin{equation}
P(\Theta|D)\propto P(D|\Theta)\, P(\Theta)
\label{eq:posterior}
\end{equation}

where $P(\Theta)$, the \emph{prior} distribution, encodes any belief we might have (prior to obtaining the current photometry) on the parameter values. In general, priors become more informative in the analysis as less data points are available to inform the model. 

The R06 models comprise a library of SED models for a broad range of YSO parameters. The fixed grid of parameter values was originally generated by randomly sampling YSO ages ($t$) and masses ($m_*$), and then assigning corresponding accretion rates and disk/envelope properties using theoretical or semi-empirical relations. The resulting SEDs (each of them generated at 10 different viewing angles) are then convolved with a library of observing filters from the UV to the far-infrared, giving model fluxes for all models in each band. To account for interstellar extinction, these ``raw'' SEDs can then be further obscured by a dust screen of optical extinction $A_V$, using the typical Galactic ISM extinction curve modified for the mid-IR extinction whose properties are derived in \citet{Indebetouw05} The models are normalized to a distance of 1~kpc, but can be easily scaled to the desired distance.

The priors on mass ($M_*$) and age ($t_*$) are set by the original pre-calculated SED grid, and are proper uniform priors $\mathcal{U}(\rm{min},\rm{max})$ defined on the logarithmic space of YSO ages and masses. Class 0 YSOs are undetected below $20~\mum$, and so we expect to detect mostly class I/II YSOs. The lifetime of class I YSO is about $10^5$~yr, whereas that of a class II YSO on average a few times $10^6$~yr.  We modify the age prior and use a normal distribution $\mathcal{N}(\mu,\sigma)$ centered at $\log t_* = 5.5~\rm{yr}$ with a broad standard deviation of 1 order of magnitude:

\begin{equation}
P(\log M_*) = \mathcal{U}(-1,3)
\end{equation}

\begin{equation}
P(\log t_*) = \mathcal{N}(5.5,\sigma=1)
\end{equation}

We use a normal prior for $A_V$. Although in principle we could use reddening data from recently released Pan-STARSS photometry \citep[e.g.][]{Green15} in order to obtain a better estimate of extinction in the line of sight and at the distance to our sources, we decide to adopt a more conservative approach: we use a normal distribution centered at $A_V = 10~\rm{mag}$, with a standard deviation of $\sigma_{Av}=0.8~\rm{mag}$. This mean is about 0.5 dex higher than the mean of the distribution for our sources, as derived from Pan-STARRS data, and is justified by the fact that at scales that are smaller than the beam of the 1.8~m Pan-STARRS telescope, we expect optical extinction to be higher than average towards regions of massive star formation. As for the inclination angle $\phi$, we assume that it is proportional to the cosine of the inclination angle, as expected from an ensemble of randomly oriented disks:

\begin{equation}
P(\log A_V) = \mathcal{N}(1.0,\sigma=0.8)
\end{equation}

\begin{equation}
P(\phi) \propto \cos \phi
\end{equation}

We now have a probabilistic model that gives the posterior probability for the YSO parameters (Eq.~\ref{eq:posterior}). Computing $P(\Theta|D)$ by brute-force, calculating the posterior at each possible point of the parameter space is computationally intractable. Even more so for the parameter space of our problem, whose dimensionality is $\rm{dim} = 4\times n_{\rm{sources}}$ (4 parameters for each source plus one degree of freedom to account to uncertainty in the distance to the sources, minus one degree of freedom due to the fact that the unresolved photometric points must satisfy the condition that their sum must equal the observed flux in each band). 

The majority of the 14 varying parameters in the R06 models are correlated to other model parameters \citep{Robitaille06, Robitaille08}. For example, the envelope accretion rate ($dM_{\rm{env}}/dt$), steeply decays for all models for ages beyond $10^5~$yr. In general, most parameters show a correlation with either the YSO mass or age. Therefore, for a cluster with 3 visible clustered YSOs (a typical case in our dataset) the dimensionality of the parameter space is 12, but dimensionality grows linearly with the number of sources. Typically, the three UKIDSS bands are observed for each of the cluster members, whereas the IRAC bands are observed for the cluster as a whole. The MIPS24 band is not available for all clusters. Therefore, for a cluster with 3 resolved UKDISS sources we have at most 14 observations available to fit the 12 parameters. 

Probability distribution functions can be efficiently sampled using stochastic methods such as Markov Chain Monte Carlo (MCMC) sampling. MCMC methods can be extremely slow at converging if the initial guess for the parameters is far from a significant peak of probability, specially in a parameter space with many dimensions, such as in our case. It is therefore a good course of action to first use an optimization algorithm to find a MAP estimate, and then use the MAP result to initialize the MCMC sampler.

\subsection{Optimization of the probabilistic model with Genetic Algorithms}
\label{sec:optimization}

We use genetic algorithms to optimize the posterior probability distribution prior to MCMC fitting. Genetic algorithms are inspired in the stochasticity of biological evolution: given an initial population of solutions (we will call these solutions \emph{individuals}), whose \emph{genes} (the parameters values) are replicated using a particular mechanism into the following generation, only the fittest solutions (those with larger posterior probability) will survive after many generations. In the context of our problem the fitness function is the posterior probability of Eq.~\ref{eq:posterior} and after each generation the resulting \emph{population} will be graded according to the average value of this fitness. Given a random initial population, with random values assigned to the genes of each individual, at each generation we perform the following operations between individuals:

\begin{itemize}

\item \textbf{Reproduction.} Given a population of individuals with different fitnesses, we update the population in such a way that the best fit individuals will have more offspring, while keeping the total number of individuals unchanged.

\item \textbf{Crossover.} We randomly exchange genes between the members of the updated population to create new individuals. Such exchange of genes is performed by cutting and then exchanging parameters between parent individuals.

\item \textbf{Mutation} Natural selection and diversity would not happen without unlikely random mutations of the genes. To simulate mutation in our population of individuals, we randomly change the value of one of the parameters in a random individual, with very low probability.

\end{itemize}

The average fitness of the new generation, as well as the fitness of the individuals is evaluated. Evolution should lead to an increase in fitness with each generation. After a sufficient number of generations, the population should be primarily composed of highly fit individuals (i.e., solutions with a higher posterior probability). By selecting the best among these descendants, we obtain the best possible solution.

The main parameters controlling the outcome of this evolutionary process within the algorithm are: 

\begin{itemize}

\item The size of the population ($N$). This parameter remains fixed along the entire simulation, and should be adjusted so that it is not too small (which would make crossover unlikely, resulting on only a small portion of the parameters space being explored), or too large (which would make the whole process very slow). We use a population of between 30 and 40 members, depending on the number of sources being fitted.
\item The probability of mutation ($p_m$), which should be kept low to avoid turning the algorithm into a random search. We have used values between 1\% and 2\%.
\item The survival rate of the best fit individuals ($p_s$), i.e., the fraction of individuals (ordered by decreasing fitness) that are selected to become parents of the next generation. Here we use between 20\% and 40\% for this parameter.
\item The probability $p_o$of an individual being selected of being selected to be a parent of the next generation even if you are not among the best fit individuals. For this parameter we use a value of 5\%.
\item The crossover probability ($p_c$). How likely it is that crossover takes place. If there is no crossover at all, children are just identical copies of parents. Here we adopt $p_c=1.0$

\end{itemize}

Our adopted values represent a reasonable estimate based on trial and error experiments and the final conclusions do not depend critically on their precise values. For each cluster in Table~\ref{tab:sources}, we optimize the posterior probability of Eq.~\ref{eq:prob} by applying the genetic algorithm with the parameters specified above, over $10^4$ generations, and use the outcome of this process to initialize the MCMC fitting. 

\subsection{MCMC: sampling the parameter posteriors}
The output parameters of the GA optimization are used as initialization parameters of the MCMC sampler. We use our own implementation of the standard Metropolis-Hastings algorithm, for which we customize the proposal distributions in order for it to work with a discrete grid of models parameters. In the $t_*-M_*$ plane, our proposal distribution consists of a uniform probability of jumping from current location $(t_{*i}, M_{*i})$ to another point $(t_{*i+1}, M_{*i+1})$ that is at a distance of at most $r_{\rm{prop}}$ from the current location, with $r_{\rm{prop}}$ tuned to obtain an acceptance rate of about 30\%. 

Similar uniform distributions are chosen for $A_V$ and the distance to the source $d$. For the inclination angle, the proposal function takes the form of a unitary jump between neighboring inclination angles. We use a total number of $n_{\rm{iter}}=10^6$ iterations. It is customary to ignore a certain number of traces at the start of the MCMC trace, because early in the chain the algorithm has not yet converged. Here we set this burn-in period to 20\% of the total number of elements in the trace. In order to check for convergence of the MCMC method, we use the Geweke test.

\subsection{A probabilistic model for unresolved images}
\label{sec:image_fitting}
We consider two sources to be unresolved in the Spitzer bands if their UKIDSS coordinates are $2^{\prime\prime}$ apart or less. Although the resolution of the IRAC images is below this number ($\sim 1.2^{\prime\prime}$), disentangling them using simple PSF fitting might be problematic, and we therefore consider them to be unresolved for the purposes of flux estimation. Yet, because the PRFs for the IRAC bands is sub-sampled every fifth of a pixel, additional information about how much each source in the cluster contributes to the unresolved fluxes is contained in the images. Here we develop an approach to image fitting that is complementary to the model-dependent SED fitting. In \S~\ref{sec:combining} we describe how the two approaches complement each other.

The idea is simple: using the models for the oversampled PRFs in each IRAC band, and the position of the cluster members in the resolved (UKIDSS) images as priors for the location of the sources in the unresolved (IRAC) images, we can build a probabilistic model that can be fit to reproduce the observed IRAC images. The model depends on the following parameters: the position of each source on the IRAC image $(x_i,y_i)$, the multiplicative scale factor $A_i$ for each individual PRF (related to the contribution of each source to the total flux in each pixel), and the background level $B$.

We assume that the observed \emph{Spitzer} images are the result of two separated processes. First, at the individual pixel level, photons hit the detector with a certain average rate that can be modeled as a Poisson process\footnote{We use a Poisson process here because we want the model to be applicable in low photon counts scenarios (e.g., X-ray astronomy). In the high photon count regime the Poisson model generalizes to a normal model}. That is, we spatially discretize the image by assuming that for each pixel, photons arrive independently of each other at a constant rate so that there is an average number of photon hits per unit time. Second, we assume that for each pixel, such process happens as many times as we have sources in the cluster. We assume that independent Poisson processes happen simultaneously in each pixel for each source that contributes to the flux in that pixel. In other words, we assume that the image is the result of a mixture of Point Response Functions (PRFs). 

Explicitly, we assume that the probability of measuring an image containing $m$ sources with flux density $N_k$ in the $k-th$ pixel, given the parameters of our model, i.e., the set of PRF scaling factors ${A_i}$, the positions $(x_i,y_i)$ of the sources in the IRAC images, and a uniform background level $B$, and assuming that the fluxes of neighboring pixels are uncorrelated, can be expressed as the product of Poisson distributions:

\begin{equation}
P(\{N_k\}|A) = \prod_{k=1}^{n_k}\frac{D_k^{N_k}e^{-D_k}}{N_k!}
\label{eq:psf_mix}
\end{equation}

where $D_k$ is the average number of photons reaching the \emph{k-th} pixel per unit time (i.e., the flux density in the \emph{k-th} pixel), and $n_k$ is the total number of pixels in the image. Note that in this equation, each of the $D_k$ is in fact a linear combination of the corresponding elements of the $m$ PRFs centered at the location of the sources (from the resolved images), plus a background term:

\begin{equation}
D_k = n_0\left[\sum_i^mA_i\times \rm{PSF}_i+B\right]
\label{eq:psf_mix1}
\end{equation}

We fit the observed images in all four bands using this probabilistic model, and for each tested model we use the UKIDSS coordinates to choose the PRF sub-sample that corresponds to the location within the IRAC pixels. Since the UKIDSS sources positions are known with a precision of $\sim 0.3^{\prime\prime}$, we allow for a variation of $\pm 1$ in the actual sub-sampled PRF used when we perform the fitting. For the scaling factors, background and source positions, we assume uniform priors, and sample the resulting posterior distribution using MCMC. The posterior probabilities for the PRF scaling factors are directly related to the flux contributed by each source to the unresolved photometry.

\subsection{Assessing the reliability of the method}
\label{sec:combining}

The SED and image fitting algorithms described in the previous section can be used iteratively in order to reduce the variance in the estimates of the physical parameters of the individual YSOs, by using the output flux posteriors of one method as priors in the other method.  Here we describe how we combine the two in order to improve the quality of our results, and then we validate the reliability of the method by applying it to a simulated cluster with known physical parameters.

\subsubsection{Iterative fitting of the SEDs and the images}
 
Given the set of resolved NIR and unresolved MIR photometry for a given cluster, we perform the fitting of the data in three steps:

\textbf{Step 1.} Using the model of Eq.~\ref{eq:prob}, we simultaneously fit the SEDs of the cluster members, and obtain posterior probabilities for the model parameters. We also obtain posterior predictive distributions for the unresolved fluxes in this step. The posterior predictive is the distribution of unobserved photometry $\tilde{D}$ conditioned on the observed data $D$. It is constructed by averaging the likelihood of new unseen data points over all possible parameter values, weighted by their posterior probability:

\begin{equation}
p(\tilde{D}|D) = \int p(\tilde{D}|\Theta) p(\Theta|D) d\Theta
\label{eq:posterior_predictive}
\end{equation}

Note that the probability of measuring a given unresolved flux for a given cluster member, given the observed data, equals the likelihood of that flux given the model parameters, times the probability of that particular choice of parameters, marginalized over all possible parameter values. We can obtain a posterior predictive for each of the unresolved bands, for each of the cluster members. Visually, the posterior predictive can be understood as the histogram of all the model fluxes in a particular band when the models are taken from the MCMCM sampled parameters.

\textbf{Step 2.} Using the image fitting model from Eqs.~\ref{eq:psf_mix} and \ref{eq:psf_mix1} we then fit the observed IRAC images and derive posterior probabilities for the fluxes in the unresolved bands. We use the posterior predictives derived in step 1 for the relative fluxes as priors for the image fitting. Note that this is consistent from a statistical point of view, since we are not using the same information in both models. While for the SED fitting we use the \emph{integrated} fluxes to constrain the unresolved fluxes, in the image fitting we are using the pixel-by-pixel fluxes. We get new posterior distributions for the fluxes as an output of the image fitting.

\textbf{Step 3.} Finally, we re-fit the SEDs but this time we use the posterior probabilities for the fluxes obtained in the previous step as individual photometric measurements in the unresolved bands. We do this for all 4 IRAC bands, but not for the MIPS 24~$\mum$ band, since at the MIPS resolution of $6^{\prime\prime}$ we would not get a good constrain on the unresolved fluxes since the UKIDSS sources are within $2^{\prime\prime}$ only. The end result of this 3-step fitting algorithm are the best fitting SED and images in each band, as well as the posterior probabilities for the relevant model parameters, evaluated for each individual source in the cluster.

\subsubsection{Simulated cluster}
\begin{figure}
\centering
\includegraphics[scale=0.5,angle=0,trim=0cm 0cm 0cm 0cm,clip=true]{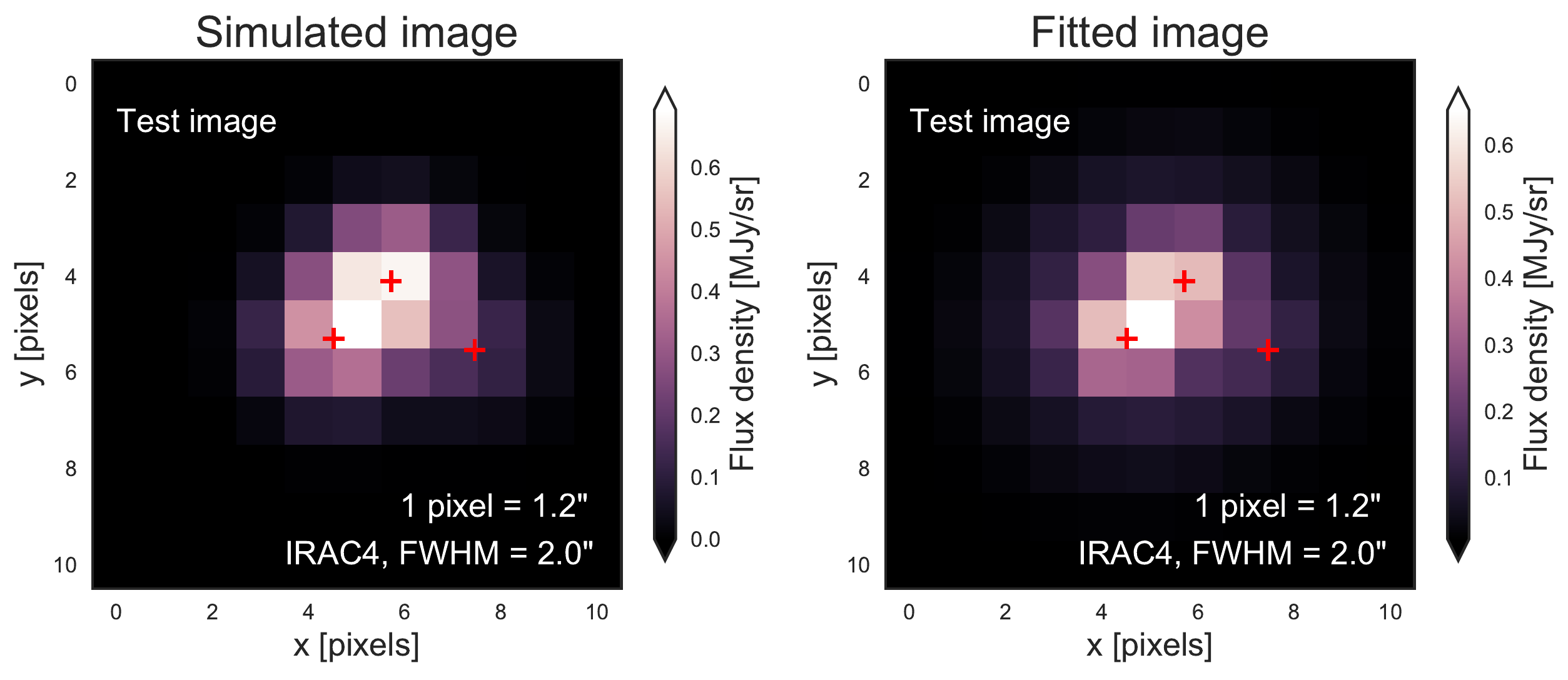}
\caption{IRAC 8~\mum\ images of a simulated confused cluster indicating the position of the individual sources with red crosses. The left panel corresponds to the originally simulated image, while the right panel shows the best fit using our method.}
\label{fig:simulated_cluster}
\end{figure}

In order to assess the reliability of our method in recovering the physical parameters of clustered YSOs, we have tested it on simulated clusters composed of three YSOs whose properties have been sampled from the library of R06 models. We have simulated the coordinates of these objects from a 2D normal probability density function with a standard deviation corresponding to the typical size of the clusters in our sample ($\sim 2^{\prime\prime}$). We assigned UKIDSS, IRAC and MIPS photometry to each source according to the corresponding R06 SEDs, and then added 10\% Gaussian noise. We also simulated IRAC images of simulated clusters by convolving the Gaussian profiles of the point sources with a model of the instrument PSF and then binning the convolved image to resemble the IRAC pixels, and matching the total fluxes with the SED fluxes. In Fig.~\ref{fig:simulated_cluster} we show an example of a simulated IRAC1 image compared with the result of our image fitting algorithm. The cluster is effectively unresolved within the IRAC beam, as is the case of many real clusters for which SED fitting of the individual members is algorithmically difficult, and if fact has not been attempted so far. Our approach allows for detailed modeling of the individual sources.

\begin{figure}
\centering
  \subfigure{
\includegraphics[scale=0.55,angle=0,trim=0cm 0cm 0cm 0cm,clip=true]{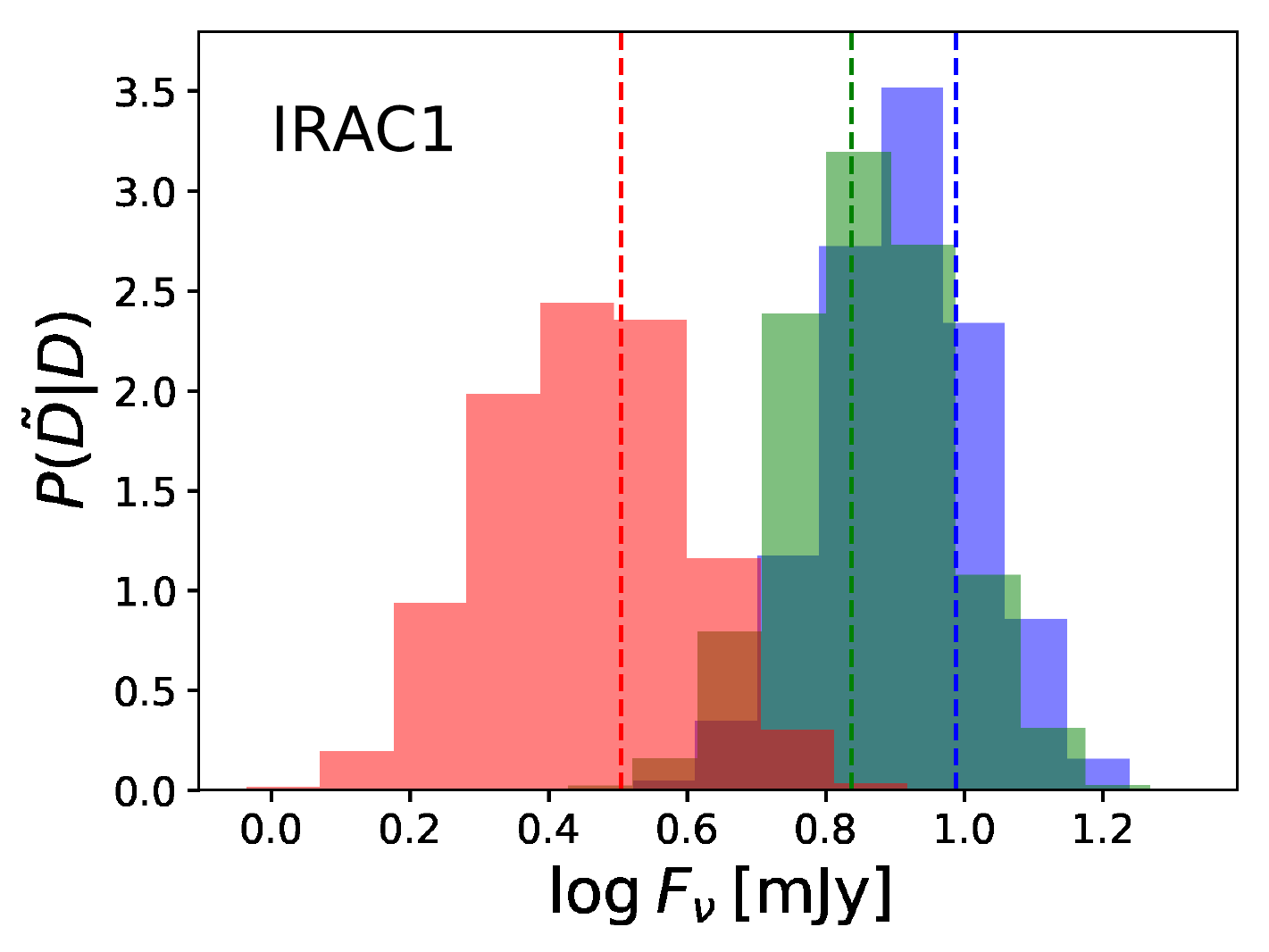}
}
  \subfigure{
\includegraphics[scale=0.55,angle=0,trim=0cm 0cm 0cm 0cm,clip=true]{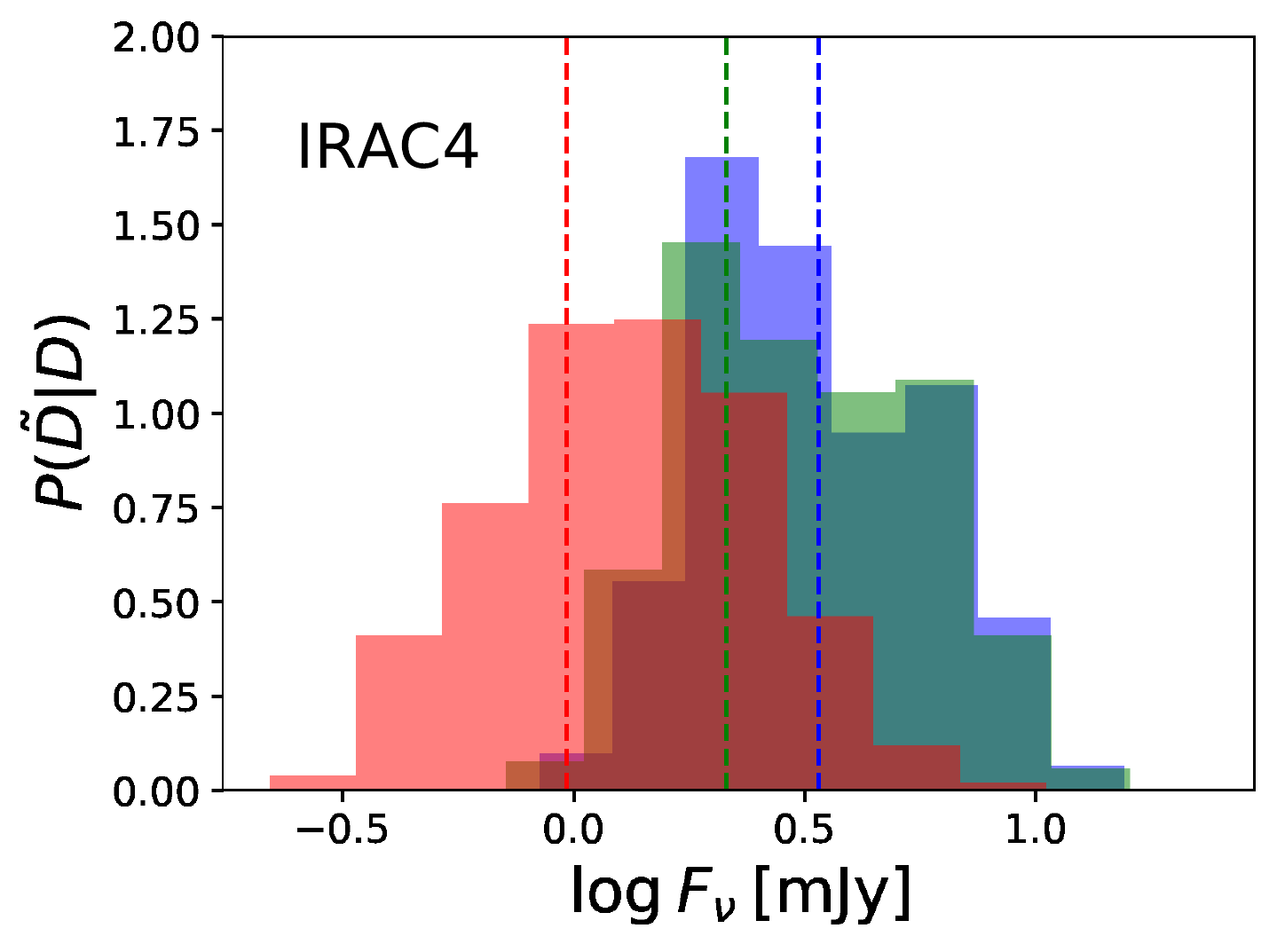}
}
\caption{The posterior predictives from our deblending SED fitting approach (shaded histograms) compared with the ground truth photometry of a simulated blended cluster, for IRAC bands 1 and 4. Each histogram corresponds to the distribution of possible fluxes for one particular source, and the dashed lines represent the corresponding true value of the photometry.}
\label{fig:comparison_posteriors}
\end{figure}

Following the steps described above, we first fit the set of resolved and unresolved photometry to obtain posterior probability distributions for the unresolved IRAC fluxes of the simulated cluster. The posterior predictives for the unresolved fluxes of each individual source are in excellent agreement with the ground truth values, as show in Fig.~\ref{fig:comparison_posteriors}. We then fit the IRAC images using those posterior probabilities as priors for the relative contributions of each source, and finally we re-fitted the SEDs using the posterior predictives from image fitting as resolved mid-infrared data points. The final results are the posterior probability distributions for the physical parameters. In Table~\ref{tab:sim_results} we compare the ground truth values for the model parameters with the credible intervals resulting from our Bayesian fitting. In our experiments with simulated clusters, the ground truth values fall within or very close to the $1\sigma$ credible intervals for all individual sources. We therefore expect our parameter estimation to be reliable within the uncertainties of the R06 models themselves.

\begin{deluxetable}{ccc}
\tablecaption{Comparison with simulated cluster \label{tab:sim_results}}
\tablecolumns{3}
\tablehead{
  \colhead{Parameter} &
  \colhead{$1\sigma$ interval} &
  \colhead{True value} 
}
\startdata
  $\log~t_*^1$ & $5.75_{-0.10}^{0.34}$ & 5.88   \\
  $\log~m_*^1$ & $0.07_{-0.12}^{0.23}$ & 0.30   \\
  $\log~A_V^1$ & $0.84_{-1.19}^{0.16}$ & 1.00   \\
  $\log~t_*^2$ & $6.21_{-0.50}^{0.44}$ & 6.37   \\
  $\log~m_*^2$ & $0.27_{-0.29}^{0.09}$ & 0.37   \\
  $\log~A_V^2$ & $0.87_{-0.17}^{0.11}$ & 1.00   \\
  $\log~t_*^3$ & $6.18_{-0.42}^{0.46}$ & 6.07   \\
  $\log~m_*^3$ & $-0.12_{-0.24}^{0.32}$ & 0.01   \\
  $\log~A_V^3$ & $0.81_{-0.43}^{0.19}$ & 1.00   
\enddata
\end{deluxetable}

\section{Results} 
\label{sec:results}
This section summarizes the results of applying the SED/imaging fitting method described in the previous section to the 70 low-mass YSO clusters listed in Table~\ref{tab:sources}. These results consist of the simultaneous fits to the blended SEDs, the best-fitting IRAC images generated by the image fitting algorithm and the derived posterior distributions for the model parameters after both methods have been combined.

\subsection{SED and image fitting}
\begin{figure}
  \centering
  \subfigure{
\includegraphics[scale=0.38,angle=0,trim=0cm 0cm 0cm 0cm,clip=true]{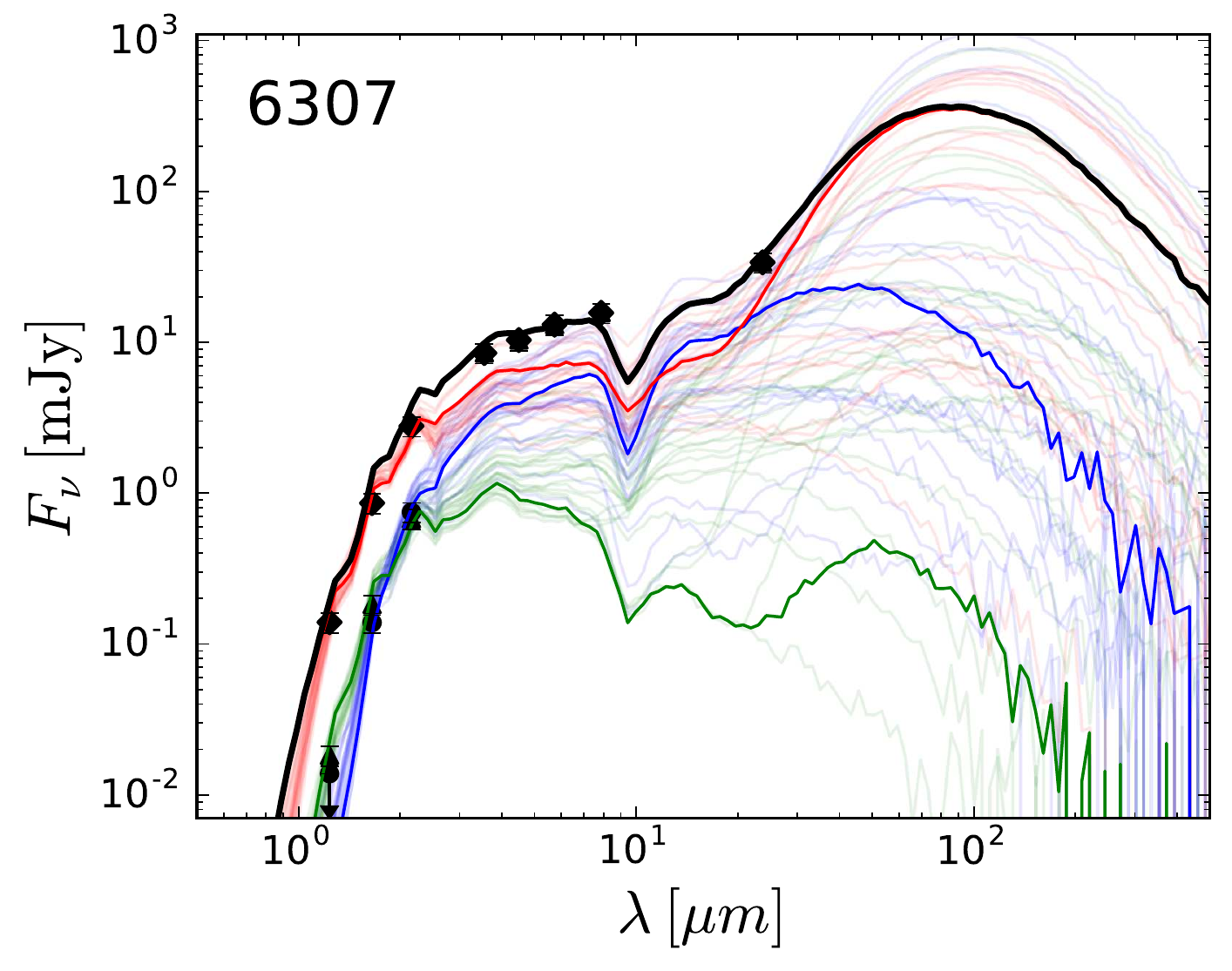}
\label{fig:sim_fit_6307_reg}
}
  \subfigure{
\includegraphics[scale=0.29,angle=0,trim=0cm 0cm 0cm 0cm,clip=true]{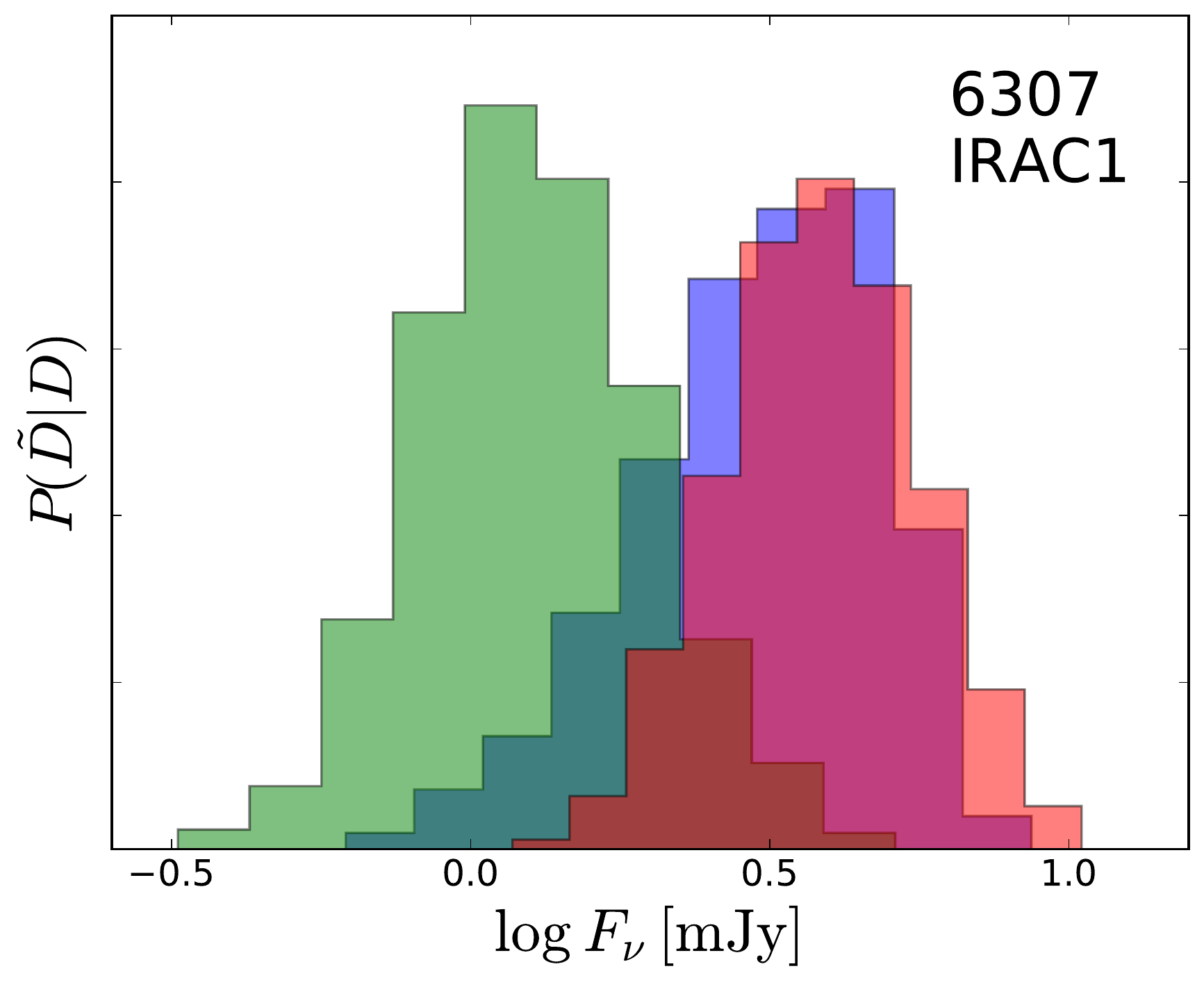}
\label{fig:post_pred_6307_reg}
}
  \subfigure{
\includegraphics[scale=0.38,angle=0,trim=0cm 0cm 0cm 0cm,clip=true]{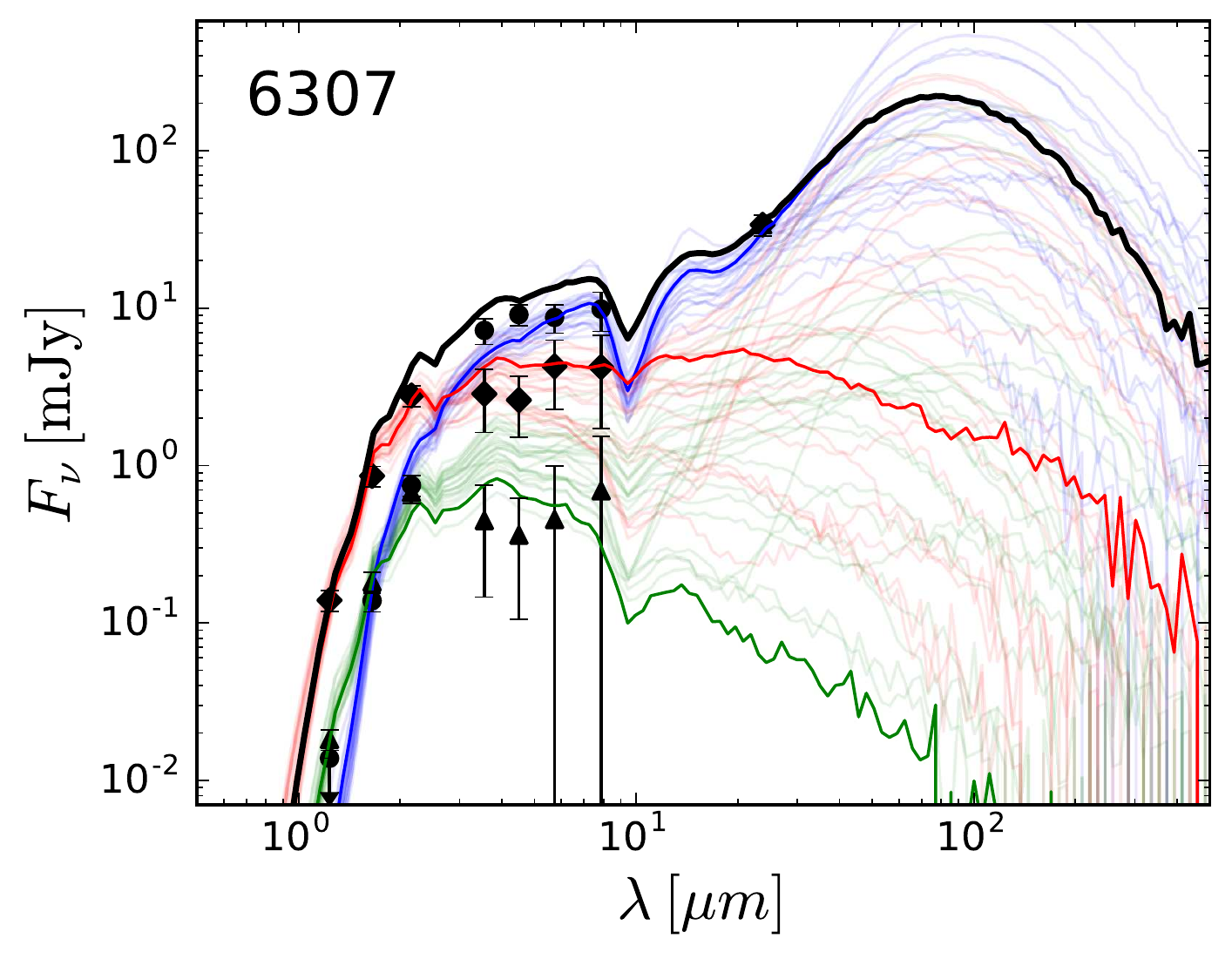}
\label{fig:sim_fit_6307_deblend}
}
  \subfigure{
\includegraphics[scale=0.38,angle=0,trim=0cm 0cm 0cm 0cm,clip=true]{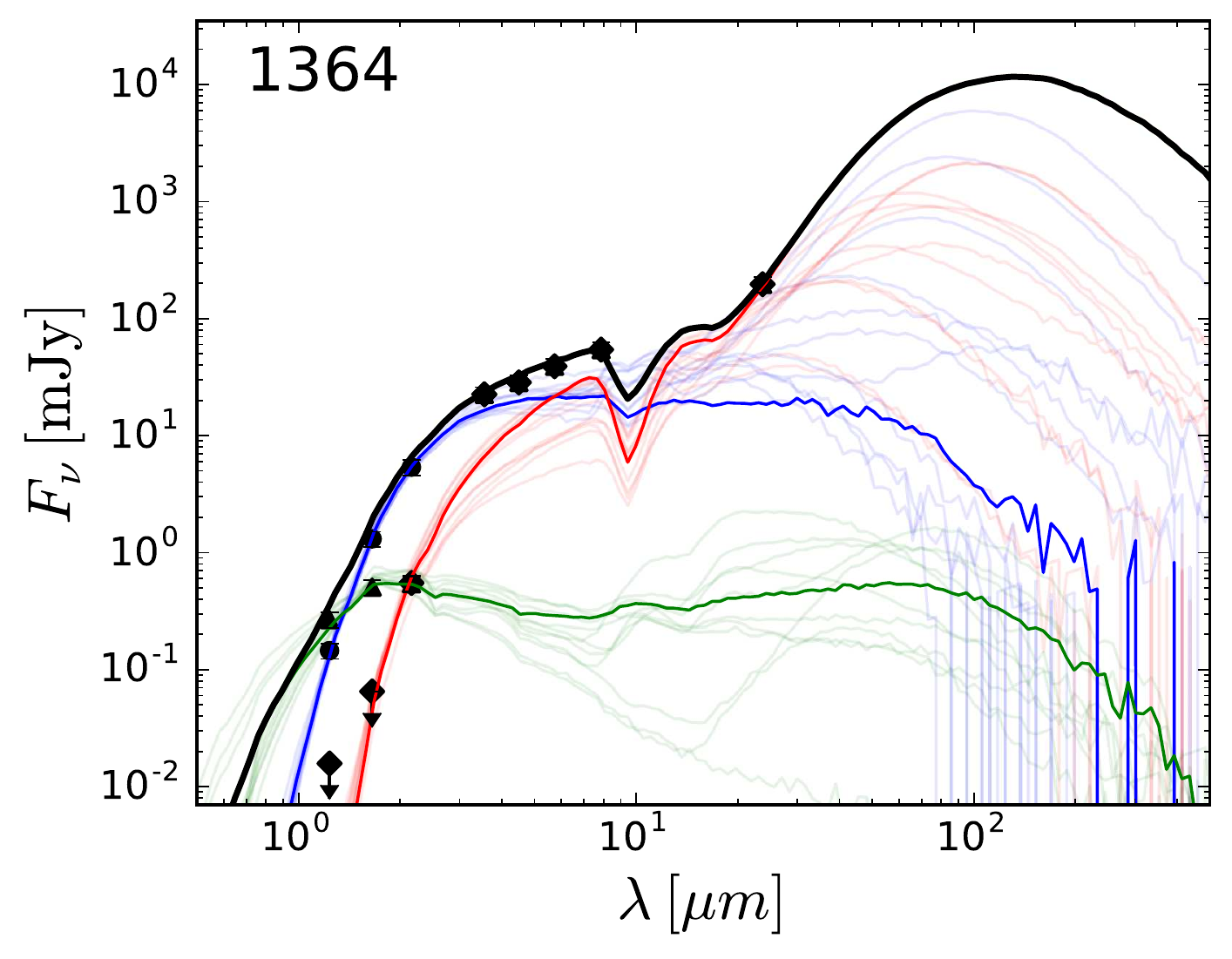}
\label{fig:sim_fit_1364_reg}
}
  \subfigure{
\includegraphics[scale=0.29,angle=0,trim=0cm 0cm 0cm 0cm,clip=true]{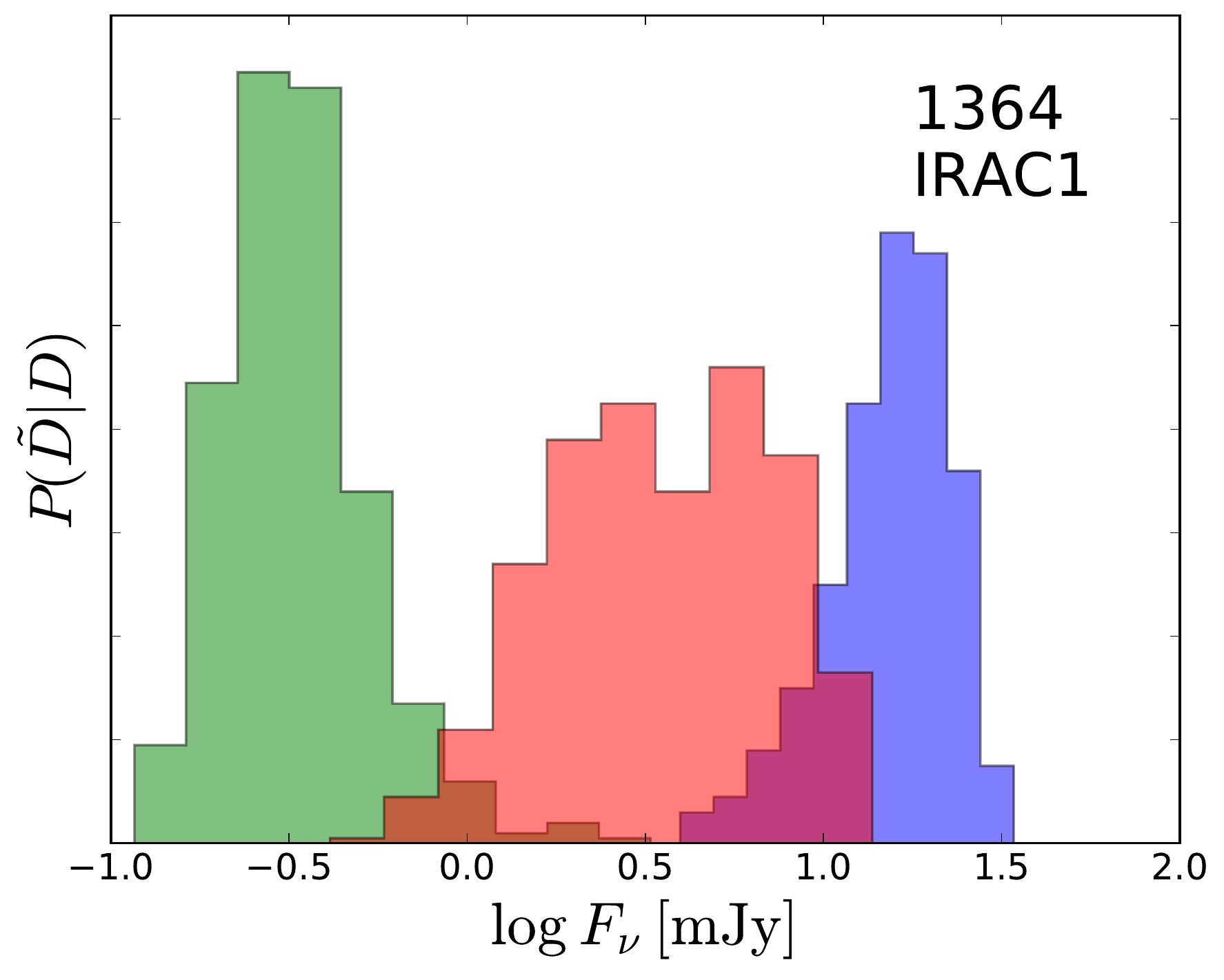}
\label{fig:post_pred_1364_reg}
}
  \subfigure{
\includegraphics[scale=0.38,angle=0,trim=0cm 0cm 0cm 0cm,clip=true]{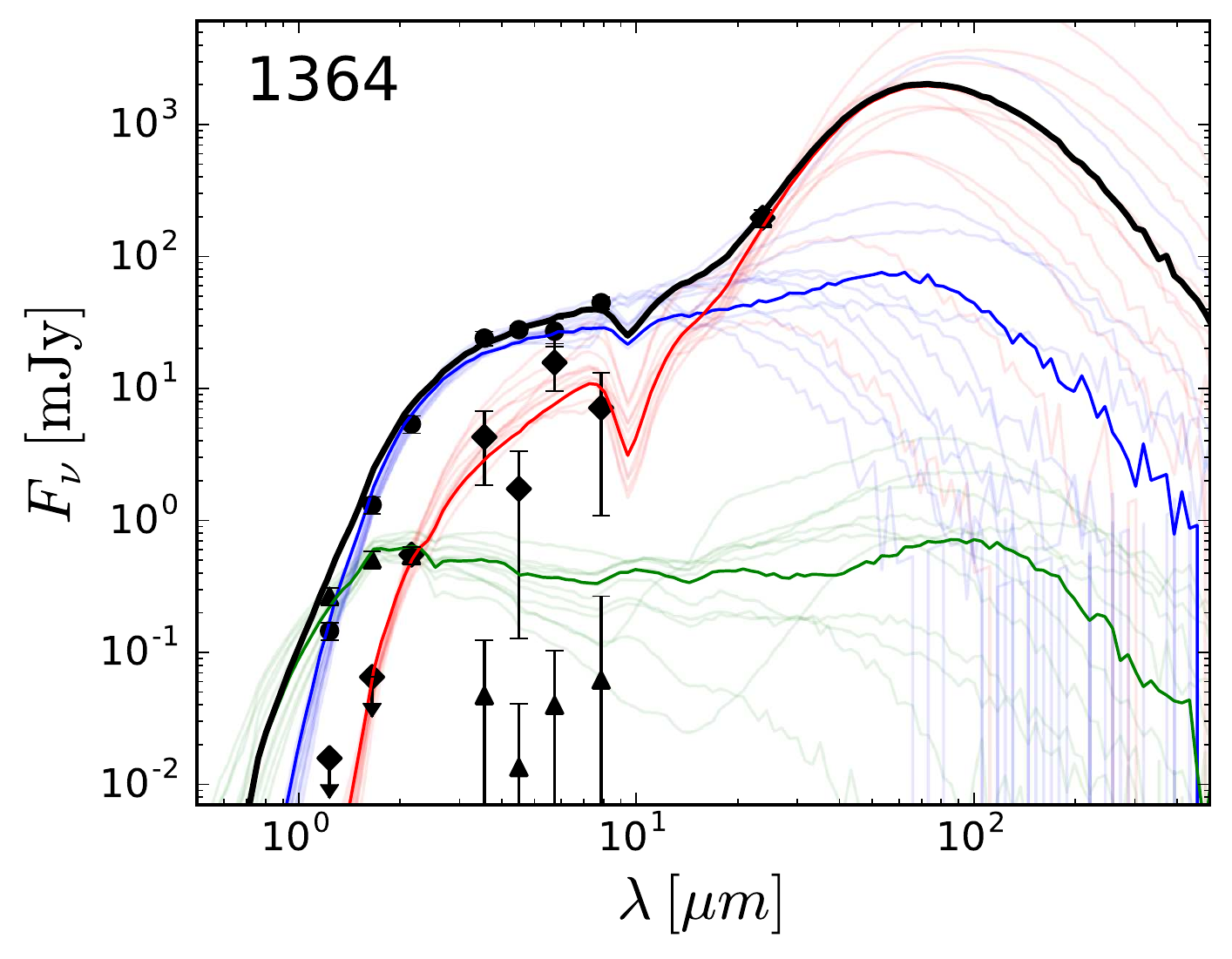}
\label{fig:sim_fit_1364_deblend}
}
  \caption{Simultaneous SED fitting of spatially blended sources in clusters 6307 (upper panels) and 1364 (lower panels). The left panels show the initial SED fit when only unresolved information is available from the IRAC bands. Each source is shown in a different color, and with different symbols for the measured photometry. The solid lines correspond to the MAP estimate, whereas the shaded lines correspond to the solutions within 1$\sigma$ of the best fit. The middle panels show the posterior predictives for the IRAC1 band fluxes derived from fitting the unresolved SEDs. The right panels show the resolved fit after constrains from the image fitting algorithm has been incorporated. The IRAC fluxes resulting from sampling the image fitting posteriors now appear as resolved measurements with associated error bars.}
\label{fig:sim_fit}
\end{figure}

To illustrate our SED fitting method, in Fig.~\ref{fig:sim_fit} we show fits to the photometry of two blended clusters (IDs 6307 and 1364), each containing three protostars. The left panels show the simultaneous fits to the resolved UKIDSS photometry and the unresolved GLIMPSE photometry of each of the individual sources in each cluster. The middle panels show the resulting posterior predictives (Eq.~\ref{eq:posterior_predictive}) for the flux in IRAC band 1. These quantify the uncertainties in the unresolved fluxes derived from our SED-fitting method that are later used as priors for the image fitting. The right panels show fits to the SEDs of the same objects once the predicted photometry from the image fitting algorithm is incorporated (step 3 in \S~\ref{sec:combining}). 

\begin{figure}[!t]
  \centering
  \subfigure{
\includegraphics[scale=0.43,angle=0,trim=0cm 0cm 0cm 0cm,clip=true]{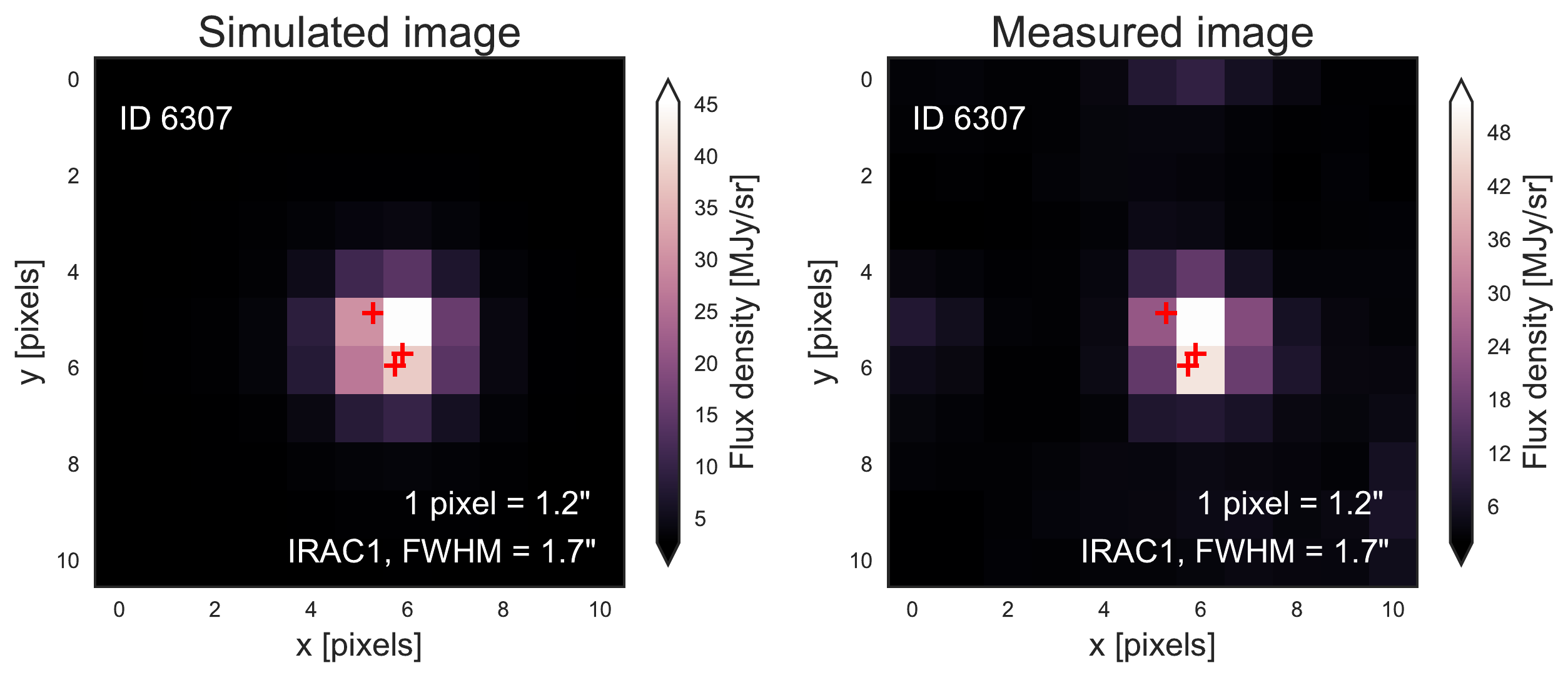}
\label{fig:post_6307}
}
  \subfigure{
\includegraphics[scale=0.32,angle=0,trim=0cm 0cm 0cm 0cm,clip=true]{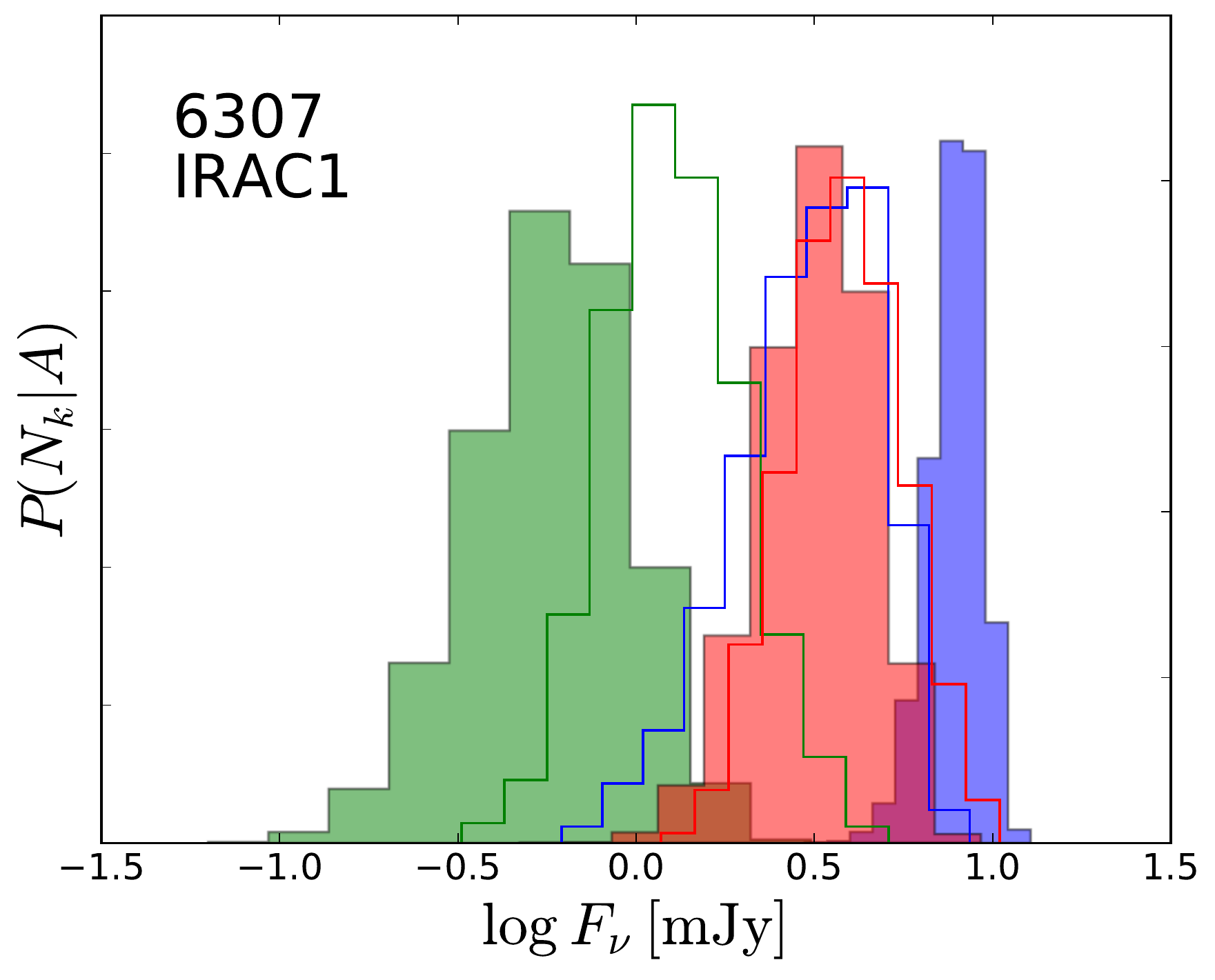}
\label{fig:pred_6307}
}
  \subfigure{
\includegraphics[scale=0.43,angle=0,trim=0cm 0cm 0cm 0cm,clip=true]{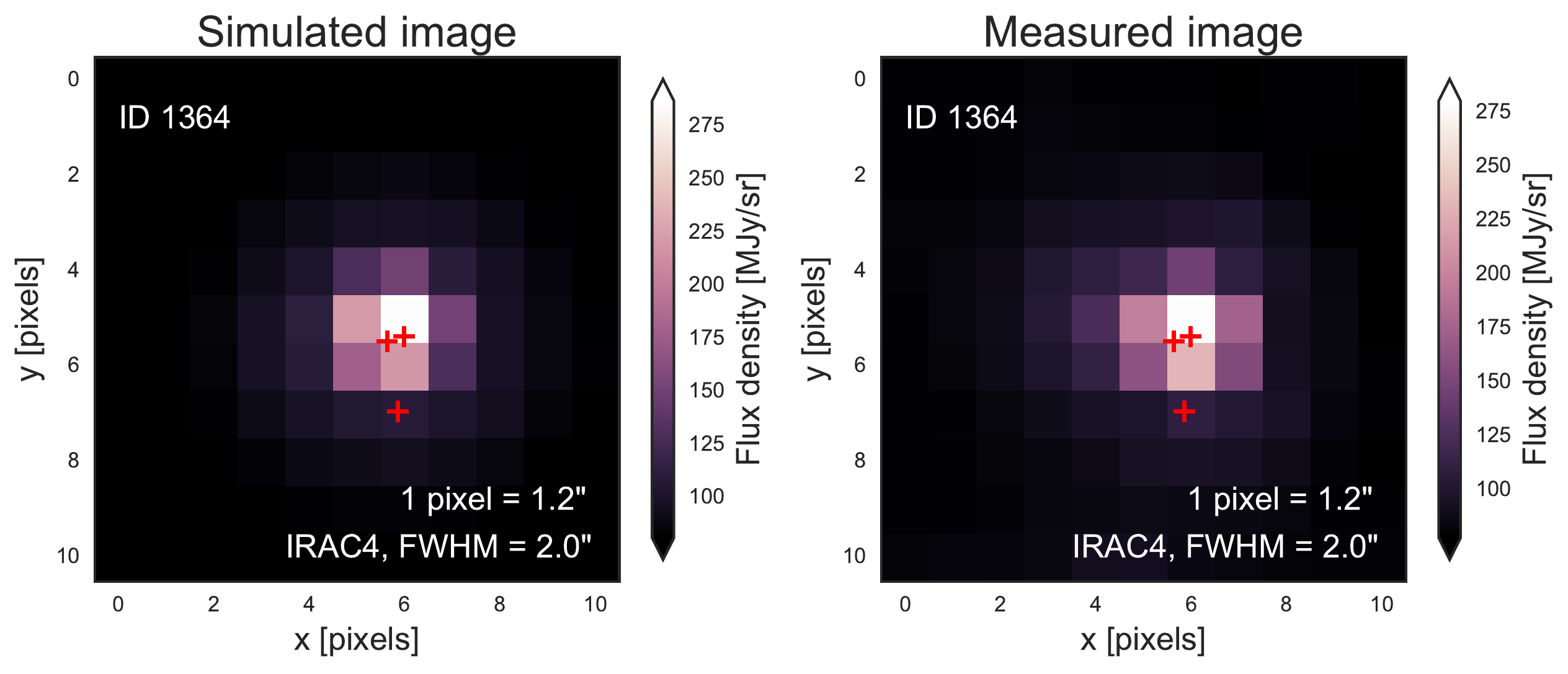}
\label{fig:post_1364}
}
  \subfigure{
\includegraphics[scale=0.32,angle=0,trim=0cm 0cm 0cm 0cm,clip=true]{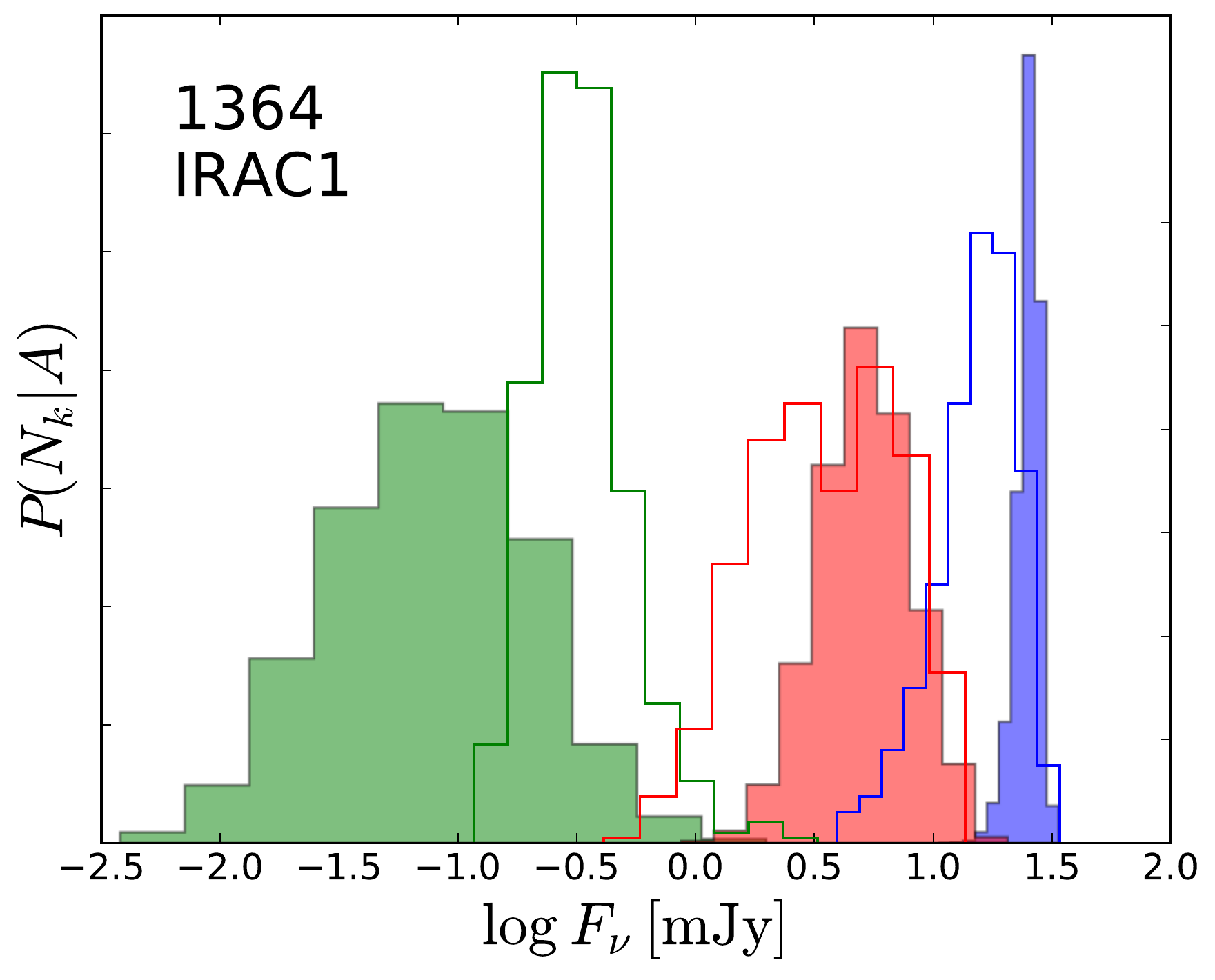}
\label{fig:pred_1364}
}
  \caption{\emph{Right panels:} MAP estimates from 2D image fitting for sources 6307 and 1364 in IRAC bands 1 and 4 respectively. \emph{Middle panels: } original measured images for the same sources in the corresponding bands, for source ID 6307 in IRAC bands 1 and 4. The red crosses indicate the location of the UKIDSS sources. \emph{Right panels:} Posterior distributions ($P(\{N_k\}|A)$) for the fluxes after image fitting. The color coding is the same as in Fig.~\ref{fig:sim_fit}. For comparison, the posteriors after the original SED fitting are shown as the solid-line histograms.}

\label{fig:simulated_images}
\end{figure}

Next, the posterior predictives in the middle panels of Fig.~\ref{fig:sim_fit} are used as priors for the 2D fitting of the IRAC images. Fig.~\ref{fig:simulated_images} shows the resulting fits to the data for sources 6307 and 1364, in IRAC bands 1 and 4 respectively. Also shown are the updated posteriors for the flux contribution for these two sources, after image fitting has been performed. The latter significantly reduces the variance in the estimated flux of the dominant sources, at the expense of larger variances for the dimmer sources, which do not contribute much to the total flux. Individual source flux densities are estimated by integrating over the modeled PRFs, for each source in a given cluster, and converted into mJy. The $1\sigma$ uncertainties in these flux densities are given by the 68\% percentiles of the posterior PDFs. The right panels of Fig.~\ref{fig:sim_fit} show the resulting fits when the updated resolved photometry is included. Similar results are obtained for most individual sources within the clusters.

The flux posteriors in the right panels of Fig.~\ref{fig:simulated_images} can now be re-interpreted as \emph{resolved} photometric measurements with associated uncertainties in each band and included in the SED fitting. The estimated fluxes in the IRAC bands and their uncertainties are listed Table~\ref{tab:sources}. Our predicted individual luminosities will soon be put to test using \emph{JWST} observations. With a spatial resolution 6 or 7 times larger than \emph{Spitzer}-IRAC, \emph{JWST}-MIRI will be able to resolve most of our clusters into individual YSOs, using filters centered at similar wavelengths, such as F560W and F770W. 

\begin{figure}
  \centering
  \subfigure[Blue SED in upper panels of Fig.~\ref{fig:sim_fit}]{
\includegraphics[scale=0.4,angle=0,trim=0cm 0cm 0cm 0cm,clip=true]{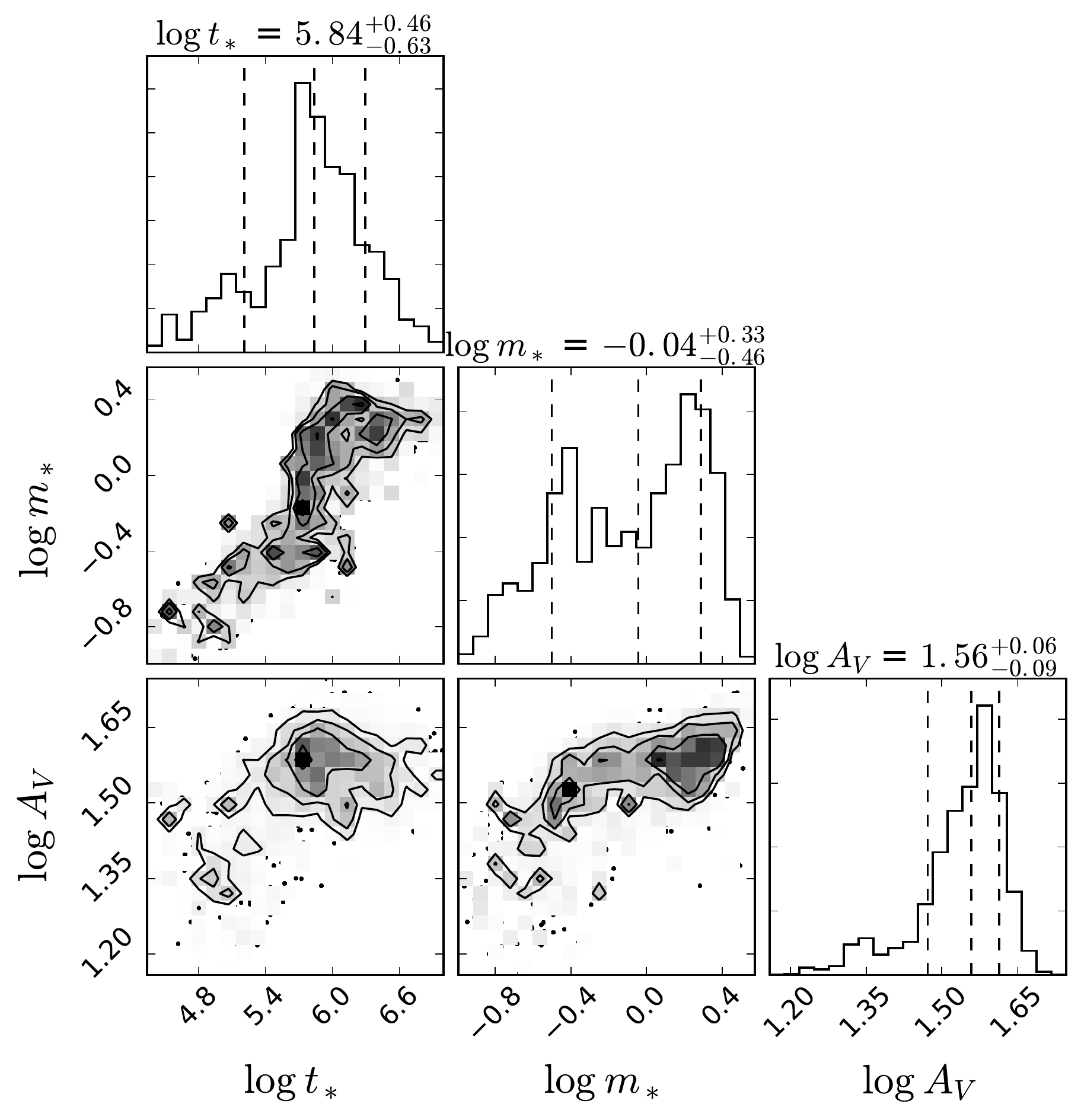}
\label{fig:corner_6307_s1_a}
}
  \subfigure[Green SED in upper panels of Fig.~\ref{fig:sim_fit}]{
\includegraphics[scale=0.4,angle=0,trim=0cm 0cm 0cm 0cm,clip=true]{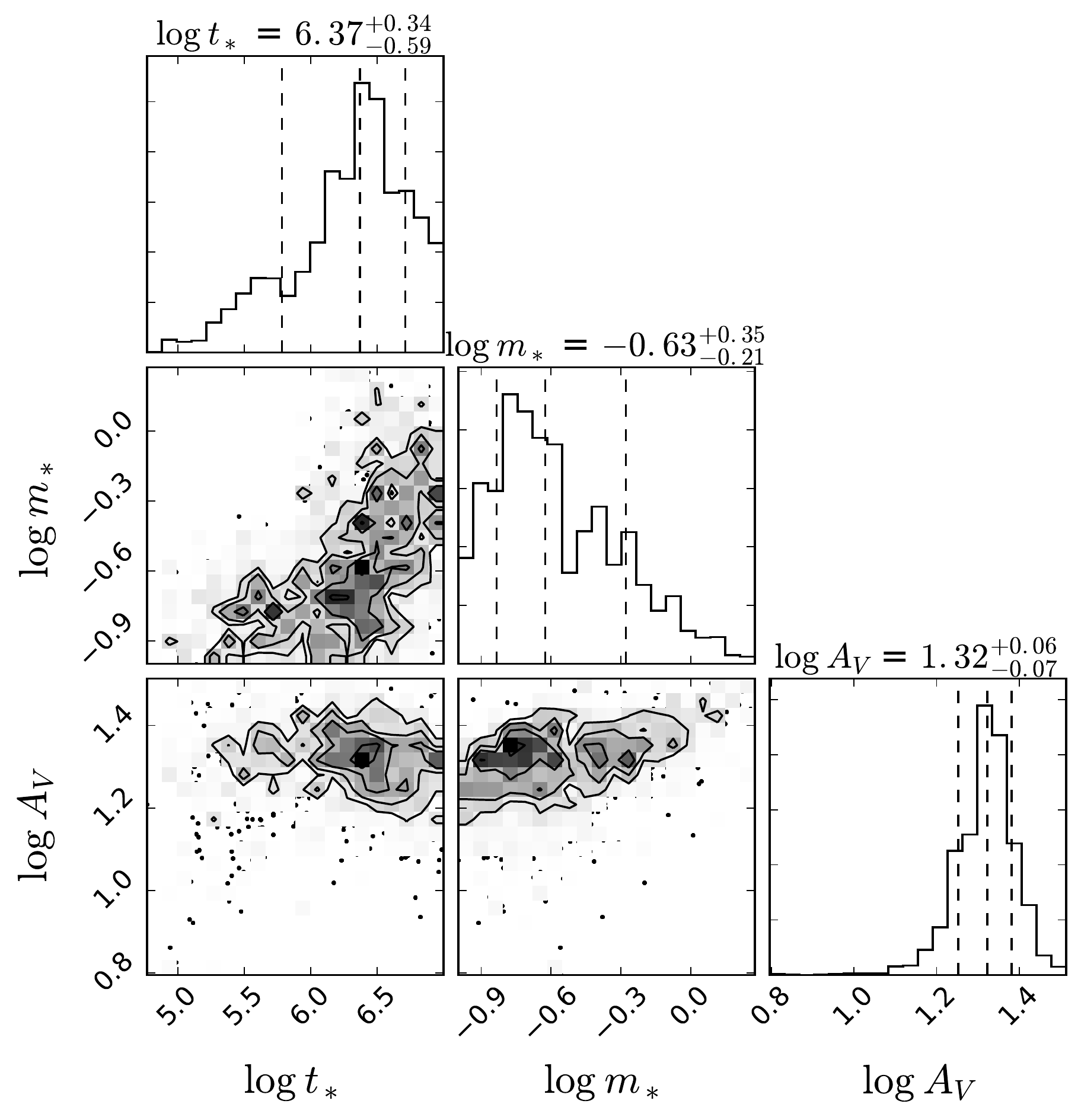}
\label{fig:corner_6307_s2_a}
}
  \subfigure[Red SED in upper panels of Fig.~\ref{fig:sim_fit}]{
\includegraphics[scale=0.4,angle=0,trim=0cm 0cm 0cm 0cm,clip=true]{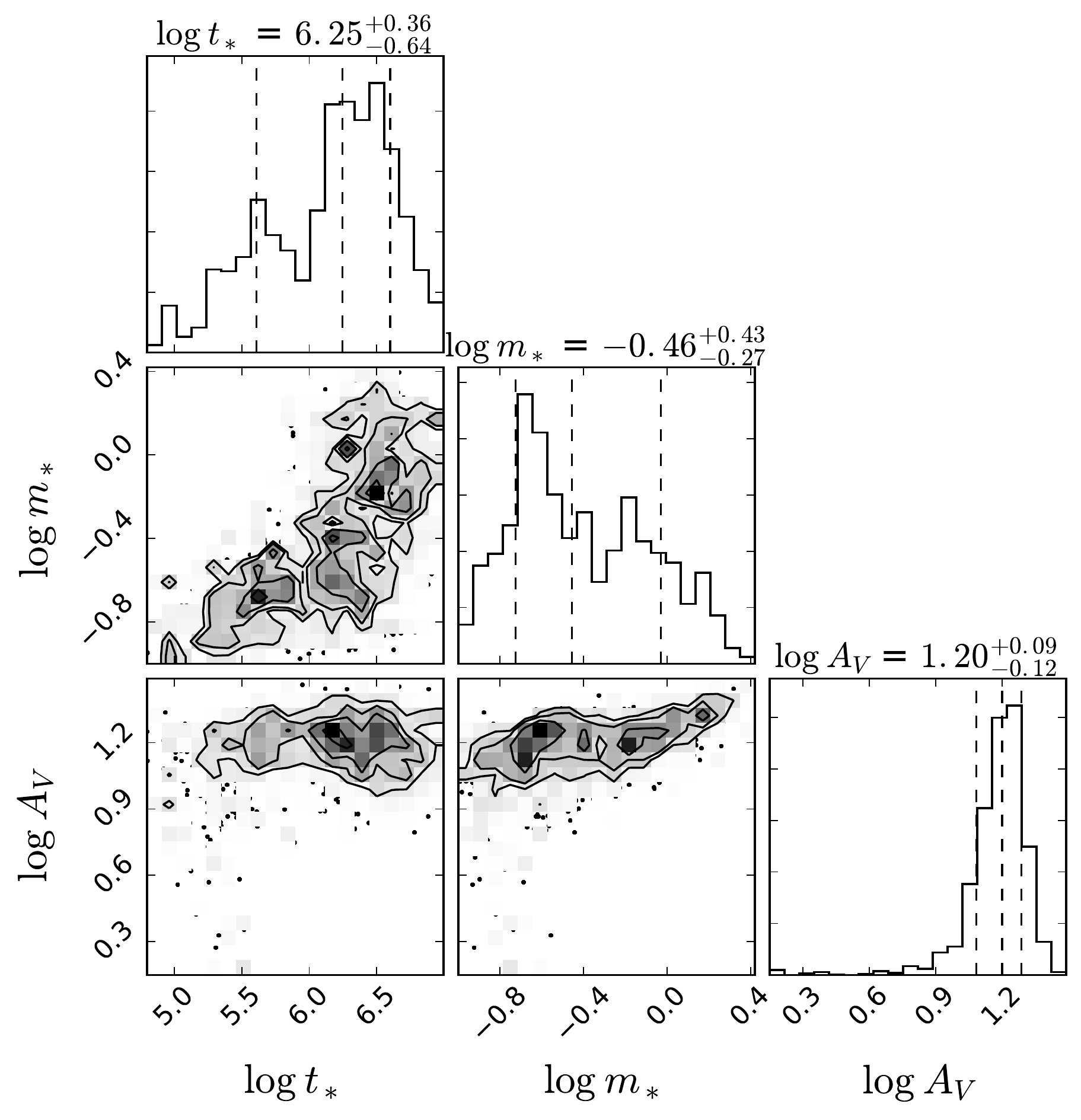}
\label{fig:corner_6307_s3_a}
}
  \caption{The posterior probability distributions of the model parameters for all three sources in cluster 6307, labeled according to the color of their SED in the upper panels of Fig.~\ref{fig:sim_fit}. These are the final posteriors after the resolved fluxes from the image analysis have been included for the IRAC bands. The gray scale matches the probability density and the marginalized probabilities are also shown. Vertical dashed lines correspond to the 0.25, 0.5 and 0.75 quantiles.}
\label{fig:2d_posteriors_a}
\end{figure}

An example of the final model parameter posteriors can be seen in Fig.~\ref{fig:2d_posteriors_a} for cluster 6307. Plotted are the posteriors for stellar mass ($m_*$), age ($t_*$), and visual extinction ($A_V$) for each of the three individual sources. For each YSO in the cluster, mass and age are degenerate, and the marginalized probabilities typically show two possible solutions, with one of the two probability maxima being significantly more prominent. It would be extremely hard to spot these two solutions for individual unresolved YSOs using conventional SED fitting methods. The estimated visual extinctions are well constrained and typically have scatters of 0.3-0.4 dex, and mean values below 40~mag.

For 6307, our results indicate similar evolutionary stages for all three YSOs composing the cluster. The MAP estimates for the individual ages are all within a 0.6~Myr range centered around 6~Myr. More generally, age uncertainties are still significant (of the order of $\pm 0.5$~Myr or more), and coeval birth of the three protostars can not be assumed based on this evidence alone. We discuss the likelihood of coeval birth within individual clusters in \S~\ref{sec:massive_first}.

For YSOs in a given cluster, the near-infrared fluxes from UKIDSS do not necessarily correlate with YSO mass. A significant fraction of the stellar luminosity is reprocessed and re-emitted at MIR wavelengths. Correctly associating NIR sources with MIR fluxes is crucial to estimate the intrinsic luminosities of each member, their masses and the amount of obscuration due to dust absorption in each case. Our method can do exactly that. For example, cluster 1364 contains a dim, embedded NIR source (red SED in the lower panels of Fig.~\ref{fig:sim_fit}) that nevertheless dominates the IRAC and MIPS bands, and which is significantly more massive than what would be expected from its UKIDSS fluxes only. Given the spatial projected proximity of this source to one other member of the cluster, it would have been very difficult to estimate its mass using conventional techniques.

\newpage
\subsection{Masses, ages, and visual extinction of clustered YSOs}
We now describe the overall statistics of the derived parameters for the entire set of 70 low-mass YSO clusters, the cluster-to-cluster variations as a function of location on the galactic disc, and the dispersion of physical properties within individual clusters.

\medskip
\medskip

\subsubsection{Completeness}
Stars with masses below a few solar masses are undetected by the current surveys for most of the heliocentric distance range considered here, and our sample is therefore incomplete. Completeness is not required, however, for the analysis that follows, and future observations with more sensitive instruments can use these same techniques to unravel even less massive protostars in clusters. Our goal here is to investigate how the properties of the most massive members in each cluster inform about how star formation proceeds in clustered environments and what the implication of the detected distributions are for studies of the IMF and the different models of star formation.

\begin{figure}
\centering
\includegraphics[scale=0.55,angle=0,trim=0cm 0cm 0cm 0cm,clip=true]{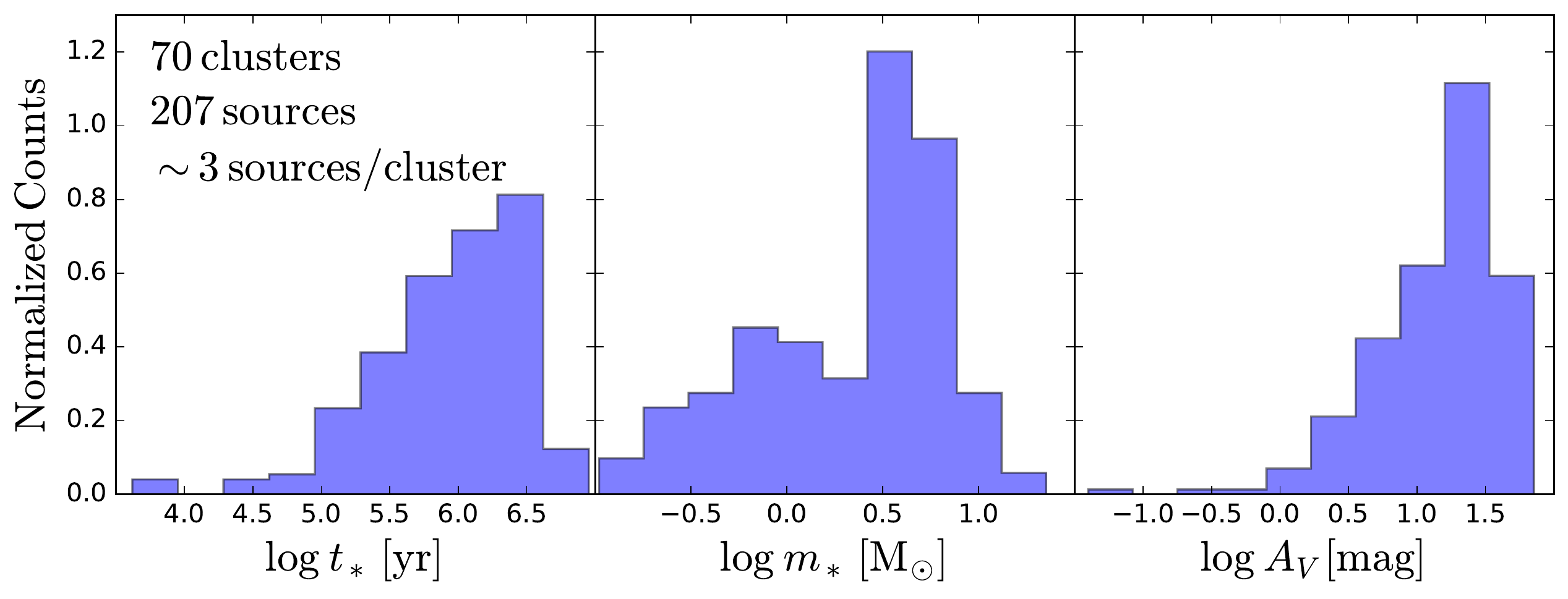}
\caption{Histograms of source properties derived from the combined SED/imaging fitting. Shown are the 50\% quantile values for YSO age, stellar mass, and visual extinction. All 207 individual YSOs making up the 70 clusters are used to generate these histograms.}
\label{fig:params_hist}
\end{figure}

Fig.~\ref{fig:params_hist} shows histograms of the YSO masses, ages, and visual extinctions of all 207 individual sources that make up the 70 studied clusters. The values shown correspond to the 50th percentile from the MCMC sample chains. On average, there are 3 detected sources per cluster, and clusters with two detected members are significantly more common (28 out of 70 clusters) than multiple clusters. The measured mass distribution is bimodal with a main peak at $m_* \sim 3.0~M_{\odot}$ and a secondary peak at  $m_* \sim 0.8~M_{\odot}$. The maximum derived mass is $m_* \sim 19~M_{\odot}$. 

The age distribution is consistent with a relatively evolved population, i.e., most sources have derived ages compatible with them being class II or class III YSOs, and only 11 out of the 207 YSOs being $10^5~$yr or younger. This is not surprising given the fact that we have selected sources with clear NIR detections. The distribution peaks at 2.5~My and is more skewed towards younger ages. There is a very steep drop-off at long ages, confirming that these young clusters had a relatively firm total age.

The large majority of clusters are located in regions of high visual extinction relative to the average $A_V$ for individual stars at the same distances, as derived from the Pan-STARRS data. For example, at distances of 15~kpc or less, \citet{Green15} derive $E(B-V)$ values in the galactic plane that are consistent with $A_V < 10$~mag, whereas our distribution of $A_V$ values peaks at about 20~mag.  Such increased extinction towards the YSO clusters is expected, and confirms that clustered star formation occurs in regions that are significantly more embedded than neighboring field stars; the broad distribution of $A_V$ for these young stars is perhaps of more interest. The fact that extinctions higher than about $A_V = 50$ are not detected is probably a selection effect, since those sources are harder to detect in the NIR.

\subsubsection{Cluster-to-cluster properties}
\begin{figure}
\centering
  \subfigure{
\includegraphics[scale=0.45,angle=0,trim=0cm 0cm 0cm 0cm,clip=true]{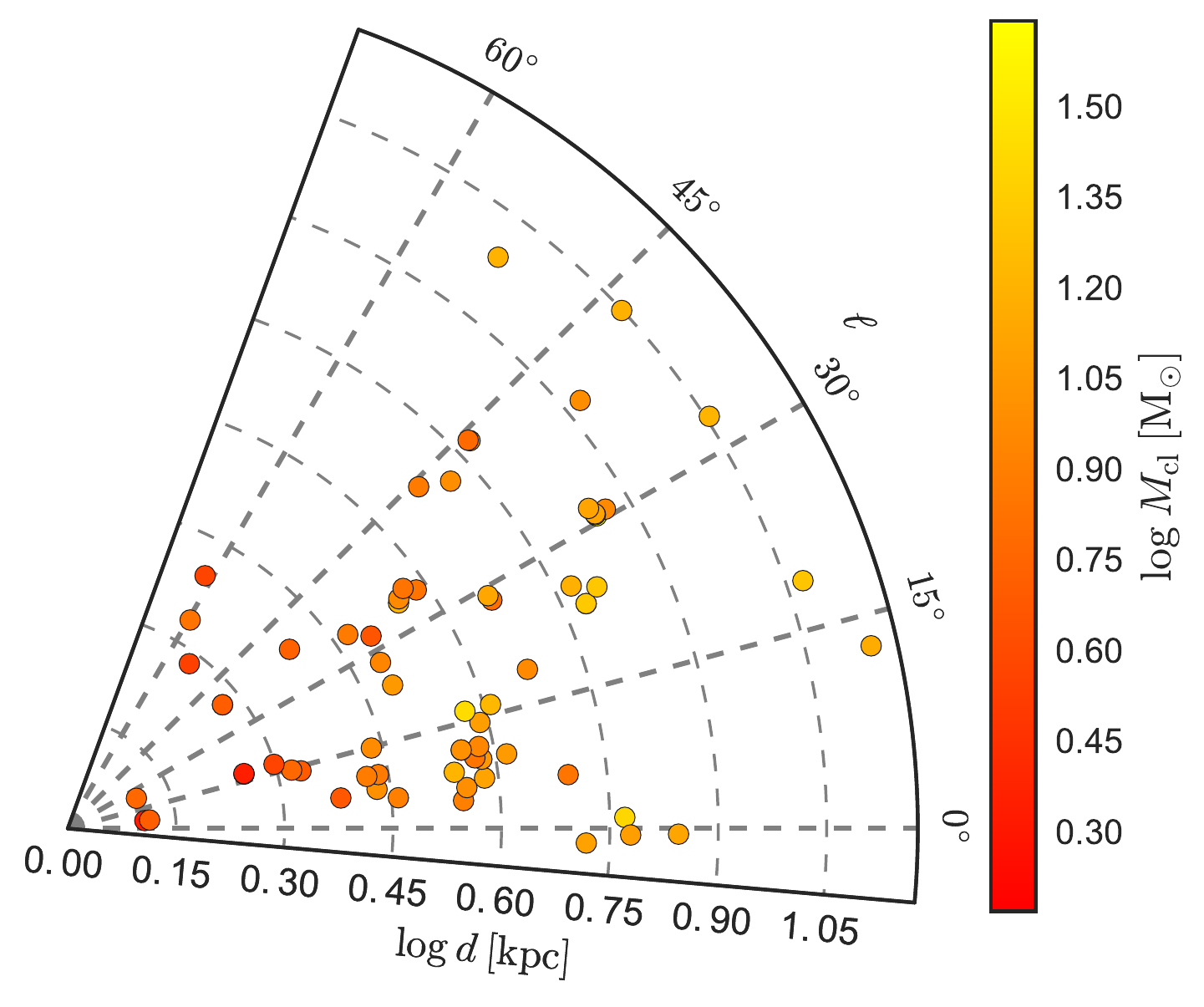}
\label{fig:mass_relation2}
}
  \subfigure{
\includegraphics[scale=0.45,angle=0,trim=0cm 0cm 0cm 0cm,clip=true]{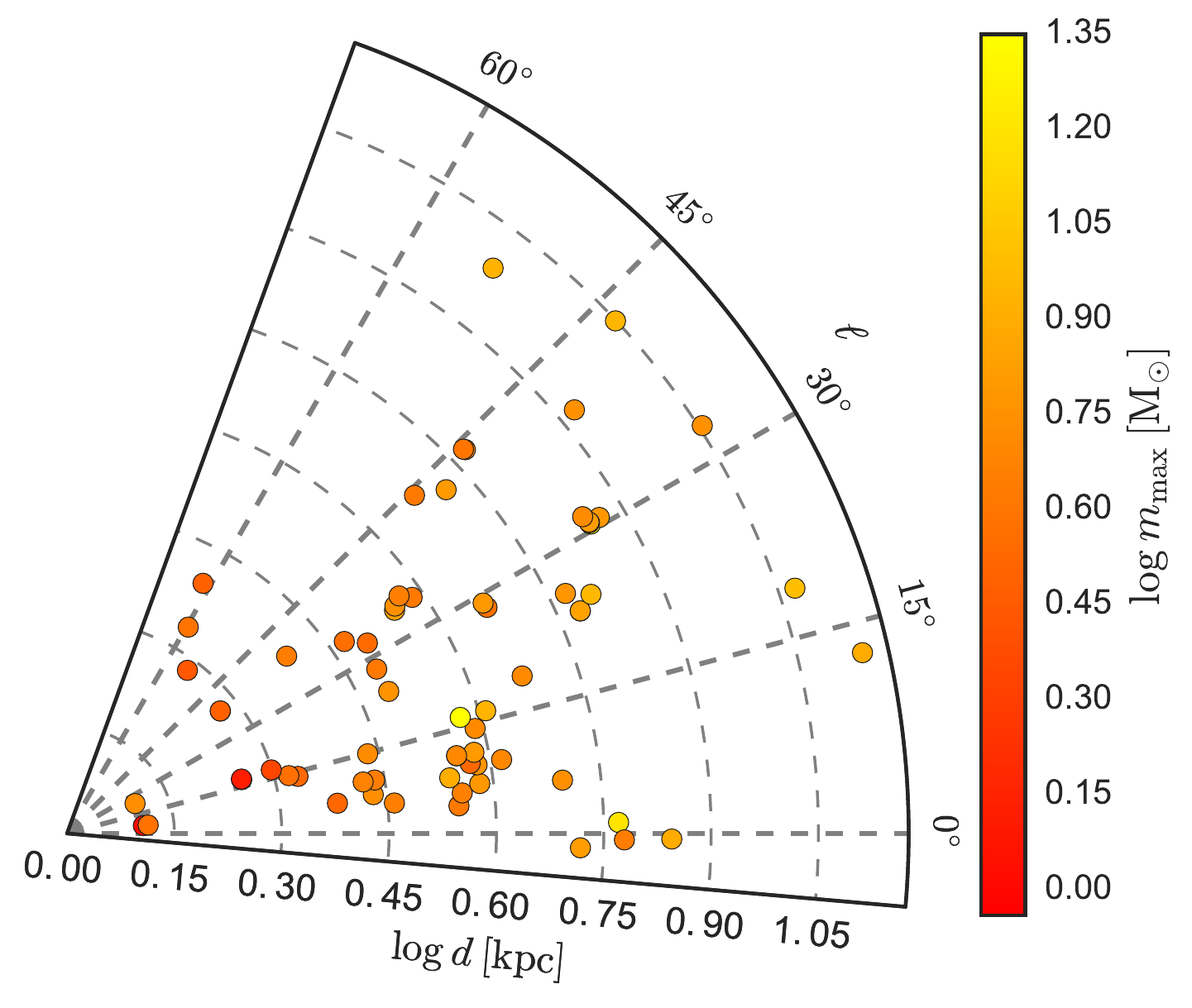}
\label{fig:mass_relation3}
}
  \subfigure{
\includegraphics[scale=0.45,angle=0,trim=0cm 0cm 0cm 0cm,clip=true]{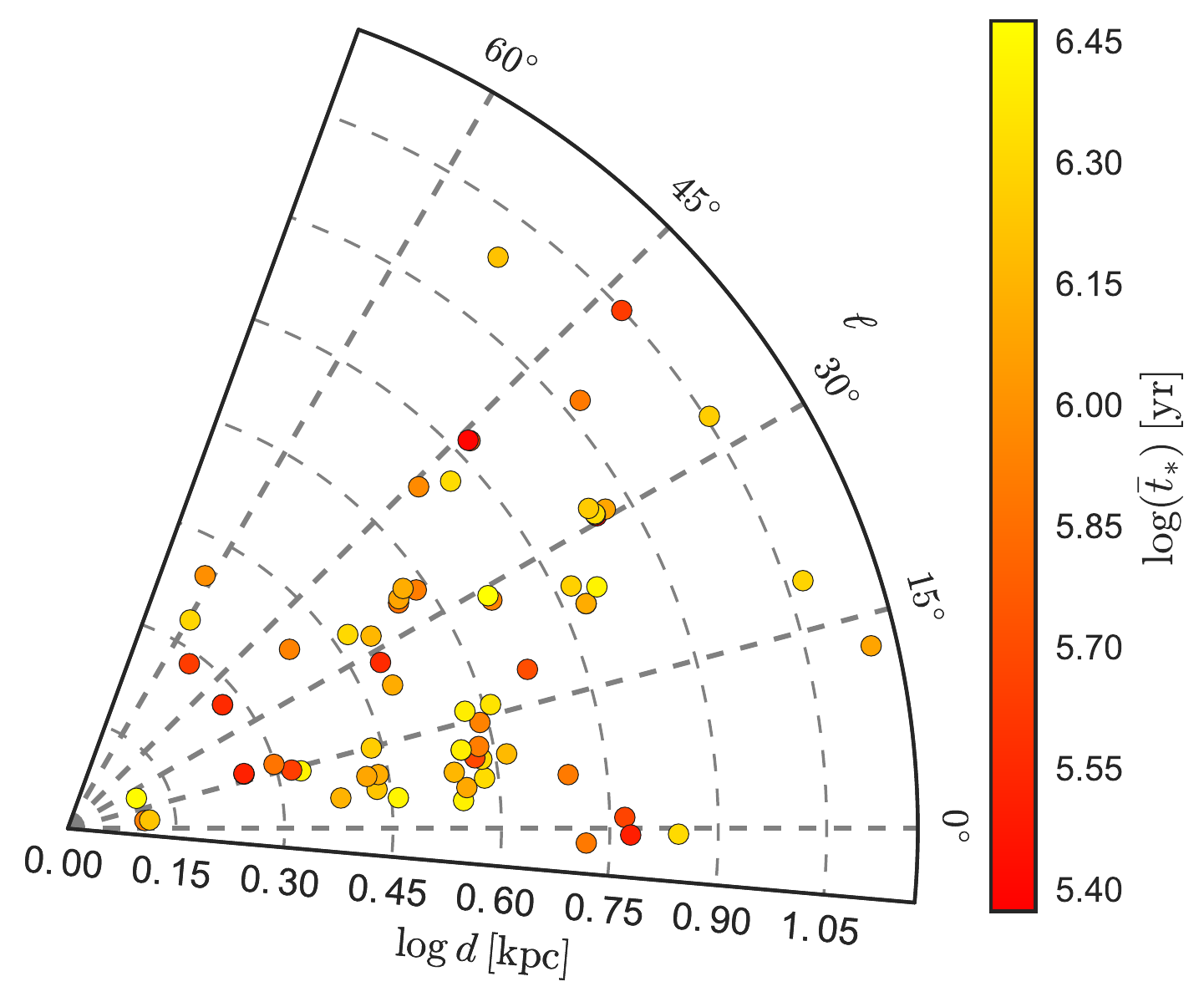}
\label{fig:mass_relation4}
}
  \subfigure{
\includegraphics[scale=0.45,angle=0,trim=0cm 0cm 0cm 0cm,clip=true]{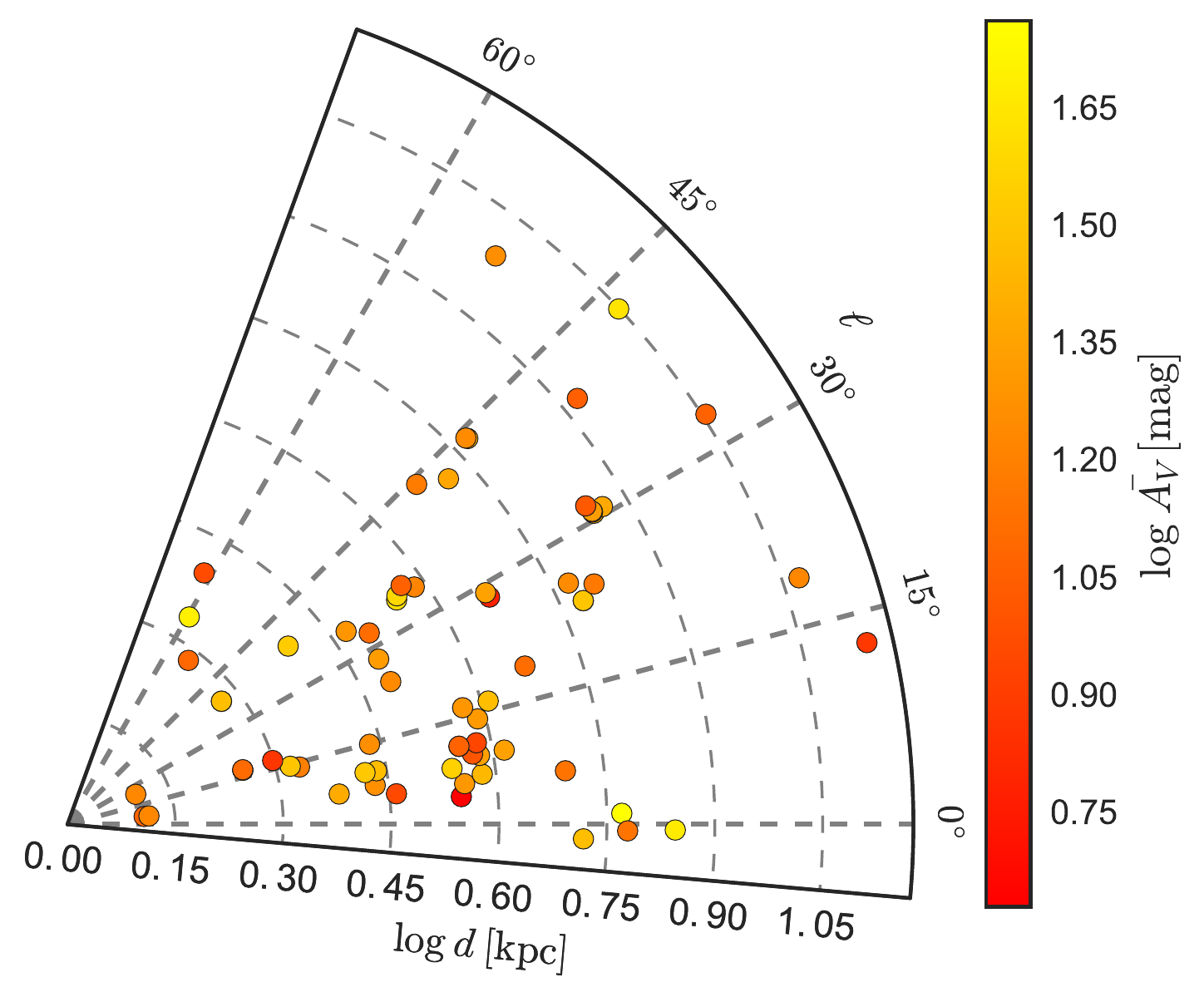}
\label{fig:mass_relation5}
}
\caption{Projected locations of the studied clusters onto the galactic plane with colors indicating different derived properties. The upper two panels show the total stellar mass of the clusters and the mass of the most massive member in each of them. The lower two panels show the average ages and the average visual extinctions for each cluster. The grid is centered on the Sun.}
\label{fig:spatial_params}
\end{figure}

Fig.~\ref{fig:spatial_params} shows the projected location of our YSO clusters onto the galactic plane, color coded respectively by the derived MAP values of \emph{a)} $\log M_{\rm{cl}}$, \emph{b)} $\log m_{\rm{max}}$, \emph{c)} $\log t_*$ and \emph{d)} $\log A_V$. These results hint at a weak correlation between the masses derived and the heliocentric distance to the sources, which is most likely the result of a bias introduced by the  sensitivity of the observations.
% * <hsmith@cfa.harvard.edu> 2018-02-13T18:34:52.099Z:
%
% ^.
About 90\% of the studied clusters are closer to us than half the distance to the most distant cluster in the sample. The most massive YSO detected (located in cluster 3568), with a mass of $22~M_{\odot}$, is at a distance of 3.7~kpc, just beyond the Sagittarius arm at a galactic longitude of $\mathcal(l)\sim 16^{\circ}$. This is relatively close compared with the most distant cluster, located at almost 15~kpc from us. This same source is also one of the least evolved Class I sources. Its 24~\mum\ flux density is almost an order of magnitude higher that that of other cluster members. The fact that the most massive source detected as part of a cluster is not particularly far away indicates that selection effects of distance on measured masses, at least within this range, are not dramatic. We nevertheless correct for this effect when we derive a $M_{\rm{cl}}-m_{\rm{max}}$ correlation in \S~\ref{sec:correlation} 

As for the spatial distribution of evolutionary stages, cluster 5976 has the youngest age estimated as the mean of the individual YSO ages. It has 5 members and is only 3 arcsecs apart from a maser source identified in \citet{Szymczak05}, in the far end of the 3kpc arm, a clear indication of massive star formation taking place in the region. The most evolved cluster is 5686, which also contains 5 members and is located in a line of sight not too far away from that to 5476, but closer to us, at about 1~kpc from the point where the Scutum-Centaurus arm meets the galactic bar. The spatial distribution of visual extinctions is also shown in Fig.~\ref{fig:spatial_params}. Cluster 360, located on the near side of the 3~kpc arm has the highest derived optical extinction ($A_V = 58$), and also neighbors ($0.53^{\prime\prime}$) a millimetric compact source containing a maser \citep{Caswell10, Urquhart13}.  
% * <hsmith@cfa.harvard.edu> 2018-02-13T18:37:05.915Z:
%
% ^.

\begin{figure}
\centering
  \subfigure{
\includegraphics[scale=0.45,angle=0,trim=0cm 0cm 0cm 0cm,clip=true]{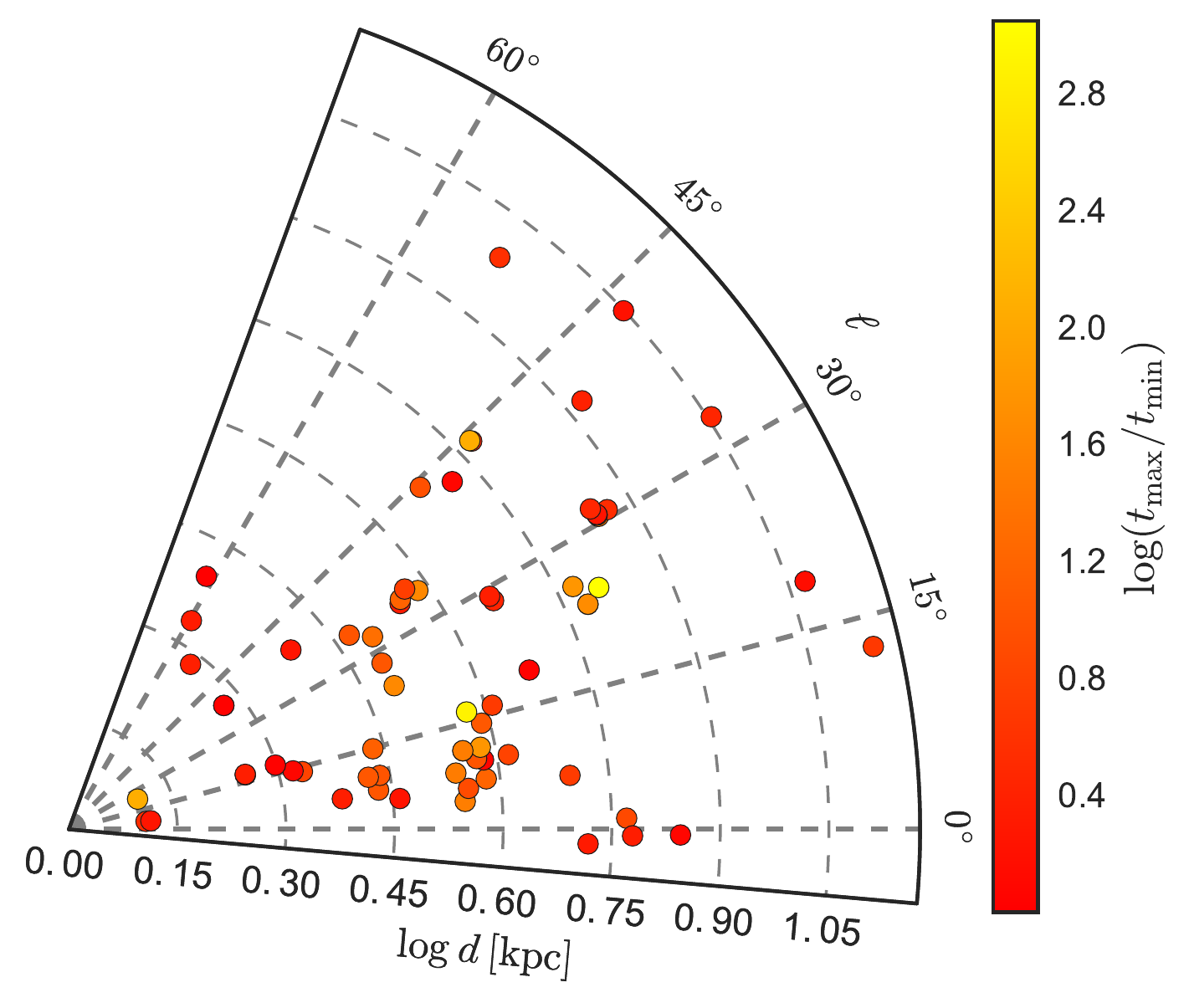}
\label{fig:mass_relation6}
}
  \subfigure{
\includegraphics[scale=0.45,angle=0,trim=0cm 0cm 0cm 0cm,clip=true]{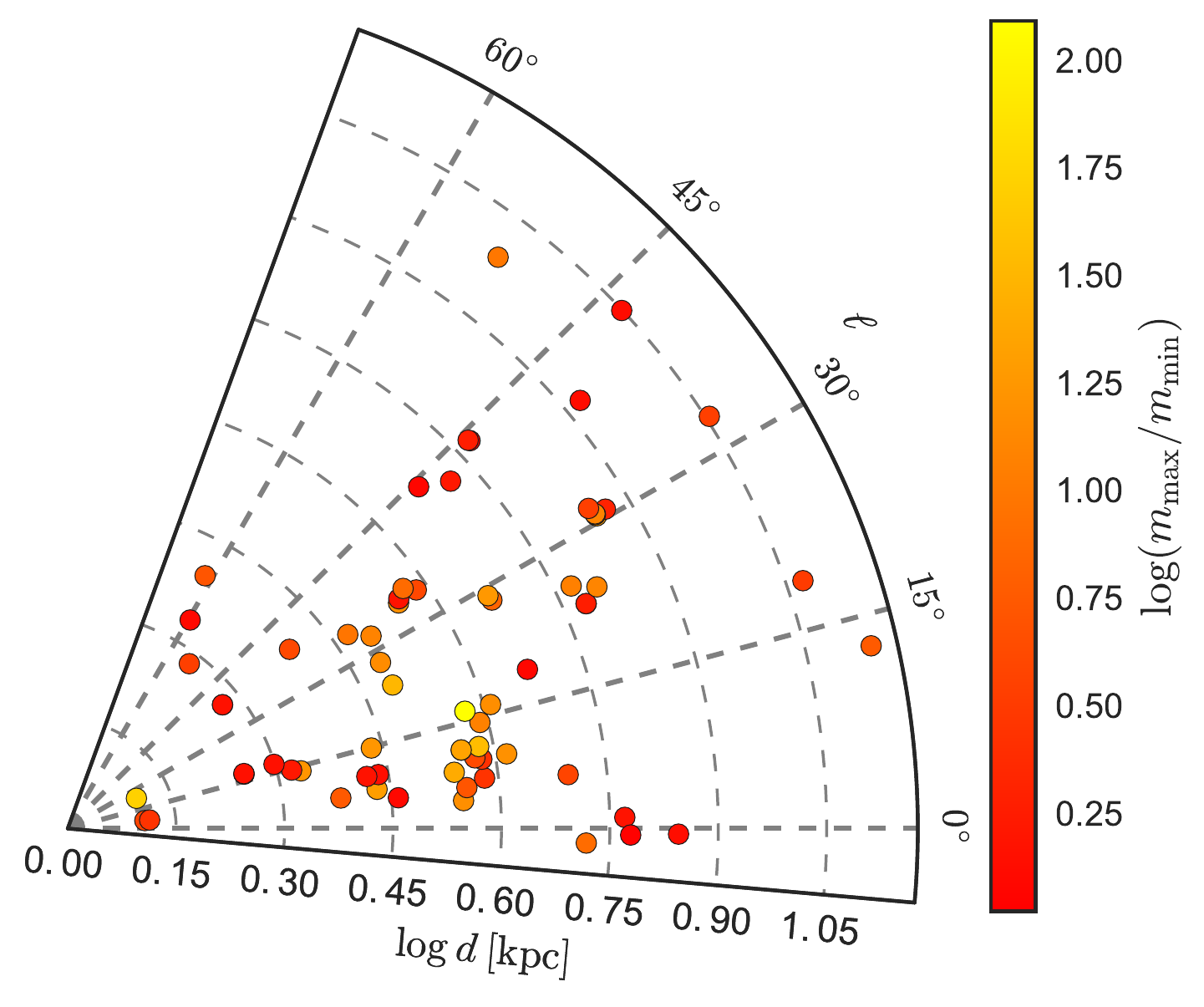}
\label{fig:mass_relation7}
}
\caption{Projected locations of the studied clusters onto the galactic plane with colors indicating ranges of stellar ages (left panel) and masses (right panel). The color code indicates the ratio between the highest and lowest value of the individual components of the cluster in each case.}
\label{fig:more_spatial_params}
\end{figure}

\subsubsection{Within-cluster gradients}
\label{sec:within_cluster}
Fig.~\ref{fig:more_spatial_params} illustrates the spatial distribution of mass and age dispersions within each cluster. The majority of clusters have small age dispersions, as measured by the difference between the ages of the oldest and the youngest star ($t_{\rm{max}}<2.5t_{\rm{min}}$), hinting at a relatively quick (although not necessarily coeval) formation of all detected members within the cluster. YSO age dispersion is less than a factor of 3 in 60\% of the clusters, and more than a factor of 10 in only 13\% of the cases. At least two clusters (4955 and 2568) show a significant age dispersion ($t_{\rm{max}}>400t_{\rm{min}}$), but those are cases where a few relatively evolved sources (a few million years old or so) are in the same cluster with a single, considerably embedded Class I young source.

\begin{figure}
\centering
  \subfigure[]{
\includegraphics[scale=0.55,angle=0,trim=0cm 0cm 0cm 0cm,clip=true]{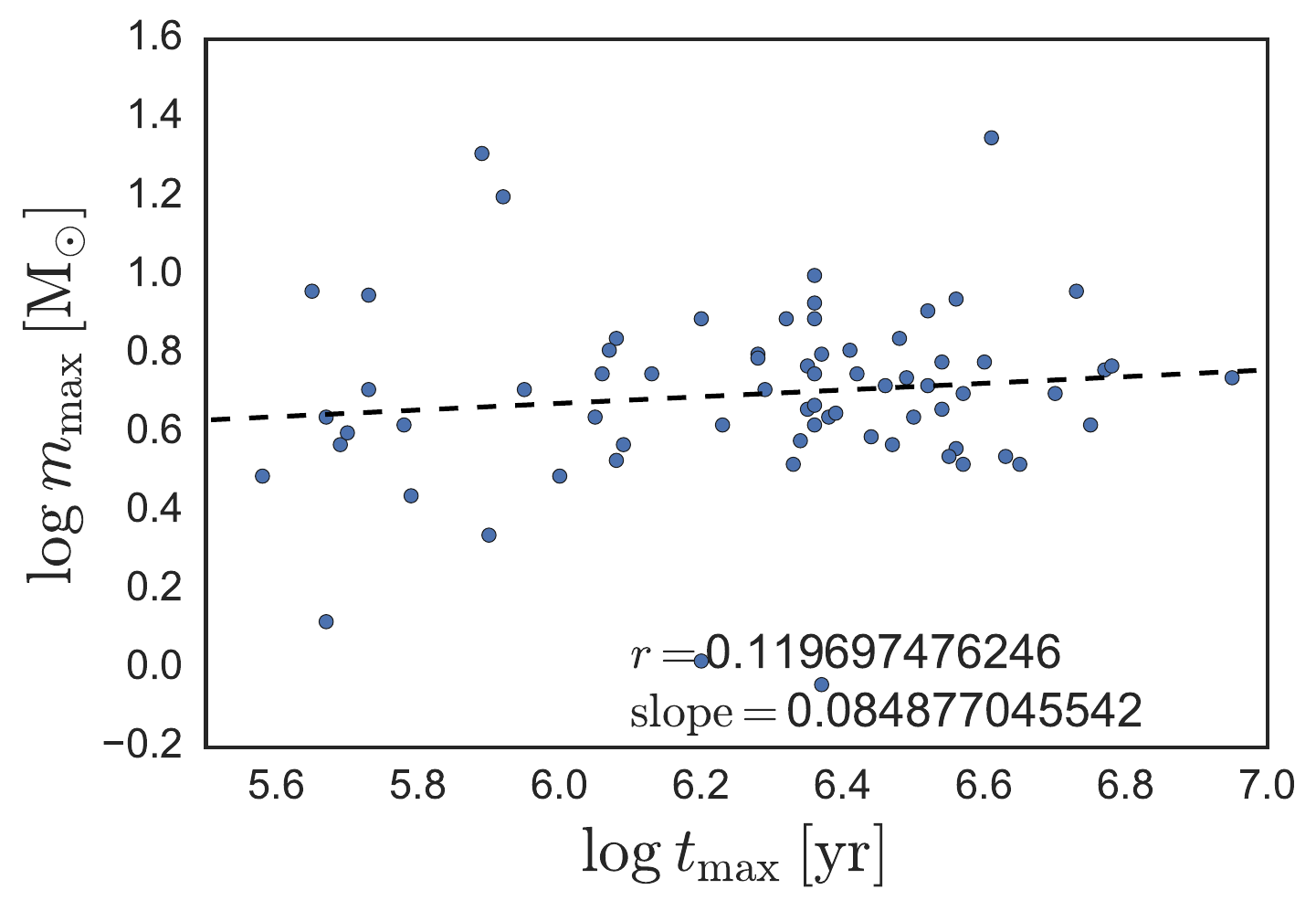}
\label{fig:trend1}
}
  \subfigure[]{
\includegraphics[scale=0.55,angle=0,trim=0cm 0cm 0cm 0cm,clip=true]{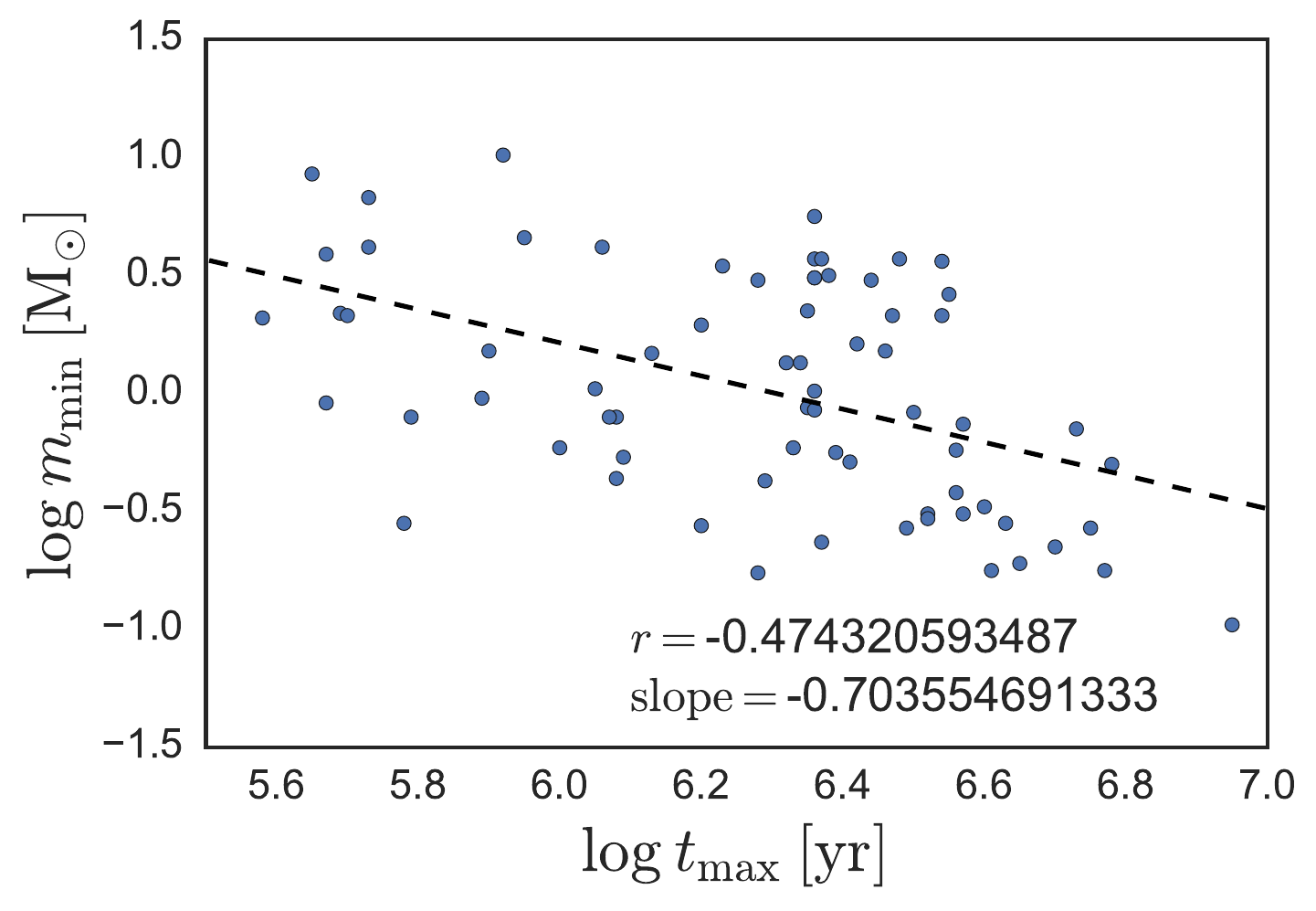}
\label{fig:trend2}
}
\caption{Relationship between the age of the oldest member in each cluster and \emph{a)} the most massive member, and \emph{b)} the least massive member. The dotted line are linear fits to the data, with parameters of the fit indicated}
\label{fig:trends}
\end{figure}

Our results from this modest sample indicate a weak positive correlation between the age of the oldest member in a given cluster ($t_{\rm{max}}$) and the mass of its most massive member ($m_{\rm{mass}}$), as well as a negative correlation between $t_{\rm{max}}$ and the mass of the cluster's least massive member, as shown in Fig.~\ref{fig:trends}. The correlations have the approximate forms $m_{\rm{max}} \propto t_{\rm{max}}^{0.1}$ and $m_{\rm{min}} \propto t_{\rm{max}}^{-0.7}$. As we will discuss in \S~\ref{sec:massive_first}, this has interesting implications in terms of the cluster accretion history and the effect of dynamical the evolution of the cluster in stopping this accretion. We point out for now that these correlations hold across 1.5 orders of magnitude in $t_{\rm{max}}$  corresponding roughly to YSOs of classes II and III.

\subsubsection{The $M_{\rm{cl}}-m_{\rm{max}}$ correlation}
\label{sec:correlation}

\begin{figure}
\centering
\includegraphics[scale=0.7,angle=0,trim=0cm 0cm 0cm 0cm,clip=true]{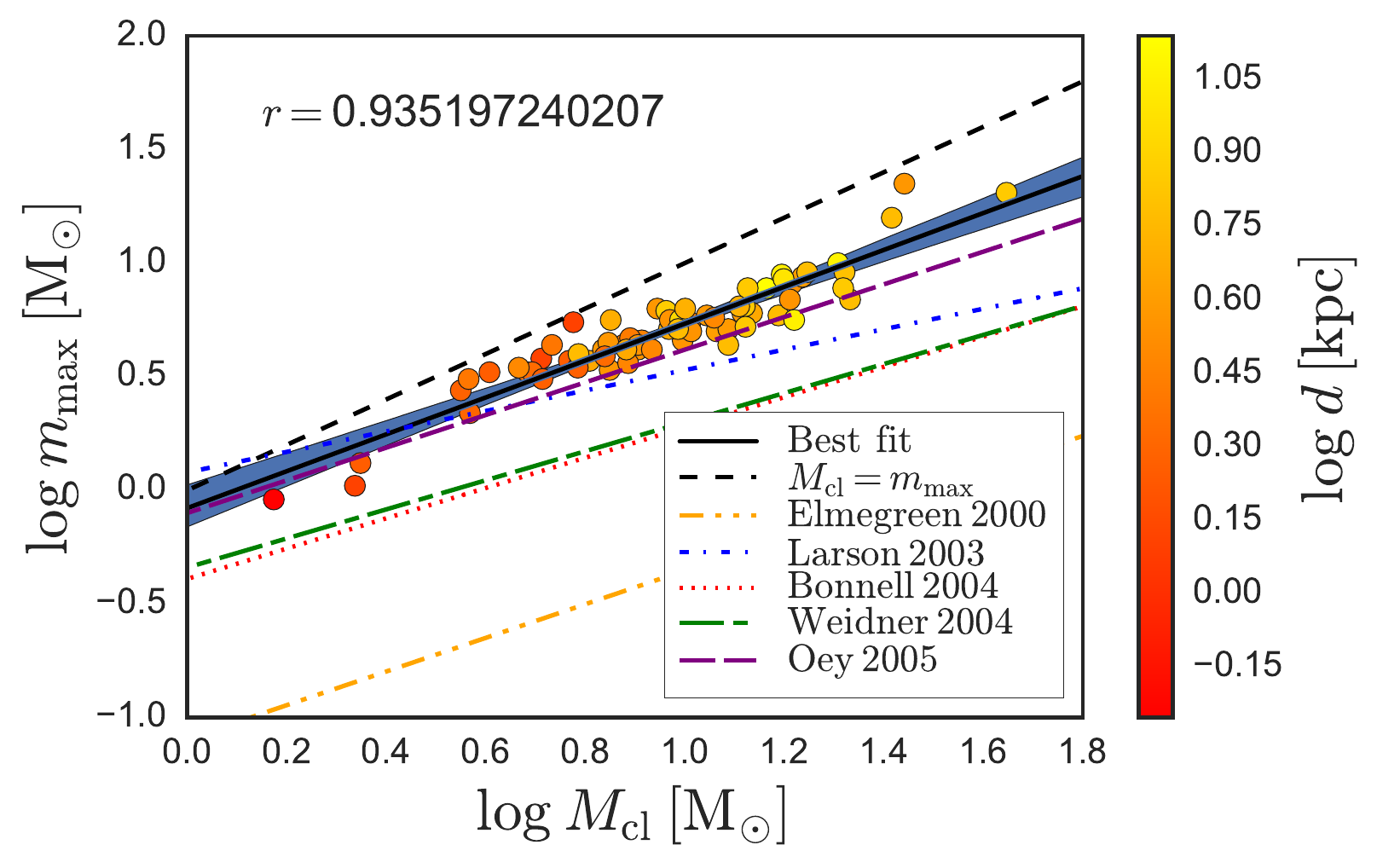}
\caption{The mass of the most massive cluster member as a function of the cluster mass. The dots are the point estimates from the MCMC sampling, and are color coded by cluster distance. The solid line corresponds to the best fit using Bayesian linear regression, and the shaded blue area is the 1-sigma credible interval. Also shown with different line styles are the theoretical and semi-empirical predictions discussed in \S~\ref{sec:correlation}. The r-value of the linear fit is also indicated.}
\label{fig:mass_relation}
\end{figure}

A key issue in star formation is whether is there any non-linear relationship between the total stellar mass of cluster and the number and mass of its individual protostars, and correspondingly, if and how the local IMF might be affected by such a nonlinearity. Fig.~\ref{fig:mass_relation} shows $M_{\rm{cl}}$ vs. $m_{\rm{max}}$ for all 70 ensembles studied here, all of which have 5 or fewer detected members with typical masses $m_*\sim 3\: \rm{M}_{\odot}$. (This value of the mass also corresponds to the completeness limit for our sample, as can be inferred by comparing the histogram of Fig.~\ref{fig:params_hist} with random samples from the canonical IMF). In a typical cluster containing hundreds or thousands of members, more than 90\% of the mass should be contained in stars less massive than about $1\, \rm{M}_{\odot}$, which means that $M_{\rm{cl}}$ estimated as the sum of the individual masses in our clusters is in fact only a fraction of the actual cluster mass. For clusters with only a few tens of members or less, as is our case, the fraction of the total cluster mass detected in individual members is considerably larger than in the case of massive clusters due to small number statistics. 

% * <hsmith@cfa.harvard.edu> 2018-02-13T18:48:04.244Z:
%
% ^.
% * <hsmith@cfa.harvard.edu> 2018-02-13T18:56:27.272Z:
%
% ^.

As we have pointed out, there is a distance selection effect included  in the correlation shown in Fig.~\ref{fig:mass_relation}, due to the fact that a star of given mass is less likely to be detected at larger heliocentric distances. Although our numbers indicate that this effect is not dramatic (see Fig.~\ref{fig:spatial_params}), there is nevertheless a trend for more massive stars being detected at the larger distances. To estimate a distance-corrected slope of the $M_{\rm{cl}}$ - $m_{\rm{max}}$ correlation, we fitted the $M_{\rm{cl}}$-$d$ and $m_{\rm{max}}$-$d$ correlations individually and evaluated the difference between the resulting slope in each case. We find that the correlation between $M_{\rm{cl}}$ and $m_{\rm{max}}$ solely due to the effect of distance is of the form $m_{\rm{max}} \propto M_{\rm{cl}}^{0.17}$. From the linear regression analysis, the uncertainty in this slope is of the order of 10\%. This contribution should be subtracted from the slope directly measured from the data points.

As a side product of this exercise, we obtain a mass-luminosity relation for YSOs in this mass range of the form: $L_* \propto m_*^{3.2}$. Our observations in this current, modest sample is satisfactorily consistent with theoretical ideas like those of \citet{Myers12}, which suggests a relation with a shallower $L_* \propto m_*^{2.2}$ . However, the mass-luminosity relation for young stars is very much more uncertain than it is for stars dominated by nuclear burning processes or Kelvin-Helmholtz contraction. Accretion processes can dominate the luminosity or young stars, and are functions of their mass and age, not to mention episodic periods determined by environmental considerations. The \citet{Myers12} relation, for example, is derived for protostars that are still accreting substantially, and of course the relation will also vary when the IMF varies from normal. All these complexities, however, point to the potential value of our new method in comparative analysis of clusters. In future papers we will explore larger samples to clarify the processes underway in clusters and help distinguish clusters, ascertain more accurately their ages, and probe their IMFs and mass cutoffs.

We can estimate how much mass we might be missing in our stellar clusters by performing Monte Carlo experiments using a Kroupa IMF corrected for binarity. A typical cluster in our sample has only one star more massive than $5\: \rm{M}_{\odot}$; a few exceptional cases have a single star more massive than $10\: \rm{M}_{\odot}$. We randomly sampled $10^5$ clusters both with a single $5\: \rm{M}_{\odot}$ star and a single $10\: \rm{M}_{\odot}$ from the Kroupa IMF and evaluated the typical number of members and total stellar mass of the resulting samples.  We find that for a cluster containing only one $M_* > 5\: \rm{M}_{\odot}$, the typical number of members is 9, and the typical total mass is $15-16\: \rm{M}_{\odot}$, whereas for a cluster with only one $M_* > 10\: \rm{M}_{\odot}$, the typical number of members is 15 and the typical total mass is $35\: \rm{M}_{\odot}$. A comparison of these numbers with those shown in Fig.~\ref{fig:mass_relation}, leads to the conclusion that between 35-45\% of the total stellar mass in our clusters is undetected in our observations. Furthermore,  given that the distribution of stellar masses in our clusters does not change dramatically with distance, we can assume that this fraction of undetected mass is similar in all of them. The slope of the correlation observed in Fig.~\ref{fig:mass_relation} is therefore unaffected by incompleteness, whereas its measured normalization is somewhat uncertain. 

In order to link the result of Fig.~\ref{fig:mass_relation} with the physics of star formation, we perform Bayesian linear regression to estimate posterior distributions for the slope and the normalization of this correlation and compare these posteriors with theoretical predictions and previous empirical findings. We assume that the data are sampled from a normal distribution, whose standard deviation is in turn sampled from a half-Cauchy distribution. The latter matches the average width of the flux posteriors measured from the SED and image fitting. 

\begin{figure}
\centering
  \subfigure{
\includegraphics[scale=0.55,angle=0,trim=0cm 0cm 0cm 0cm,clip=true]{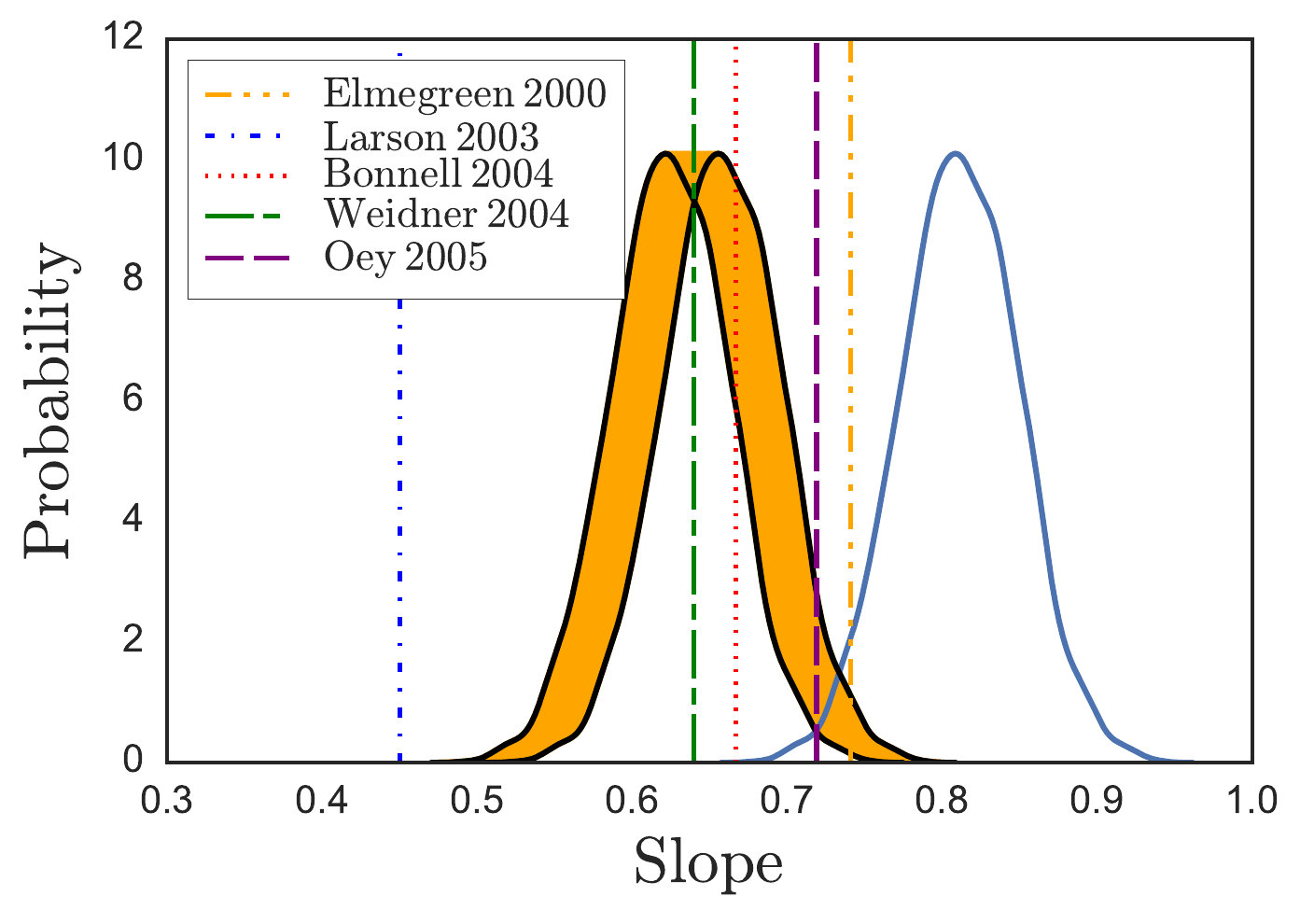}
\label{fig:mass_relation2a}
}
  \subfigure{
\includegraphics[scale=0.55,angle=0,trim=0cm 0cm 0cm 0cm,clip=true]{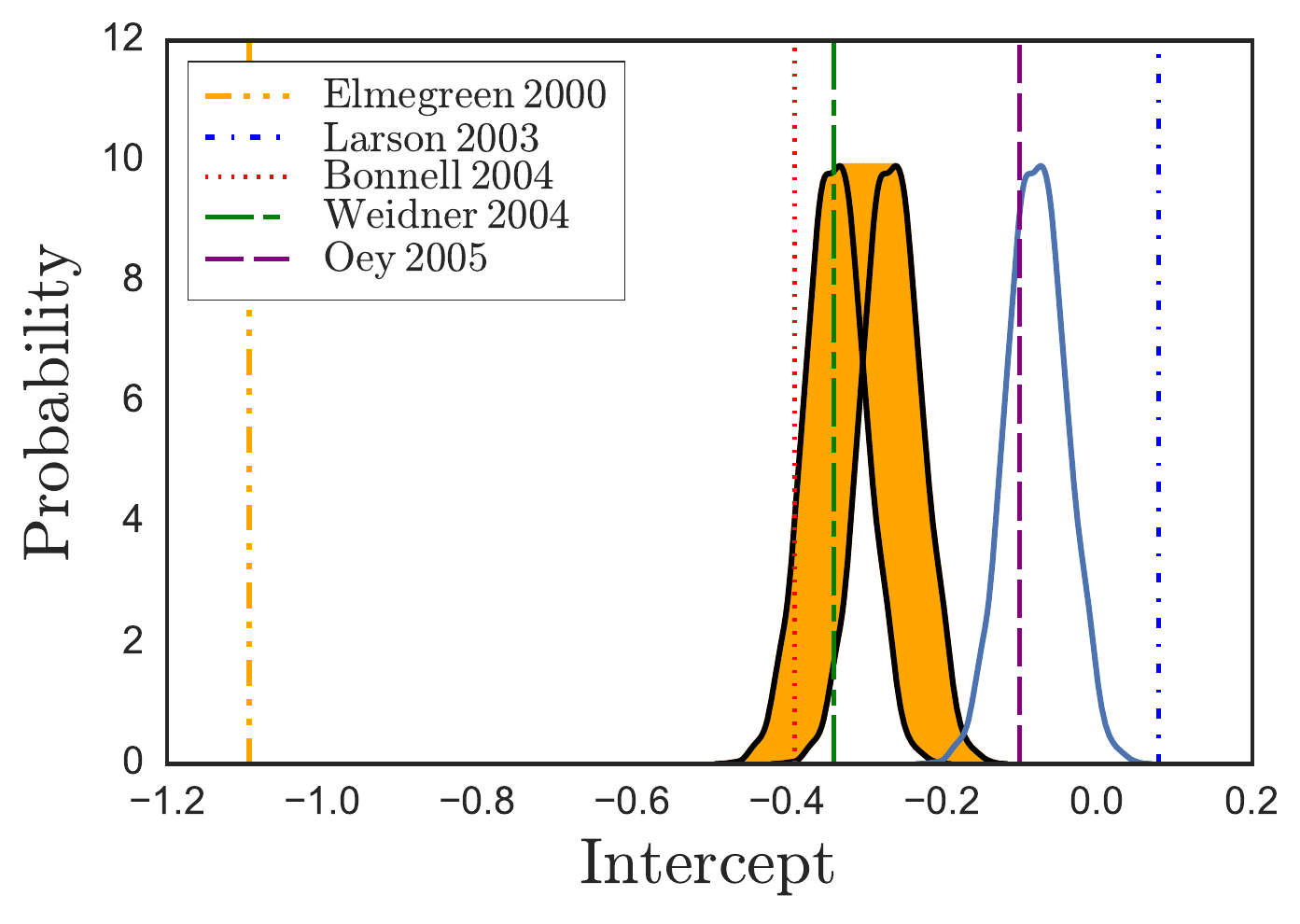}
\label{fig:mass_relation3a}
}
\caption{The posterior distributions for the slope (left) and intercept (right) of the $M_{\rm{cl}}-m_{\rm{max}}$ correlation as derived from Bayesian linear regression, compared with the values predicted by several theoretical end empirical studies (vertical lines). The blue solid line is the posterior measured from the data, while the orange shaded areas indicate the range of possible posteriors once they have been corrected respectively by distance effects and mass completeness. The results clearly agree with some models and disagree with others (see text).}
\label{fig:posteriors_correlation}
\end{figure}

The results of Bayesian linear regression are show in Fig.~\ref{fig:posteriors_correlation}. The left panel shows the posterior for the slope as measured directly from the SED fitted parameters, an also the range of possible corrected posteriors, once the effect of varying distance to the clusters has been taken into account. The shaded orange range corresponds to the 10\% uncertainty estimated for the slope variations due to distance effect. Similarly, the right panel shows the posterior for the intercept as measured from the data (blue line), and the range of possible posteriors corrected by mass incompleteness, assuming that we miss between 35\% and 45\% of the mass.

In both figures, the posteriors are compared to the following models or empirical relations:

% * <hsmith@cfa.harvard.edu> 2018-02-13T21:07:24.125Z:
%
% ^.
\begin{itemize}
\item \emph{Salpeter IMF}:  \citet{Elmegreen00} model distributions are constructed from a single slope power-law Salpeter IMF;  their combined luminosity exceeds the binding energy of the molecular cloud. The gravitational fate of the cluster is determined by the star formation efficiency, and the mass of the most massive star is set by the total number of stars, i.e., random sampling applies, and the formation of isolated massive stars is possible ( the double dot-dashed correlation in Fig.~\ref{fig:mass_relation}
\item \emph{Empirical}: \citet{Larson82} and \citet{Larson03} models compare the properties of several molecular clouds with the stellar populations of the clusters within (the $\rho$ Ophiucus cluster, the Orion Nebula cluster, the Quintuplet and the R136 clusters), and derive the empirical relation shown as a dash-dotted line in Fig.~\ref{fig:mass_relation}. The slope of this correlation (0.45) is shallower that what is predicted by a nominal Salpeter slope;  according to the authors, this is the result of a lower star formation efficiency in the high mass end due to feedback.
\item \emph{Competitive accretion}:  \citep{Bonnell03} model stars in a young cluster accreting from a shared reservoir of gas. In gas dominated regions of the cluster, usually in peripheral regions, the accretion is limited by tidal interactions, whereas in the cluster core the high relative velocities between stars results in Bondi-Hoyle accretion. The latter results in a fragmented IMF that is steeper for the high-mass stars that form in the cluster core with respect to the shallower IMF for low-mass stars that form in gas-dominated regions. This naturally results in the $M_{\rm{cl}}-m_{\rm{max}}$ correlation shown as the dotted line in Fig.~\ref{fig:mass_relation}.
\item \emph{Random sampling}: \citet{Oey05} analytically derive the the correlation between $M_{\rm{cl}}$ and $m_{\rm{max}}$ assuming that the stars are randomly produced according to a Salpeter IMF. The correlations they obtain is shown as the long-dashed line in Fig.~\ref{fig:mass_relation}. Their study, which includes results for a sample of young, nearby OB associations, concludes that there is a fundamental upper mass limit that truncates the IMF, and estimates a very low probability for optimal sampling that depends on the cluster mass.
\item \emph{Analytic - Random sampling}:  \citet{WeidnerKroupa04} assume in their study that a fundamental upper mass limit exists at $m_{\rm{max}} = 150\: \rm{M}_{\odot}$ and use the canonical multi-part Kroupa IMF to find the correlation shown as the short-dashed-log-dashed line in Fig.~\ref{fig:mass_relation}.
\end{itemize}

From the figure, it is clear that both the slope and intercept of the sources in our present study agree well with the \citet{Bonnell01} and the \citet{WeidnerKroupa04} analyses, and that the \citet{Larson03} models do not fit these results. 

\section{Discussion}
\label{sec:discusion}
In this section we explore how the physical properties  we derive for individual YSOs in low-mass ($M_{\rm{cl}}<100~\rm{M}_{\odot}$) clusters reflect the physical processes of star formation in individual clusters of various morphological types.    
% * <hsmith@cfa.harvard.edu> 2018-02-13T21:23:07.975Z:
%
% ^.

\subsection{Construction of the IMF for low mass clusters}

Are all the stellar masses in the galaxy produced  via a single universal IMF, or does the distribution of masses depend on the environment, making the  integrated galactic initial mass function of stars (IGIMF; Kroupa and Weidner 2003) different from the canonical IMF? (The IGIMF is usually discussed in the context of massive stars, but here we consider it in its broadest sense).  If it depends on environment, how different can it be, and why? The answers to these questions are intimately linked to the processes early star formation and the universality of the IMF \citep{Kroupa01, Kroupa02, Chabrier03,Elmegreen08,Bastian10,Kroupa13}. 

If the IMF  results from a random sampling process in any given cluster, then star formation is agnostic to the conditions of the environment including the total mass of the birth cluster; no self-regulation is at play. If, on the other hand, the stellar masses in a cluster are preferentially determined starting with the most massive of its members, in such a way the the resulting IGIMF has no gaps, then this implies that star formation is self-regulated, and that the mass of the most massive star depends on the available resources in the cluster. The latter also implies that there should exist a correlation between the stellar mass of the cluster ($M_{\rm{cl}}$) and the mass of its most massive star ($m_{\rm{max}}$). Different models of star formation predict a different slope and cut-off for this correlation.

The results from Figs.~\ref{fig:mass_relation}, and \ref{fig:posteriors_correlation} probe the $M_{\rm{cl}}-m_{\rm{max}}$ for clusters with masses below $100~\rm{M}_{\odot}$. They indicate that even in the low-mass cluster regime there is a complex correlation, with a slope in log-log space incompatible with simple linearity. After corrections have been applied to correct for the effects of undetected stellar mass and varying distance to the clusters, the posterior distributions for the correlation parameters are most compatible with the \citet{WeidnerKroupa04} statistical model that assumes a fundamental mass limit for stars, $m_{\rm{max}}=150~\rm{M}_{\odot}$ and assembles stellar masses randomly from the multi-part Kroupa IMF.  Our results show that such random sampling of a truncated IMF is compatible with the observational evidence in the cluster mass range considered here. This in turn implies that no significant suppression of high- or medium-mass stars occurs within the considered range of cluster masses. On the other hand, our results do not rule out optimal sampling at larger cluster masses;  competitive accretion remains a plausible model for the formation of low-mass clusters as well. We confirm previous studies by \citet{Maschberger08} and  \citet{Weidner10} both of which conclude that no suppression of high stellar masses can be inferred in clusters below the $100~\rm{M}_{\odot}$ limit, and that the distributions of stellar masses in these clusters are compatible with random sampling. However, our results do impose useful constraints on the possible mass distributions of clusters below the $100~\rm{M}_{\odot}$ limit.

Our result show that individual stellar masses in galactic clusters with masses below $100~\rm{M}_{\odot}$ are statistically determined at random from a Kroupa IMF. We find that stars with masses $m_*>5~\rm{M}_{\odot}$ can be formed in clusters with as few as 9 members, and with total stellar masses as low as $M_{\rm{cl}} = 15~\rm{M}_{\odot}$. Stars of the order of $m_*>10~\rm{M}_{\odot}$ can form in clusters with 15 members and total stellar masses of $M_{\rm{cl}} = 35~\rm{M}_{\odot}$. The occurrence of such stars is both consistent with random sampling and confirmed observationally in our sample of low-mass clusters. 

\subsection{Implications for competing theories of Star Formation}
Competitive accretion is not ruled out for  low mass stars, according to our above analysis and given the uncertainties, but it is unlikely to affect the IMF of low-mass clusters significantly.  The theory of competitive accretion \citep{Bonnell04} results in optimal sampling, and predicts a slope of the $M_{\rm{cl}}$-$m_{\rm{max}}$ correlation within $1\sigma$ of our measured slope, and an intercept within $2\sigma$ of our measured value.  However, the competition for a limited reservoir of gas as envisioned in this model is more likely to be seen in more massive clusters than those studied here.   We can, nevertheless, rule out some of the proposed models of star formation for low-mass clusters. 
The empirical correlation found by \citet{Larson03} that holds for more massive clusters such as the $\rho$ Ophiuchus cluster, the Orion Nebula cluster and R136, breaks down for low-mass clusters. This correlation results from an increasing difficulty in forming progressively more massive stars due to the effects of radiation pressure and winds, and translates into a steeper IMF than the one observed in the present work. At the mass range studied here, however, the IMF appears to be self-similar, implying that radiation effects in low-mass clusters do not prevent the formation of the most massive stars allowed by the IMF. Likewise, the \citet{Elmegreen00} model, which assumes a dominant role for gravity in limiting growth, is excluded by our results.
% * <hsmith@cfa.harvard.edu> 2018-02-13T21:57:32.867Z:
%
% ^.

Consistency between  the observed photometry and the models informs the mode of star formation. For example, the spherical geometries with accreting material assumed by the Robitaille SED models are consistent with monolithic collapse \citep{McKee03}, but inconsistent with the stellar merger model \citep{Bonnell98}, according to which massive stars do not form via accretion, but rather as the result of mergers of smaller stars. Also, if competitive accretion were the preferred mode of SF in the Milky Way, we would expect to see many more clusters than isolated single cores, and the luminosity of the cluster members should depend on their location with respect to the gravitational potential of the cluster. \emph{Consistency} is not the same as \emph{proof}, however, and so our results are mostly valuable from a statistical point of view, and to provide a robust starting point for more detailed modeling. 

\subsection{Dynamical stopping of accretion and the IMF}
\label{sec:massive_first}
Radiation hydrodynamical simulations have shown that in clustered environments, the IMF originates from competition between accretion and the dynamical interactions that terminate this accretion \citep{Bate12}. In these simulations, low-mass and high-mass stars form via the same process, but in the case of massive stars the dynamical termination of the accretion occurs later. Building along these lines, \citet{Myers11} proposes a model for competitive accretion in the dense regions of young clusters. According to this model, which assumes a constant birth rate for the protostars as opposed to coeval birth, the maximum protostar luminosity in a cluster indicates the age and mass of its oldest accreting protostar. The distribution of protostar masses evolves in time as the least massive stars undergo early accretion stopping while the massive stars continue accreting.

The positive, although marginal, correlation between $t_{\rm{max}}$ and $m_{\rm{max}}$, as well as the anti-correlation between $t_{\rm{max}}$ and $m_{\rm{min}}$ shown in Fig.~\ref{fig:trends} support this accretion-driven,  mass-evolving scenario: dynamical effects stop the accretion of individual stars, but the termination of accretion occurs later for massive stars. As the most massive stars in the cluster continue feeding from the surrounding gas and accretion has been terminated for stars of lower mass, the mass of the most massive star in the cluster continues to increase. But with a finite amount of gas available, longer accretion periods for the massive stars also means shorter accretion periods for the low mass stars, which explains the fact that the clusters that accrete for longer also have the lowest mass protostars (see Fig.~\ref{fig:trend2}). This is consistent with the dynamical termination of the accretion scenario described in \citet{Bate12}, which translates into a time evolution of the distribution of masses as accretion stops.

We interpret the shallow slope of the $t_{\rm{max}}$-$m_{\rm{max}}$ correlation as the result of most clusters being close to the final, time-independent mass distribution. Our hypothesis is that this is a stage in their evolution that lasts for a relatively long time so that our sample merely reflects this time-weighted distribution. The last YSOs still accreting are reaching the dynamical termination of their accreting phases, and the mass distribution is settling down, resulting in a slow increase of the maximum mass with age. This picture is consistent with the distribution of ages shown in Fig~\ref{fig:params_hist}, with the majority of YSOs being older than 1~Myr.

\subsection{Accelerating Star Formation and Cascade Fragmentation}

We pointed out in \S~\ref{sec:within_cluster} that at least two clusters,  4955 and 2568, contain highly embedded YSOs next to more evolved Class III objects. This association is not uncommon. In nearby embedded clusters such as the Serpens cloud core and NGC~1333, \citet{Winston09} have reported a significant age spread of the YSO populations, and report spatial segregation of young stars of different ages. Additionally, very many studies \citet{Willis13} have shown that star formation in GMCs spatially progresses across regions, with younger, more embedded sources, typically clustering in the central regions of the cluster. One possible scenario to explain age spreads of this magnitude is the accelerating star formation proposed in \citet{Huff06} for the Orion Nebula cluster, according to which the parent cloud rapidly contracts before dissipating, creating an event of accelerated star formation. Contamination from field stars is less likely for the spatially compacts clusters studied here, but cannot be completely ruled out. It is nevertheless hard to asses the validity of this theory with such a small sample.

Another scenario to consider here is the turbulent fragmentation cascade \citep{Joncour17}, in which the initial fragmentation of a dense core into a wide pair will lead to further fragmentation of each of the members of the pair, the extent of which depends on the initial separation between both fragments. The physical size of practically all of the small clusters considered here is less than 0.15~pc, about the typical width of the interstellar filaments identified by \emph{Herschel} \citep{Andre14}. This implies that all members of a cluster can be associated with a single initial core within a filament. A significant fraction of the multiple systems in our sample are wide pairs, with separations larger than $10^4$~AU. 

According to the fragmentation cascade scenario, it is likely that this wide configuration is an imprint of their spatial correlation at birth. Further fragmentation of the pair is predicted by the theory, and it is a possibility that fragmentation at smaller scales is not resolved by our NIR observations. This is something that we will be able to test with \emph{JWST}. Meanwhile, our observations are consistent with these clusters being formed as the result of the fragmentation of a single initial core. Since protostellar multiplicity is higher that the multiplicity of field stars \citep{Duchene13}, suggesting that early dynamical evolution disrupts these young clusters, the ages derived by our analysis impose a lower limit for the cluster age at which this dynamical disruption ends. This age is of the order of the oldest individual age derived here, which is just below 10~Myr.

\section{Conclusions}
\label{sec:conclusions}
Most stars form in clusters, and studies of star formation processes are inhibited by the fact that individual sources are often blended in spatially unresolved young clusters. This long-standing problem in the study of YSOs has hindered our ability to characterize star formation processes more precisely.  In this paper we offer a new Bayesian statistical method to address the issue. We  analyze the infrared SEDs of a sample of 70 \emph{Spitzer}-selected, low-mass ($M_{\rm{cl}}<100~\rm{M}_{\odot}$) young clusters in the galactic plane whose individual members appear blended together within the \emph{Spitzer} beam.  The technique allows us to model the probable SEDs of \emph{individual} YSOS using all available bands, included those where the cluster is spatial unresolved. Starting with prior information from the highest resolution images, our method estimates the most likely flux for each individual member in each band, by sequentially fitting the unresolved SEDs and images, and for each individual YSO it recovers the posterior probability distributions for the fundamental physical parameters: stellar mass, evolutionary stage, and optical extinction. 

The combined information obtained on  individual YSO properties, and the average properties of the low-mass clusters,  allows us to investigate how star formation proceeds in clustered environments containing tens of stars, and to assess whether the IMF is populated randomly in this mass range, or if self-regulating mechanisms lead to optimal sampling. Our main conclusions to date are based on a modest but representative sample of 70 clusters selected from \textit{Spitzer} surveys, and are aimed in part at illustrating the power of this method. 

\begin{itemize}
\item We have extracted the most probable photometry for YSO members of unresolved low-mass clusters across the galactic plane; definitive measurements require higher spatial resolution than is available. We present the method and compare the results against a variety of theoretical scenarios.  The method is very general, and can be applied to young protostellar clusters with even larger-beam, longer wavelength datasets including WISE and \textit{Herschel}.  The FIR measurements in particular can  also constrain the total dust masses and temperatures of the clusters.  The predictions will soon be testable using the observational capabilities of \emph{JWST}.  

\item For clusters with total stellar masses below $100~\rm{M}_{\odot}$, we find that there exists a non-trivial $M_{\rm{cl}}$-$m_{\rm{max}}$ correlation. The measured slope and intersect of this correlation are most compatible with random sampling of a Kroupa IMF with a fundamental high-mass limit of $150~\rm{M}_{\odot}$. Sampling of the IMF due to self-regulated star formation is \textit{not} readily inferred from our results. This is perhaps not surprising as we expect the effects of self-regulation to be detectable in the much denser environments of massive $M_{\rm{cl}}>10^3~\rm{M}_{\odot}$ clusters, but we nevertheless might have expected to see some modest influences.

\item Random sampling of the IMF in low-mass clusters is able to produce intermediate-mass stars ($m_*\sim 5~\rm{M}_{\odot}$) in clusters with as few as 9 members that have total stellar $M_{\rm{cl}} \sim 15~\rm{M}_{\odot}$. Similarly, stars of the order of $m_*\sim 10~\rm{M}_{\odot}$ can form in clusters with about 15 members and total stellar masses of $M_{\rm{cl}} = 35~\rm{M}_{\odot}$. 

\item The age of the oldest star in a star-forming cluster marginally scales with the mass of its most massive star, and anti-correlates with the mass of its least massive star. These relations are in support of the putative effects of dynamical stopping in an accretion scenario, as derived from several SPH simulations. In this scenario, low-mass and high-mass stars form via the same accretion mechanism, however the dynamical termination of accretion occurs later in the case of massive stars which therefore end up accreting for longer times.

\item Stellar mass growth due to accretion in stars that are born at a constant rate produces a time-dependent distribution of stellar masses. This distribution evolves as the least massive stars undergo an early termination of accretion while the massive stars continue accreting for longer times. When  accretion stops  for these massive stars, an equilibrium distribution is reached. Our results indicate that in clusters with masses below $100~\rm{M}_{\rm{\odot}}$, this equilibrium distribution is reached when the cluster age reaches 1~Myr.

\item The masses of all clusters studied here are compatible with their having formed from the fragmentation of a large core in a molecular filament. For those systems which are binary, cascade fragmentation suggests that multiplicity can increase at smaller scales, beyond our resolution limit. \emph{JWST} will be able to test this hypothesis in detail.

\item Using SED fitting, we have identified two sites of early massive star formation in the vicinity of maser emission sources. These sources contain some of the most embedded YSOs in our sample and are located on opposite sites of the 3kpc arm.

\end{itemize}

\acknowledgments

The authors are grateful to the referee for very insightful comments that have improved this paper. The authors would like to acknowledge partial support from NASA grants NNX12AI55G and NNX10AD68G for the succesful completion of this work. We would also like to thank Phil Myers and Lynn Carlson for a critical reading of the manuscript. Some of the calculations for the present work were done using the Odyssey cluster at Harvard University's Faculty of Arts \& Sciences.

\clearpage
%\begin{landscape}
%\LongTables
\begin{longrotatetable}
\begin{deluxetable}{cccccccccccc}
\tablecaption{GLIMPSE sources, derived fluxes, and physical parameters \label{tab:sources}}
\tablecolumns{9}
\tablehead{
  \colhead{Source ID} &
  \colhead{$\alpha$} &
  \colhead{$\delta$} &
  \colhead{$d\: (kpc)$} &
  \colhead{$\delta d\: (kpc)$} &
  \colhead{$F_{3.6}\: (mJy)$} &
  \colhead{$F_{4.5}\: (mJy)$} &
  \colhead{$F_{5.8}\: (mJy)$} &
  \colhead{$F_{8.0}\: (mJy)$} &
  \colhead{$\log\: t_*\: (\rm{yr})$} &
  \colhead{$\log\: M_*\: (\rm{M}_{\odot})$} &
  \colhead{$\log\: A_V (\rm{mag})$} 
}
\startdata
  360 & 267.2023 & -28.0199 &   5.9 &  2.0 & $116.87 \pm 3.70$  &  $332.32 \pm 8.52$  &  $582.80 \pm 14.93$  &  $582.80 \pm 14.93$  & $5.04_{-0.32}^{+0.11}$  & $1.20_{-0.01}^{+0.06}$  &  $1.76_{-0.02}^{+0.02}$  \\
      &          &          &       &      & $12.43 \pm 3.13$  &  $58.17 \pm 8.37$  &  $201.23 \pm 14.18$  &  $201.23 \pm 14.18$  &  $5.92_{-0.40}^{+0.25}$  & $1.01_{-0.05}^{+0.14}$  &  $1.77_{-0.05}^{+0.03}$  \\
 1062 & 268.3317 & -25.2607 &  3.54 & 1.36 & $19.09 \pm 5.53$  &  $5.17 \pm 1.14$  &  $8.05 \pm 1.72$  &  $8.05 \pm 1.72$  &         $5.23_{-0.63}^{+0.59}$  & $0.56_{-0.26}^{+0.32}$  &  $0.75_{-0.43}^{+0.24}$  \\
      &          &          &       &      & $30.76 \pm 4.78$  &  $33.24 \pm 1.65$  &  $29.70 \pm 1.93$  &  $29.70 \pm 1.93$  &      $6.31_{-1.24}^{+0.33}$  & $0.62_{-0.07}^{+0.12}$  &  $0.74_{-0.45}^{+0.22}$  \\
      &          &          &       &      & $0.01 \pm 0.02$  &  $0.00 \pm 0.02$  &  $0.00 \pm 0.00$  &  $0.00 \pm 0.00$  &          $6.75_{-0.18}^{+0.14}$  & $-0.57_{-0.18}^{+0.19}$  &  $0.25_{-0.42}^{+0.28}$  \\
 1277 & 269.4151 & -24.3469 &  2.88 & 1.35 & $21.23 \pm 1.95$  &  $11.63 \pm 1.61$  &  $9.25 \pm 1.76$  &  $9.25 \pm 1.76$  &        $6.30_{-0.93}^{+0.35}$  & $0.66_{-0.11}^{+0.05}$  &  $1.08_{-0.15}^{+0.08}$  \\
      &          &          &       &      & $24.83 \pm 2.02$  &  $22.40 \pm 1.89$  &  $22.42 \pm 1.93$  &  $22.42 \pm 1.93$  &      $6.54_{-0.55}^{+0.27}$  & $0.56_{-0.06}^{+0.11}$  &  $0.77_{-0.23}^{+0.19}$  \\
 1364 & 269.9419 & -24.0042 &  3.59 & 0.96 & $24.06 \pm 3.02$  &  $27.88 \pm 1.84$  &  $27.14 \pm 6.37$  &  $27.14 \pm 6.37$  &      $6.35_{-0.44}^{+0.26}$  & $0.66_{-0.05}^{+0.08}$  &  $1.28_{-0.08}^{+0.06}$  \\
      &          &          &       &      & $0.05 \pm 0.08$  &  $0.01 \pm 0.03$  &  $0.04 \pm 0.06$  &  $0.04 \pm 0.06$  &          $6.11_{-0.35}^{+0.41}$  & $-0.06_{-0.33}^{+0.27}$  &  $0.59_{-0.49}^{+0.35}$  \\
      &          &          &       &      & $4.28 \pm 2.44$  &  $1.74 \pm 1.61$  &  $15.70 \pm 6.20$  &  $15.70 \pm 6.20$  &        $5.44_{-0.38}^{+1.10}$  & $0.65_{-0.33}^{+0.10}$  &  $1.58_{-0.16}^{+0.08}$  \\
 1369 & 270.2175 & -24.1208 &  1.28 & 0.09 & $4.97 \pm 1.02$  &  $3.21 \pm 0.78$  &  $3.78 \pm 1.16$  &  $3.78 \pm 1.16$  &          $5.66_{-0.61}^{+0.42}$  & $0.02_{-0.37}^{+0.35}$  &  $1.25_{-0.16}^{+0.08}$  \\
      &          &          &       &      & $2.77 \pm 1.24$  &  $3.35 \pm 0.85$  &  $4.07 \pm 1.48$  &  $4.07 \pm 1.48$  &          $5.72_{-0.82}^{+0.66}$  & $-0.07_{-0.48}^{+0.44}$  &  $0.90_{-0.38}^{+0.21}$  \\
      &          &          &       &      & $0.10 \pm 0.16$  &  $0.04 \pm 0.10$  &  $0.02 \pm 0.08$  &  $0.02 \pm 0.08$  &          $6.20_{-1.08}^{+0.50}$  & $-0.56_{-0.26}^{+0.39}$  &  $1.01_{-0.29}^{+0.14}$  \\
 1396 & 270.7117 &  -24.299 &   1.3 &  0.1 & $17.50 \pm 3.09$  &  $21.33 \pm 1.17$  &  $12.59 \pm 3.64$  &  $12.59 \pm 3.64$  &      $6.34_{-0.11}^{+0.32}$  & $0.58_{-0.07}^{+0.07}$  &  $1.47_{-0.02}^{+0.02}$  \\
      &          &          &       &      & $8.51 \pm 3.52$  &  $1.03 \pm 0.48$  &  $8.71 \pm 8.48$  &  $8.71 \pm 8.48$  &          $6.07_{-0.40}^{+0.38}$  & $0.13_{-0.36}^{+0.22}$  &  $0.68_{-0.46}^{+0.21}$  \\
 1437 & 270.2374 &  -23.825 &  4.97 & 0.26 & $22.71 \pm 1.25$  &  $34.14 \pm 1.41$  &  $47.96 \pm 2.84$  &  $47.96 \pm 2.84$  &      $6.13_{-0.12}^{+0.28}$  & $0.75_{-0.10}^{+0.06}$  &  $1.21_{-0.74}^{+0.13}$  \\
      &          &          &       &      & $0.10 \pm 0.21$  &  $0.18 \pm 0.34$  &  $1.99 \pm 4.40$  &  $1.99 \pm 4.40$  &          $5.41_{-0.45}^{+0.44}$  & $0.17_{-0.50}^{+0.32}$  &  $1.08_{-0.26}^{+0.13}$  \\
 1493 & 270.0784 & -23.4725 &  2.4  &      & $0.13 \pm 0.11$  &  $0.08 \pm 0.11$  &  $0.11 \pm 0.21$  &  $0.11 \pm 0.21$  &          $6.08_{-0.41}^{+0.40}$  & $-0.03_{-0.43}^{+0.34}$  &  $1.23_{-0.16}^{+0.13}$  \\
      &          &          &       &      & $1.47 \pm 0.52$  &  $2.89 \pm 0.58$  &  $5.43 \pm 1.08$  &  $5.43 \pm 1.08$  &          $6.33_{-0.57}^{+0.36}$  & $0.52_{-0.15}^{+0.12}$  &  $1.60_{-0.08}^{+0.07}$  \\
      &          &          &       &      & $0.09 \pm 0.11$  &  $0.06 \pm 0.14$  &  $0.15 \pm 0.30$  &  $0.15 \pm 0.30$  &          $5.96_{-0.55}^{+0.38}$  & $-0.23_{-0.34}^{+0.49}$  &  $1.23_{-0.17}^{+0.13}$  \\
 1566 & 270.5074 & -23.0983 &  3.81 & 0.77 & $49.15 \pm 3.04$  &  $66.30 \pm 4.01$  &  $67.15 \pm 5.25$  &  $67.15 \pm 5.25$  &      $6.54_{-0.29}^{+0.14}$  & $0.78_{-0.07}^{+0.04}$  &  $1.43_{-0.05}^{+0.05}$  \\
      &          &          &       &      & $9.03 \pm 2.95$  &  $13.54 \pm 4.22$  &  $23.22 \pm 5.22$  &  $23.22 \pm 5.22$  &       $6.43_{-0.21}^{+0.19}$  & $0.69_{-0.08}^{+0.06}$  &  $1.52_{-0.10}^{+0.04}$  \\
      &          &          &       &      & $0.52 \pm 0.38$  &  $0.92 \pm 0.82$  &  $5.03 \pm 2.05$  &  $5.03 \pm 2.05$  &          $5.33_{-0.36}^{+0.45}$  & $0.33_{-0.56}^{+0.24}$  &  $1.38_{-0.14}^{+0.13}$  \\
 1629 & 270.7366 & -22.8421 &   2.7 &  0.5 & $25.14 \pm 1.43$  &  $38.91 \pm 1.49$  &  $59.06 \pm 2.04$  &  $59.06 \pm 2.04$  &      $6.49_{-0.18}^{+0.21}$  & $0.74_{-0.07}^{+0.08}$  &  $1.53_{-0.07}^{+0.06}$  \\
      &          &          &       &      & $0.02 \pm 0.01$  &  $0.01 \pm 0.01$  &  $0.01 \pm 0.01$  &  $0.01 \pm 0.01$  &          $6.45_{-0.42}^{+0.33}$  & $-0.57_{-0.21}^{+0.25}$  &  $0.65_{-0.30}^{+0.30}$  \\
      &          &          &       &      & $0.17 \pm 0.11$  &  $0.05 \pm 0.08$  &  $0.10 \pm 0.27$  &  $0.10 \pm 0.27$  &          $6.10_{-0.39}^{+0.32}$  & $0.44_{-0.27}^{+0.33}$  &  $1.51_{-0.15}^{+0.11}$  \\
      &          &          &       &      & $0.10 \pm 0.10$  &  $0.10 \pm 0.11$  &  $0.04 \pm 0.06$  &  $0.04 \pm 0.06$  &          $5.99_{-0.57}^{+0.38}$  & $-0.47_{-0.32}^{+0.44}$  &  $0.92_{-0.27}^{+0.14}$  \\
      &          &          &       &      & $2.61 \pm 0.64$  &  $3.84 \pm 0.62$  &  $4.78 \pm 1.04$  &  $4.78 \pm 1.04$  &          $5.45_{-0.16}^{+0.34}$  & $0.15_{-0.13}^{+0.33}$  &  $1.17_{-0.30}^{+0.16}$  \\
 1807 & 270.5608 & -21.5428 &  3.47 & 0.75 & $6.91 \pm 1.24$  &  $0.12 \pm 0.17$  &  $0.13 \pm 0.21$  &  $0.13 \pm 0.21$  &          $5.03_{-0.26}^{+0.33}$  & $0.64_{-0.49}^{+0.19}$  &  $1.47_{-0.20}^{+0.11}$  \\
      &          &          &       &      & $2.67 \pm 1.05$  &  $34.76 \pm 1.38$  &  $72.74 \pm 2.15$  &  $72.74 \pm 2.15$  &       $6.04_{-0.54}^{+0.27}$  & $0.91_{-0.06}^{+0.02}$  &  $1.85_{-0.07}^{+0.02}$  \\
      &          &          &       &      & $0.01 \pm 0.01$  &  $0.00 \pm 0.00$  &  $0.00 \pm 0.00$  &  $0.00 \pm 0.00$  &          $6.52_{-0.46}^{+0.21}$  & $-0.51_{-0.29}^{+0.27}$  &  $0.81_{-0.28}^{+0.09}$  \\
      &          &          &       &      & $2.69 \pm 0.85$  &  $1.14 \pm 0.38$  &  $1.85 \pm 0.66$  &  $1.85 \pm 0.66$  &          $6.14_{-0.78}^{+0.47}$  & $0.56_{-0.09}^{+0.13}$  &  $1.58_{-0.05}^{+0.10}$  \\
 2073 & 272.3163 &  -21.053 &  3.81 & 0.59 & $1.65 \pm 0.55$  &  $1.68 \pm 0.54$  &  $1.61 \pm 0.70$  &  $1.61 \pm 0.70$  &          $6.30_{-0.41}^{+0.09}$  & $0.47_{-0.12}^{+0.02}$  &  $1.45_{-0.10}^{+0.02}$  \\
      &          &          &       &      & $12.36 \pm 1.04$  &  $12.84 \pm 0.91$  &  $12.29 \pm 1.09$  &  $12.29 \pm 1.09$  &      $6.28_{-0.21}^{+0.20}$  & $0.77_{-0.09}^{+0.04}$  &  $1.34_{-0.04}^{+0.00}$  \\
      &          &          &       &      & $0.39 \pm 0.29$  &  $0.03 \pm 0.04$  &  $0.57 \pm 0.42$  &  $0.57 \pm 0.42$  &          $6.35_{-0.49}^{+0.10}$  & $0.35_{-0.11}^{+0.10}$  &  $1.24_{-0.03}^{+0.10}$  \\
 2090 & 272.5295 & -20.9922 &  4.13 & 0.51 & $0.05 \pm 0.06$  &  $0.03 \pm 0.04$  &  $0.02 \pm 0.03$  &  $0.02 \pm 0.03$  &          $6.09_{-0.51}^{+0.49}$  & $0.06_{-0.54}^{+0.23}$  &  $0.54_{-0.38}^{+0.28}$  \\
      &          &          &       &      & $0.00 \pm 0.00$  &  $0.00 \pm 0.00$  &  $0.00 \pm 0.00$  &  $0.00 \pm 0.00$  &          $6.57_{-1.34}^{+0.25}$  & $-0.51_{-0.26}^{+0.28}$  &  $0.75_{-0.26}^{+0.20}$  \\
      &          &          &       &      & $0.02 \pm 0.05$  &  $0.00 \pm 0.01$  &  $0.02 \pm 0.08$  &  $0.02 \pm 0.08$  &          $5.79_{-0.98}^{+0.64}$  & $-0.12_{-0.32}^{+0.52}$  &  $1.29_{-0.18}^{+0.16}$  \\
      &          &          &       &      & $3.89 \pm 0.98$  &  $8.31 \pm 2.40$  &  $10.40 \pm 3.10$  &  $10.40 \pm 3.10$  &        $5.75_{-0.53}^{+0.72}$  & $0.64_{-0.15}^{+0.13}$  &  $1.52_{-0.07}^{+0.10}$  \\
      &          &          &       &      & $1.25 \pm 1.11$  &  $8.60 \pm 2.12$  &  $11.92 \pm 2.96$  &  $11.92 \pm 2.96$  &        $6.16_{-0.52}^{+0.37}$  & $0.70_{-0.08}^{+0.15}$  &  $1.69_{-0.07}^{+0.09}$  \\
 2130 & 272.5325 & -20.7702 &  3.73 & 0.59 & $0.02 \pm 0.02$  &  $0.01 \pm 0.01$  &  $0.00 \pm 0.00$  &  $0.00 \pm 0.00$  &          $6.08_{-0.77}^{+0.44}$  & $-0.10_{-0.49}^{+0.23}$  &  $1.03_{-0.27}^{+0.08}$  \\
      &          &          &       &      & $6.06 \pm 2.51$  &  $4.87 \pm 1.04$  &  $16.42 \pm 2.14$  &  $16.42 \pm 2.14$  &        $5.06_{-0.65}^{+0.78}$  & $0.53_{-0.21}^{+0.17}$  &  $0.57_{-0.33}^{+0.27}$  \\
      &          &          &       &      & $1.11 \pm 1.25$  &  $0.44 \pm 0.40$  &  $1.02 \pm 1.06$  &  $1.02 \pm 1.06$  &          $5.14_{-0.46}^{+1.69}$  & $0.46_{-0.57}^{+0.13}$  &  $1.21_{-0.14}^{+0.07}$  \\
 2133 & 271.9143 & -20.4448 &  2.73 & 0.84 & $8.19 \pm 2.33$  &  $7.81 \pm 1.83$  &  $7.63 \pm 2.48$  &  $7.63 \pm 2.48$  &          $5.36_{-0.31}^{+0.87}$  & $0.50_{-0.26}^{+0.14}$  &  $1.34_{-0.14}^{+0.07}$  \\
      &          &          &       &      & $8.05 \pm 2.48$  &  $10.88 \pm 2.01$  &  $16.72 \pm 3.22$  &  $16.72 \pm 3.22$  &       $6.38_{-0.35}^{+0.29}$  & $0.64_{-0.08}^{+0.08}$  &  $1.57_{-0.09}^{+0.05}$  \\
 2141 & 271.9798 & -20.4531 &  2.63 & 0.87 & $19.98 \pm 1.84$  &  $26.22 \pm 3.38$  &  $33.12 \pm 2.85$  &  $33.12 \pm 2.85$  &      $6.36_{-0.33}^{+0.26}$  & $0.67_{-0.05}^{+0.09}$  &  $1.55_{-0.05}^{+0.05}$  \\
      &          &          &       &      & $1.94 \pm 1.52$  &  $1.68 \pm 1.88$  &  $3.12 \pm 2.33$  &  $3.12 \pm 2.33$  &          $5.31_{-0.28}^{+0.67}$  & $0.49_{-0.34}^{+0.16}$  &  $1.49_{-0.10}^{+0.07}$  \\
 2412 & 272.6466 & -19.2146 &  3.59 & 0.56 & $20.03 \pm 1.66$  &  $19.40 \pm 1.37$  &  $17.45 \pm 2.50$  &  $17.45 \pm 2.50$  &      $6.50_{-0.22}^{+0.26}$  & $0.64_{-0.08}^{+0.12}$  &  $1.28_{-0.08}^{+0.07}$  \\
      &          &          &       &      & $1.61 \pm 1.04$  &  $2.75 \pm 1.22$  &  $3.57 \pm 2.48$  &  $3.57 \pm 2.48$  &          $5.71_{-0.44}^{+0.96}$  & $0.29_{-0.42}^{+0.22}$  &  $1.11_{-0.18}^{+0.11}$  \\
      &          &          &       &      & $0.15 \pm 0.13$  &  $0.08 \pm 0.07$  &  $0.02 \pm 0.03$  &  $0.02 \pm 0.03$  &          $6.14_{-0.40}^{+0.33}$  & $-0.08_{-0.31}^{+0.26}$  &  $0.29_{-0.62}^{+0.31}$  \\
      &          &          &       &      & $0.29 \pm 0.28$  &  $0.26 \pm 0.34$  &  $0.14 \pm 0.23$  &  $0.14 \pm 0.23$  &          $5.75_{-0.44}^{+0.47}$  & $-0.04_{-0.37}^{+0.37}$  &  $1.05_{-0.35}^{+0.15}$  \\
 2416 & 273.0237 & -19.3811 &  3.80 & 0.52 & $0.00 \pm 0.00$  &  $0.00 \pm 0.00$  &  $0.00 \pm 0.00$  &  $0.00 \pm 0.00$  &          $5.39_{-0.22}^{+0.07}$  & $-0.67_{-0.11}^{+0.11}$  &  $0.12_{-0.24}^{+0.10}$  \\
      &          &          &       &      & $0.09 \pm 0.06$  &  $0.00 \pm 0.00$  &  $0.03 \pm 0.01$  &  $0.03 \pm 0.01$  &          $6.28_{-0.13}^{+0.13}$  & $0.18_{-0.09}^{+0.10}$  &  $1.14_{-0.05}^{+0.15}$  \\
      &          &          &       &      & $0.01 \pm 0.01$  &  $0.00 \pm 0.00$  &  $0.01 \pm 0.01$  &  $0.01 \pm 0.01$  &          $5.54_{-0.04}^{+0.36}$  & $-0.76_{-0.24}^{+0.03}$  &  $-0.45_{-0.26}^{+0.06}$  \\
      &          &          &       &      & $0.01 \pm 0.01$  &  $3.62 \pm 3.77$  &  $6.82 \pm 4.17$  &  $6.82 \pm 4.17$  &          $4.48_{-0.07}^{+0.08}$  & $-0.22_{-0.23}^{+0.15}$  &  $0.24_{-0.24}^{+0.03}$  \\
      &          &          &       &      & $63.49 \pm 1.95$  &  $95.97 \pm 3.10$  &  $128.19 \pm 3.81$  &  $128.19 \pm 3.81$  &    $6.21_{-0.09}^{+0.07}$  & $0.80_{-0.03}^{+0.04}$  &  $1.43_{-0.04}^{+0.04}$  \\
 2419 & 272.6617 &  -19.188 &  3.59 & 0.56 & $7.05 \pm 2.02$  &  $2.83 \pm 1.51$  &  $6.73 \pm 3.60$  &  $6.73 \pm 3.60$  &          $6.51_{-0.15}^{+0.16}$  & $0.70_{-0.15}^{+0.07}$  &  $1.46_{-0.05}^{+0.07}$  \\
      &          &          &       &      & $5.04 \pm 2.04$  &  $21.93 \pm 2.29$  &  $43.57 \pm 4.27$  &  $43.57 \pm 4.27$  &       $5.20_{-0.15}^{+0.33}$  & $0.61_{-0.12}^{+0.07}$  &  $1.09_{-0.15}^{+0.19}$  \\
      &          &          &       &      & $0.19 \pm 0.15$  &  $0.19 \pm 0.16$  &  $0.25 \pm 0.23$  &  $0.25 \pm 0.23$  &          $6.26_{-0.39}^{+0.51}$  & $0.00_{-0.28}^{+0.20}$  &  $0.73_{-0.11}^{+0.44}$  \\
      &          &          &       &      & $0.01 \pm 0.00$  &  $0.00 \pm 0.00$  &  $0.00 \pm 0.00$  &  $0.00 \pm 0.00$  &          $6.70_{-0.11}^{+0.12}$  & $-0.65_{-0.19}^{+0.20}$  &  $0.00_{-0.39}^{+0.29}$  \\
 2713 & 273.0283 & -17.6483 & 13.80 & 0.77 & $0.02 \pm 0.07$  &  $0.00 \pm 0.00$  &  $0.00 \pm 0.00$  &  $0.00 \pm 0.00$  &          $5.64_{-0.53}^{+0.61}$  & $0.13_{-0.44}^{+0.34}$  &  $1.08_{-0.20}^{+0.20}$  \\
      &          &          &       &      & $24.19 \pm 1.89$  &  $2.06 \pm 1.98$  &  $30.38 \pm 2.35$  &  $30.38 \pm 2.35$  &       $6.32_{-0.17}^{+0.21}$  & $0.89_{-0.05}^{+0.08}$  &  $0.85_{-0.51}^{+0.16}$  \\
      &          &          &       &      & $4.09 \pm 1.61$  &  $19.74 \pm 21.43$  &  $3.28 \pm 2.07$  &  $3.28 \pm 2.07$  &        $6.06_{-1.07}^{+0.51}$  & $0.74_{-0.08}^{+0.07}$  &  $0.57_{-0.36}^{+0.23}$  \\
 2985 &   274.06 & -16.9622 &  2.15 & 0.73 & $0.01 \pm 0.01$  &  $0.00 \pm 0.01$  &  $0.01 \pm 0.01$  &  $0.01 \pm 0.01$  &          $6.54_{-0.96}^{+0.23}$  & $-0.41_{-0.29}^{+0.34}$  &  $1.20_{-0.16}^{+0.13}$  \\
      &          &          &       &      & $0.00 \pm 0.00$  &  $0.00 \pm 0.00$  &  $0.00 \pm 0.00$  &  $0.00 \pm 0.00$  &          $6.65_{-0.38}^{+0.22}$  & $-0.72_{-0.20}^{+0.27}$  &  $0.69_{-0.38}^{+0.21}$  \\
      &          &          &       &      & $0.58 \pm 0.60$  &  $0.42 \pm 0.45$  &  $1.54 \pm 1.41$  &  $1.54 \pm 1.41$  &          $5.84_{-0.30}^{+0.47}$  & $0.02_{-0.36}^{+0.33}$  &  $1.36_{-0.09}^{+0.09}$  \\
      &          &          &       &      & $7.84 \pm 1.03$  &  $8.17 \pm 0.79$  &  $8.53 \pm 1.37$  &  $8.53 \pm 1.37$  &          $6.28_{-0.49}^{+0.39}$  & $0.52_{-0.07}^{+0.06}$  &  $1.29_{-0.06}^{+0.05}$  \\
 3156 & 274.2054 & -16.4183 &  3.88 & 0.43 & $0.00 \pm 0.00$  &  $0.00 \pm 0.00$  &  $0.00 \pm 0.00$  &  $0.00 \pm 0.00$  &          $5.94_{-0.91}^{+0.64}$  & $-0.37_{-0.33}^{+0.44}$  &  $1.17_{-0.39}^{+0.21}$  \\
      &          &          &       &      & $1.43 \pm 0.61$  &  $3.18 \pm 0.73$  &  $3.65 \pm 1.16$  &  $3.65 \pm 1.16$  &          $5.22_{-0.89}^{+1.03}$  & $0.58_{-0.40}^{+0.22}$  &  $1.36_{-0.52}^{+0.24}$  \\
      &          &          &       &      & $4.07 \pm 0.82$  &  $1.32 \pm 0.51$  &  $1.21 \pm 0.66$  &  $1.21 \pm 0.66$  &          $5.70_{-0.54}^{+0.85}$  & $0.46_{-0.40}^{+0.10}$  &  $0.94_{-0.24}^{+0.18}$  \\
      &          &          &       &      & $7.44 \pm 1.06$  &  $12.37 \pm 1.08$  &  $14.61 \pm 1.55$  &  $14.61 \pm 1.55$  &       $6.29_{-0.52}^{+0.32}$  & $0.71_{-0.10}^{+0.17}$  &  $1.56_{-0.09}^{+0.07}$  \\ 
 3227 & 274.8098 & -16.5083 &  2.09 & 0.71 & $6.21 \pm 1.33$  &  $6.73 \pm 1.20$  &  $9.71 \pm 2.05$  &  $9.71 \pm 2.05$  &          $5.69_{-0.82}^{+0.90}$  & $0.57_{-0.18}^{+0.14}$  &  $1.58_{-0.09}^{+0.05}$  \\
      &          &          &       &      & $1.85 \pm 1.00$  &  $2.50 \pm 1.11$  &  $3.44 \pm 1.80$  &  $3.44 \pm 1.80$  &          $5.60_{-0.49}^{+0.65}$  & $0.34_{-0.39}^{+0.20}$  &  $1.43_{-0.10}^{+0.08}$  \\
 3282 & 274.4012 & -16.0383 &  2.72 & 0.61 & $0.00 \pm 0.00$  &  $0.00 \pm 0.00$  &  $0.00 \pm 0.00$  &  $0.00 \pm 0.00$  &          $6.52_{-0.25}^{+0.30}$  & $-0.48_{-0.32}^{+0.29}$  &  $1.08_{-0.12}^{+0.12}$  \\
      &          &          &       &      & $20.91 \pm 1.80$  &  $24.15 \pm 2.14$  &  $28.73 \pm 3.11$  &  $28.73 \pm 3.11$  &      $6.17_{-1.37}^{+0.37}$  & $0.72_{-0.08}^{+0.15}$  &  $1.50_{-0.04}^{+0.05}$  \\
      &          &          &       &      & $0.07 \pm 0.07$  &  $0.04 \pm 0.05$  &  $0.05 \pm 0.10$  &  $0.05 \pm 0.10$  &          $6.36_{-0.30}^{+0.31}$  & $-0.53_{-0.25}^{+0.28}$  &  $0.27_{-0.53}^{+0.39}$  \\
      &          &          &       &      & $2.98 \pm 1.45$  &  $6.73 \pm 1.86$  &  $6.53 \pm 2.68$  &  $6.53 \pm 2.68$  &          $5.36_{-0.43}^{+0.62}$  & $0.57_{-0.42}^{+0.18}$  &  $1.45_{-0.15}^{+0.10}$  \\
 3538 & 275.2079 & -14.7652 &  4.07 & 0.38 & $13.08 \pm 3.25$  &  $8.87 \pm 4.12$  &  $8.21 \pm 5.67$  &  $8.21 \pm 5.67$  &         $6.45_{-0.17}^{+0.15}$  & $0.70_{-0.08}^{+0.10}$  &  $1.48_{-0.04}^{+0.05}$  \\
      &          &          &       &      & $0.01 \pm 0.01$  &  $0.05 \pm 0.04$  &  $0.07 \pm 0.11$  &  $0.07 \pm 0.11$  &          $5.85_{-0.22}^{+0.74}$  & $-0.24_{-0.29}^{+0.29}$  &  $1.28_{-0.14}^{+0.02}$  \\
      &          &          &       &      & $11.25 \pm 3.72$  &  $29.01 \pm 3.33$  &  $38.00 \pm 5.11$  &  $38.00 \pm 5.11$  &      $6.32_{-0.16}^{+0.08}$  & $0.94_{-0.17}^{+0.15}$  &  $1.62_{-0.11}^{+0.09}$  \\
      &          &          &       &      & $1.40 \pm 2.11$  &  $0.62 \pm 0.94$  &  $1.28 \pm 1.94$  &  $1.28 \pm 1.94$  &          $6.56_{-0.66}^{+0.16}$  & $0.47_{-0.34}^{+0.20}$  &  $1.48_{-0.14}^{+0.05}$  \\
 3568 & 275.7657 & -14.9323 &  3.74 & 0.42 & $0.00 \pm 0.00$  &  $0.00 \pm 0.00$  &  $0.00 \pm 0.00$  &  $0.00 \pm 0.00$  &          $6.61_{-0.24}^{+0.26}$  & $-0.29_{-0.28}^{+0.07}$  &  $0.57_{-0.23}^{+0.13}$  \\
      &          &          &       &      & $0.00 \pm 0.00$  &  $0.00 \pm 0.00$  &  $0.00 \pm 0.00$  &  $0.00 \pm 0.00$  &          $6.46_{-0.48}^{+0.06}$  & $-0.75_{-0.13}^{+0.35}$  &  $0.77_{-0.29}^{+0.18}$  \\
      &          &          &       &      & $2.18 \pm 0.85$  &  $3.33 \pm 0.98$  &  $17.19 \pm 6.20$  &  $17.19 \pm 6.20$  &        $6.50_{-0.27}^{+0.23}$  & $0.66_{-0.14}^{+0.14}$  &  $1.69_{-0.09}^{+0.05}$  \\
      &          &          &       &      & $3.97 \pm 1.03$  &  $9.40 \pm 1.20$  &  $9.44 \pm 3.19$  &  $9.44 \pm 3.19$  &          $3.70_{-0.38}^{+0.49}$  & $1.35_{-0.09}^{+0.05}$  &  $1.32_{-0.14}^{+0.11}$  \\
 3749 & 274.7755 &  -13.598 &   1.8 &  0.1 & $1.68 \pm 1.44$  &  $1.37 \pm 1.07$  &  $2.17 \pm 1.52$  &  $2.17 \pm 1.52$  &          $5.59_{-0.66}^{+0.44}$  & $-0.13_{-0.30}^{+0.49}$  &  $1.29_{-0.14}^{+0.08}$  \\
      &          &          &       &      & $19.21 \pm 1.73$  &  $18.01 \pm 1.48$  &  $16.01 \pm 1.85$  &  $16.01 \pm 1.85$  &      $6.57_{-0.30}^{+0.24}$  & $0.52_{-0.05}^{+0.07}$  &  $1.38_{-0.05}^{+0.04}$  \\
 3756 & 274.6668 &   -13.51 &   1.8 &  0.1 & $7.57 \pm 1.23$  &  $10.29 \pm 1.53$  &  $12.99 \pm 2.42$  &  $12.99 \pm 2.42$  &       $6.47_{-0.27}^{+0.31}$  & $0.57_{-0.07}^{+0.13}$  &  $1.66_{-0.04}^{+0.05}$  \\
      &          &          &       &      & $1.82 \pm 1.48$  &  $5.82 \pm 1.72$  &  $8.27 \pm 2.60$  &  $8.27 \pm 2.60$  &          $5.84_{-0.33}^{+0.63}$  & $0.33_{-0.37}^{+0.17}$  &  $1.55_{-0.07}^{+0.05}$  \\
 3776 & 275.2404 & -13.7086 &  1.99 & 0.64 & $5.75 \pm 1.39$  &  $7.43 \pm 1.58$  &  $8.50 \pm 2.21$  &  $8.50 \pm 2.21$  &          $5.86_{-0.34}^{+0.66}$  & $0.34_{-0.43}^{+0.13}$  &  $0.90_{-0.19}^{+0.12}$  \\
      &          &          &       &      & $3.92 \pm 1.54$  &  $5.96 \pm 1.69$  &  $3.75 \pm 2.35$  &  $3.75 \pm 2.35$  &          $5.90_{-0.34}^{+0.51}$  & $0.18_{-0.47}^{+0.21}$  &  $0.85_{-0.29}^{+0.16}$  \\
 3786 & 274.8371 & -13.4591 &   1.8 &  0.1 & $9.58 \pm 1.10$  &  $8.31 \pm 0.90$  &  $9.03 \pm 1.37$  &  $9.03 \pm 1.37$  &          $5.30_{-0.45}^{+0.37}$  & $0.12_{-0.24}^{+0.36}$  &  $1.05_{-0.36}^{+0.19}$  \\
      &          &          &       &      & $1.88 \pm 0.82$  &  $1.42 \pm 0.77$  &  $2.35 \pm 1.40$  &  $2.35 \pm 1.40$  &          $5.67_{-0.60}^{+0.61}$  & $-0.04_{-0.34}^{+0.38}$  &  $1.16_{-0.23}^{+0.12}$  \\
 4013 & 276.3031 &  -12.741 & 11.84 & 0.40 & $14.25 \pm 1.57$  &  $6.07 \pm 1.30$  &  $9.95 \pm 1.70$  &  $9.95 \pm 1.70$  &         $6.36_{-0.21}^{+0.26}$  & $0.86_{-0.13}^{+0.08}$  &  $1.12_{-0.08}^{+0.05}$  \\
      &          &          &       &      & $0.02 \pm 0.02$  &  $0.05 \pm 0.12$  &  $0.02 \pm 0.05$  &  $0.02 \pm 0.05$  &          $6.19_{-1.38}^{+0.31}$  & $0.49_{-0.55}^{+0.28}$  &  $1.31_{-0.24}^{+0.13}$  \\
      &          &          &       &      & $39.26 \pm 1.95$  &  $40.06 \pm 1.77$  &  $40.69 \pm 2.08$  &  $40.69 \pm 2.08$  &      $6.31_{-0.15}^{+0.16}$  & $1.00_{-0.09}^{+0.15}$  &  $1.24_{-0.11}^{+0.05}$  \\
 4101 &  276.713 & -12.4251 &  4.71 & 0.31 & $17.11 \pm 3.07$  &  $24.36 \pm 5.02$  &  $21.13 \pm 4.14$  &  $21.13 \pm 4.14$  &      $5.69_{-0.95}^{+0.81}$  & $0.71_{-0.11}^{+0.29}$  &  $1.03_{-0.33}^{+0.12}$  \\
      &          &          &       &      & $8.14 \pm 4.62$  &  $9.37 \pm 8.35$  &  $6.81 \pm 5.46$  &  $6.81 \pm 5.46$  &          $5.73_{-0.97}^{+0.86}$  & $0.62_{-0.15}^{+0.23}$  &  $1.19_{-0.19}^{+0.09}$  \\
 4745 & 278.6758 &  -8.5279 &  6.05 & 0.36 & $7.06 \pm 1.21$  &  $8.06 \pm 1.37$  &  $8.93 \pm 2.45$  &  $8.93 \pm 2.45$  &          $5.43_{-0.87}^{+0.89}$  & $0.93_{-0.12}^{+0.22}$  &  $1.42_{-0.11}^{+0.10}$  \\
      &          &          &       &      & $3.03 \pm 1.13$  &  $4.81 \pm 1.27$  &  $6.12 \pm 2.26$  &  $6.12 \pm 2.26$  &          $5.65_{-0.73}^{+0.56}$  & $0.96_{-0.10}^{+0.15}$  &  $1.56_{-0.12}^{+0.10}$  \\
 4750 & 278.6704 &  -8.5073 &  6.05 & 0.36 & $0.01 \pm 0.02$  &  $0.00 \pm 0.00$  &  $0.20 \pm 0.20$  &  $0.20 \pm 0.20$  &          $4.80_{-0.68}^{+0.40}$  & $0.63_{-0.18}^{+0.19}$  &  $0.95_{-0.30}^{+0.19}$  \\
      &          &          &       &      & $3.34 \pm 1.49$  &  $3.65 \pm 1.74$  &  $8.05 \pm 1.99$  &  $8.05 \pm 1.99$  &          $6.02_{-0.19}^{+0.34}$  & $0.82_{-0.08}^{+0.12}$  &  $1.73_{-0.08}^{+0.05}$  \\
      &          &          &       &      & $3.07 \pm 1.25$  &  $4.45 \pm 3.79$  &  $3.93 \pm 2.14$  &  $3.93 \pm 2.14$  &          $6.48_{-1.11}^{+0.24}$  & $0.57_{-0.13}^{+0.14}$  &  $1.27_{-0.11}^{+0.05}$  \\
      &          &          &       &      & $0.42 \pm 0.50$  &  $0.58 \pm 0.87$  &  $0.90 \pm 1.15$  &  $0.90 \pm 1.15$  &          $6.06_{-0.64}^{+0.28}$  & $0.84_{-0.13}^{+0.09}$  &  $1.68_{-0.07}^{+0.14}$  \\
 4796 & 278.6437 &  -8.1738 & $<1.27$ &      & $0.75 \pm 0.54$  &  $0.62 \pm 0.57$  &  $1.38 \pm 0.91$  &  $1.38 \pm 0.91$  &          $4.86_{-0.18}^{+0.21}$  & $-0.84_{-0.10}^{+0.08}$  &  $0.58_{-0.15}^{+0.14}$  \\
      &          &          &       &      & $5.92 \pm 1.73$  &  $3.31 \pm 1.73$  &  $0.67 \pm 1.54$  &  $0.67 \pm 1.54$  &          $5.95_{-0.13}^{+0.00}$  & $-0.64_{-0.02}^{+0.25}$  &  $1.19_{-0.03}^{+0.01}$  \\
      &          &          &       &      & $2.15 \pm 2.43$  &  $1.59 \pm 1.48$  &  $0.80 \pm 0.74$  &  $0.80 \pm 0.74$  &          $6.95_{-0.16}^{+0.05}$  & $-0.98_{-0.01}^{+0.22}$  &  $1.10_{-0.16}^{+0.10}$  \\
      &          &          &       &      & $15.60 \pm 1.78$  &  $27.84 \pm 2.78$  &  $42.27 \pm 2.09$  &  $42.27 \pm 2.09$  &      $6.22_{-0.00}^{+0.12}$  & $0.74_{-0.07}^{+0.02}$  &  $1.57_{-0.03}^{+0.00}$  \\
 4816 & 278.5914 &  -8.0471 &  3.10 & 0.44 & $1.13 \pm 0.84$  &  $0.87 \pm 0.64$  &  $1.02 \pm 0.72$  &  $1.02 \pm 0.72$  &          $5.51_{-0.33}^{+0.35}$  & $0.12_{-0.40}^{+0.38}$  &  $1.31_{-0.14}^{+0.11}$  \\
      &          &          &       &      & $8.95 \pm 1.69$  &  $2.47 \pm 1.44$  &  $0.41 \pm 0.58$  &  $0.41 \pm 0.58$  &          $5.12_{-0.33}^{+0.54}$  & $0.42_{-0.35}^{+0.39}$  &  $0.79_{-0.17}^{+0.18}$  \\
      &          &          &       &      & $0.01 \pm 0.00$  &  $0.02 \pm 0.01$  &  $0.27 \pm 0.21$  &  $0.27 \pm 0.21$  &          $6.77_{-0.25}^{+0.13}$  & $-0.75_{-0.16}^{+0.15}$  &  $0.82_{-0.12}^{+0.18}$  \\
      &          &          &       &      & $6.12 \pm 1.34$  &  $9.90 \pm 1.47$  &  $14.83 \pm 1.69$  &  $14.83 \pm 1.69$  &        $5.13_{-0.85}^{+1.30}$  & $0.76_{-0.13}^{+0.26}$  &  $1.49_{-0.18}^{+0.10}$  \\
      &          &          &       &      & $1.55 \pm 0.78$  &  $2.46 \pm 0.86$  &  $0.24 \pm 0.52$  &  $0.24 \pm 0.52$  &          $5.18_{-0.31}^{+0.59}$  & $0.20_{-0.26}^{+0.35}$  &  $1.38_{-0.12}^{+0.08}$  \\
 4955 & 278.5693 &  -7.2357 &  6.38 & 0.48 & $7.04 \pm 2.30$  &  $11.97 \pm 2.27$  &  $16.42 \pm 4.43$  &  $16.42 \pm 4.43$  &       $6.73_{-0.06}^{+0.00}$  & $0.65_{-0.11}^{+0.00}$  &  $-1.40_{-0.00}^{+0.00}$  \\
      &          &          &       &      & $0.14 \pm 0.12$  &  $0.01 \pm 0.00$  &  $0.00 \pm 0.01$  &  $0.00 \pm 0.01$  &          $6.63_{-0.00}^{+0.00}$  & $-0.15_{-0.00}^{+0.08}$  &  $1.03_{-0.00}^{+0.02}$  \\
      &          &          &       &      & $1.27 \pm 0.64$  &  $0.32 \pm 0.35$  &  $3.91 \pm 4.59$  &  $3.91 \pm 4.59$  &          $3.67_{-0.05}^{+0.00}$  & $-0.11_{-0.00}^{+0.06}$  &  $0.14_{-0.00}^{+0.00}$  \\
      &          &          &       &      & $33.09 \pm 3.01$  &  $43.68 \pm 2.71$  &  $56.12 \pm 3.68$  &  $56.12 \pm 3.68$  &      $6.23_{-0.00}^{+0.00}$  & $0.96_{-0.00}^{+0.04}$  &  $1.50_{-0.00}^{+0.00}$  \\
      &          &          &       &      & $20.85 \pm 4.01$  &  $10.51 \pm 3.71$  &  $12.77 \pm 7.44$  &  $12.77 \pm 7.44$  &      $6.32_{-0.03}^{+0.00}$  & $0.77_{-0.00}^{+0.00}$  &  $1.48_{-0.00}^{+0.00}$  \\
 5201 & 279.6925 &  -6.4813 &  5.93 & 0.42 & $9.58 \pm 0.99$  &  $12.46 \pm 0.93$  &  $14.59 \pm 1.42$  &  $14.59 \pm 1.42$  &       $4.98_{-0.07}^{+0.03}$  & $0.76_{-0.05}^{+0.04}$  &  $1.41_{-0.03}^{+0.04}$  \\
      &          &          &       &      & $0.05 \pm 0.07$  &  $0.08 \pm 0.14$  &  $0.83 \pm 0.61$  &  $0.83 \pm 0.61$  &          $5.04_{-0.18}^{+0.15}$  & $-0.30_{-0.16}^{+0.48}$  &  $1.29_{-0.11}^{+0.10}$  \\
      &          &          &       &      & $0.02 \pm 0.02$  &  $0.01 \pm 0.03$  &  $0.01 \pm 0.01$  &  $0.01 \pm 0.01$  &          $6.09_{-0.20}^{+0.41}$  & $-0.18_{-0.09}^{+0.26}$  &  $1.02_{-0.20}^{+0.06}$  \\
      &          &          &       &      & $0.70 \pm 0.29$  &  $1.08 \pm 0.31$  &  $1.79 \pm 0.65$  &  $1.79 \pm 0.65$  &          $6.32_{-0.03}^{+0.00}$  & $0.77_{-0.00}^{+0.00}$  &  $1.48_{-0.00}^{+0.00}$  \\
      &          &          &       &      & $0.00 \pm 0.00$  &  $0.00 \pm 0.00$  &  $0.00 \pm 0.00$  &  $0.00 \pm 0.00$  &          $6.78_{-0.27}^{+0.04}$  & $0.42_{-0.03}^{+0.14}$  &  $0.48_{-0.50}^{+0.08}$  \\
 5504 & 280.9674 &  -4.6098 &  3.09 & 0.43 & $0.01 \pm 0.01$  &  $0.01 \pm 0.01$  &  $0.01 \pm 0.01$  &  $0.01 \pm 0.01$  &          $4.74_{-0.01}^{+0.00}$  & $-0.55_{-0.05}^{+0.12}$  &  $-0.39_{-0.00}^{+0.44}$  \\
      &          &          &       &      & $6.24 \pm 1.51$  &  $10.33 \pm 2.29$  &  $10.98 \pm 1.93$  &  $10.98 \pm 1.93$  &       $5.55_{-0.00}^{+0.85}$  & $-0.47_{-0.20}^{+0.00}$  &  $1.33_{-0.00}^{+0.12}$  \\
      &          &          &       &      & $20.71 \pm 1.93$  &  $21.69 \pm 2.11$  &  $26.02 \pm 2.36$  &  $26.02 \pm 2.36$  &      $5.78_{-0.62}^{+0.00}$  & $0.58_{-0.37}^{+0.09}$  &  $1.51_{-0.13}^{+0.00}$  \\
      &          &          &       &      & $0.01 \pm 0.01$  &  $0.00 \pm 0.02$  &  $0.00 \pm 0.01$  &  $0.00 \pm 0.01$  &          $5.65_{-0.68}^{+0.11}$  & $0.62_{-0.34}^{+0.03}$  &  $1.50_{-0.11}^{+0.03}$  \\
 5559 & 281.0159 &  -4.2557 &  4.64 & 0.39 & $0.34 \pm 0.37$  &  $0.24 \pm 0.37$  &  $0.21 \pm 0.40$  &  $0.21 \pm 0.40$  &          $6.00_{-0.98}^{+0.55}$  & $-0.27_{-0.37}^{+0.46}$  &  $0.98_{-0.30}^{+0.17}$  \\
      &          &          &       &      & $7.57 \pm 1.02$  &  $8.20 \pm 0.81$  &  $10.52 \pm 1.23$  &  $10.52 \pm 1.23$  &        $5.64_{-0.56}^{+0.62}$  & $0.34_{-0.40}^{+0.25}$  &  $0.66_{-0.45}^{+0.27}$  \\
      &          &          &       &      & $7.95 \pm 1.40$  &  $17.46 \pm 2.27$  &  $17.34 \pm 3.66$  &  $17.34 \pm 3.66$  &       $6.09_{-1.02}^{+0.56}$  & $0.57_{-0.11}^{+0.15}$  &  $0.60_{-0.44}^{+0.26}$  \\
 5686 & 280.9778 &  -3.4684 &  4.62 & 0.40 & $0.00 \pm 0.00$  &  $0.00 \pm 0.00$  &  $0.00 \pm 0.00$  &  $0.00 \pm 0.00$  &          $6.53_{-0.28}^{+0.12}$  & $0.78_{-0.13}^{+0.15}$  &  $1.60_{-0.10}^{+0.05}$  \\
      &          &          &       &      & $16.52 \pm 1.79$  &  $16.37 \pm 2.37$  &  $29.28 \pm 3.78$  &  $29.28 \pm 3.78$  &      $6.48_{-0.43}^{+0.22}$  & $-0.17_{-0.38}^{+0.21}$  &  $0.91_{-0.13}^{+0.16}$  \\
      &          &          &       &      & $0.01 \pm 0.02$  &  $0.00 \pm 0.00$  &  $0.00 \pm 0.00$  &  $0.00 \pm 0.00$  &          $6.23_{-0.07}^{+0.16}$  & $0.77_{-0.17}^{+0.20}$  &  $1.78_{-0.07}^{+0.08}$  \\
      &          &          &       &      & $0.00 \pm 0.00$  &  $0.00 \pm 0.00$  &  $0.00 \pm 0.00$  &  $0.00 \pm 0.00$  &          $6.60_{-0.32}^{+0.19}$  & $-0.13_{-0.30}^{+0.19}$  &  $0.26_{-0.60}^{+0.32}$  \\
      &          &          &       &      & $0.42 \pm 0.58$  &  $0.23 \pm 0.32$  &  $0.90 \pm 1.15$  &  $0.90 \pm 1.15$  &          $6.48_{-0.18}^{+0.19}$  & $-0.48_{-0.33}^{+0.16}$  &  $0.28_{-0.30}^{+0.40}$  \\
 5976 &  281.904 &  -2.1462 &  7.08 & 0.50 & $23.42 \pm 3.07$  &  $31.14 \pm 3.41$  &  $35.26 \pm 5.47$  &  $35.26 \pm 5.47$  &      $4.39_{-0.90}^{+0.28}$  & $-0.02_{-0.29}^{+1.40}$  &  $1.35_{-0.39}^{+0.80}$  \\
      &          &          &       &      & $19.45 \pm 2.55$  &  $26.97 \pm 2.74$  &  $35.48 \pm 4.61$  &  $35.48 \pm 4.61$  &      $5.40_{-0.77}^{+0.33}$  & $0.92_{-0.06}^{+0.55}$  &  $1.44_{-0.12}^{+0.16}$  \\
      &          &          &       &      & $2.23 \pm 1.45$  &  $2.56 \pm 1.64$  &  $8.13 \pm 2.30$  &  $8.13 \pm 2.30$  &          $4.56_{-0.24}^{+0.11}$  & $1.31_{-0.25}^{+0.15}$  &  $1.42_{-0.16}^{+0.10}$  \\
      &          &          &       &      & $2.03 \pm 1.23$  &  $3.10 \pm 1.49$  &  $5.21 \pm 1.38$  &  $5.21 \pm 1.38$  &          $5.02_{-0.62}^{+1.36}$  & $0.95_{-0.31}^{+0.29}$  &  $1.41_{-0.30}^{+0.23}$  \\
      &          &          &       &      & $9.77 \pm 1.02$  &  $16.49 \pm 1.15$  &  $28.63 \pm 1.91$  &  $28.63 \pm 1.91$  &       $5.89_{-0.33}^{+0.29}$  & $0.76_{-0.08}^{+0.15}$  &  $1.44_{-0.07}^{+0.14}$  \\
 5997 & 282.0054 &  -2.0609 &  7.33 & 0.25 & $0.28 \pm 0.48$  &  $0.40 \pm 0.76$  &  $0.68 \pm 0.92$  &  $0.68 \pm 0.92$  &          $6.28_{-0.10}^{+0.21}$  & $0.79_{-0.05}^{+0.08}$  &  $1.45_{-0.15}^{+0.08}$  \\
      &          &          &       &      & $4.45 \pm 1.69$  &  $4.01 \pm 1.48$  &  $2.91 \pm 1.03$  &  $2.91 \pm 1.03$  &          $5.74_{-0.69}^{+1.42}$  & $0.48_{-0.60}^{+0.09}$  &  $1.30_{-0.25}^{+0.11}$  \\
 6004 & 282.0535 &  -2.0414 &  7.07 & 0.50 & $19.32 \pm 2.26$  &  $17.49 \pm 1.98$  &  $17.39 \pm 3.14$  &  $17.39 \pm 3.14$  &      $5.28_{-0.39}^{+0.26}$  & $0.74_{-0.17}^{+0.32}$  &  $0.98_{-0.21}^{+0.18}$  \\
      &          &          &       &      & $0.06 \pm 0.07$  &  $0.17 \pm 0.17$  &  $0.16 \pm 0.16$  &  $0.16 \pm 0.16$  &          $6.12_{-0.19}^{+0.20}$  & $0.75_{-0.08}^{+0.06}$  &  $0.76_{-0.19}^{+0.11}$  \\
      &          &          &       &      & $39.90 \pm 2.21$  &  $32.65 \pm 1.81$  &  $32.19 \pm 2.74$  &  $32.19 \pm 2.74$  &      $5.38_{-0.56}^{+1.22}$  & $0.29_{-0.35}^{+0.28}$  &  $1.03_{-0.51}^{+0.20}$  \\
      &          &          &       &      & $0.01 \pm 0.01$  &  $0.00 \pm 0.00$  &  $0.00 \pm 0.00$  &  $0.00 \pm 0.00$  &          $6.20_{-0.34}^{+0.23}$  & $0.89_{-0.08}^{+0.08}$  &  $1.42_{-0.05}^{+0.04}$  \\
 6028 &  281.718 &  -1.7779 &  7.07 & 0.95 & $3.10 \pm 0.86$  &  $5.58 \pm 0.99$  &  $4.92 \pm 1.33$  &  $4.92 \pm 1.33$  &          $6.41_{-0.22}^{+0.17}$  & $-0.29_{-0.34}^{+0.15}$  &  $0.33_{-0.60}^{+0.19}$  \\
      &          &          &       &      & $4.43 \pm 0.95$  &  $6.90 \pm 0.98$  &  $6.59 \pm 1.46$  &  $6.59 \pm 1.46$  &          $6.31_{-0.22}^{+0.20}$  & $0.73_{-0.13}^{+0.17}$  &  $1.58_{-0.10}^{+0.09}$  \\
      &          &          &       &      & $0.01 \pm 0.02$  &  $0.00 \pm 0.00$  &  $0.00 \pm 0.00$  &  $0.00 \pm 0.00$  &          $6.40_{-0.15}^{+0.22}$  & $0.81_{-0.16}^{+0.15}$  &  $1.62_{-0.07}^{+0.06}$  \\
      &          &          &       &      & $8.77 \pm 2.20$  &  $7.03 \pm 1.50$  &  $7.45 \pm 2.35$  &  $7.45 \pm 2.35$  &          $6.17_{-0.25}^{+0.55}$  & $-0.06_{-0.24}^{+0.27}$  &  $0.51_{-0.18}^{+0.40}$  \\
 6148 & 282.1367 &  -1.1541 &  7.01 & 1.02 & $0.04 \pm 0.05$  &  $0.03 \pm 0.05$  &  $0.01 \pm 0.03$  &  $0.01 \pm 0.03$  &          $6.35_{-0.22}^{+0.28}$  & $0.72_{-0.08}^{+0.12}$  &  $1.28_{-0.08}^{+0.08}$  \\
      &          &          &       &      & $0.23 \pm 0.26$  &  $0.18 \pm 0.24$  &  $0.10 \pm 0.18$  &  $0.10 \pm 0.18$  &          $5.99_{-0.73}^{+0.42}$  & $0.18_{-0.41}^{+0.21}$  &  $1.20_{-0.17}^{+0.14}$  \\
      &          &          &       &      & $7.35 \pm 2.59$  &  $4.71 \pm 1.38$  &  $4.95 \pm 1.75$  &  $4.95 \pm 1.75$  &          $6.03_{-0.38}^{+0.47}$  & $0.40_{-0.19}^{+0.12}$  &  $0.25_{-0.44}^{+0.36}$  \\
      &          &          &       &      & $0.56 \pm 0.29$  &  $0.65 \pm 0.46$  &  $1.58 \pm 1.10$  &  $1.58 \pm 1.10$  &          $6.46_{-0.43}^{+0.30}$  & $0.60_{-0.09}^{+0.12}$  &  $0.79_{-0.35}^{+0.22}$  \\
 6256 &  282.466 &  -0.4041 &  3.14 & 0.45 & $0.01 \pm 0.02$  &  $0.00 \pm 0.01$  &  $0.01 \pm 0.02$  &  $0.01 \pm 0.02$  &          $5.29_{-0.77}^{+0.74}$  & $-0.22_{-0.19}^{+0.65}$  &  $0.89_{-0.43}^{+0.22}$  \\
      &          &          &       &      & $0.91 \pm 0.34$  &  $2.82 \pm 0.66$  &  $4.42 \pm 1.32$  &  $4.42 \pm 1.32$  &          $6.05_{-0.50}^{+0.72}$  & $-0.55_{-0.18}^{+0.38}$  &  $0.85_{-0.41}^{+0.14}$  \\
      &          &          &       &      & $0.00 \pm 0.00$  &  $0.00 \pm 0.00$  &  $0.00 \pm 0.00$  &  $0.00 \pm 0.00$  &          $6.63_{-0.54}^{+0.15}$  & $0.54_{-0.13}^{+0.11}$  &  $1.55_{-0.05}^{+0.06}$  \\
      &          &          &       &      & $2.15 \pm 0.74$  &  $3.17 \pm 0.69$  &  $6.28 \pm 1.22$  &  $6.28 \pm 1.22$  &          $5.49_{-0.38}^{+0.22}$  & $-0.54_{-0.21}^{+0.43}$  &  $0.22_{-0.35}^{+0.34}$  \\
 6286 & 282.8543 &  -0.2419 & 11.35 & 0.46 & $0.01 \pm 0.01$  &  $0.00 \pm 0.00$  &  $0.00 \pm 0.01$  &  $0.00 \pm 0.01$  &          $6.42_{-0.24}^{+0.19}$  & $0.66_{-0.07}^{+0.16}$  &  $0.45_{-0.74}^{+0.27}$  \\
      &          &          &       &      & $3.41 \pm 4.18$  &  $1.21 \pm 2.05$  &  $17.61 \pm 16.33$  &  $17.61 \pm 16.33$  &      $5.96_{-0.64}^{+0.70}$  & $0.21_{-0.43}^{+0.17}$  &  $0.64_{-0.36}^{+0.36}$  \\
      &          &          &       &      & $2.50 \pm 5.54$  &  $8.22 \pm 25.34$  &  $12.25 \pm 36.49$  &  $12.25 \pm 36.49$  &     $6.22_{-0.20}^{+0.28}$  & $0.75_{-0.10}^{+0.18}$  &  $1.40_{-0.13}^{+0.11}$  \\
      &          &          &       &      & $7.22 \pm 1.34$  &  $9.09 \pm 1.38$  &  $8.72 \pm 1.78$  &  $8.72 \pm 1.78$  &          $6.32_{-0.16}^{+0.36}$  & $0.68_{-0.11}^{+0.12}$  &  $1.17_{-0.13}^{+0.12}$  \\
 6307 & 282.4882 &   0.0925 &  0.56 & 0.54 & $0.45 \pm 0.30$  &  $0.36 \pm 0.26$  &  $0.45 \pm 0.54$  &  $0.45 \pm 0.54$  &          $5.84_{-0.63}^{+0.46}$  & $-0.04_{-0.46}^{+0.33}$  &  $1.56_{-0.09}^{+0.06}$  \\
      &          &          &       &      & $2.85 \pm 1.23$  &  $2.60 \pm 1.09$  &  $4.27 \pm 2.00$  &  $4.27 \pm 2.00$  &          $6.37_{-0.59}^{+0.34}$  & $-0.63_{-0.21}^{+0.35}$  &  $1.32_{-0.07}^{+0.06}$  \\
      &          &          &       &      & $11.46 \pm 2.22$  &  $9.67 \pm 1.24$  &  $18.98 \pm 4.53$  &  $18.98 \pm 4.53$  &       $6.25_{-0.64}^{+0.36}$  & $-0.46_{-0.27}^{+0.43}$  &  $1.20_{-0.09}^{+0.12}$  \\
 6512 & 283.3616 &    1.264 &  3.58 & 0.46 & $0.05 \pm 0.06$  &  $0.02 \pm 0.03$  &  $0.01 \pm 0.01$  &  $0.01 \pm 0.01$  &          $5.82_{-1.06}^{+0.20}$  & $0.74_{-0.15}^{+0.20}$  &  $1.55_{-0.10}^{+0.09}$  \\
      &          &          &       &      & $2.30 \pm 8.13$  &  $13.22 \pm 1.27$  &  $12.58 \pm 17.48$  &  $12.58 \pm 17.48$  &     $5.75_{-0.87}^{+0.67}$  & $-0.10_{-0.42}^{+0.60}$  &  $0.80_{-0.38}^{+0.27}$  \\
      &          &          &       &      & $0.00 \pm 0.00$  &  $0.58 \pm 0.43$  &  $0.82 \pm 0.59$  &  $0.82 \pm 0.59$  &          $6.08_{-0.37}^{+0.32}$  & $0.84_{-0.11}^{+0.14}$  &  $1.76_{-0.09}^{+0.08}$  \\
      &          &          &       &      & $0.13 \pm 0.19$  &  $0.21 \pm 0.32$  &  $0.42 \pm 0.67$  &  $0.42 \pm 0.67$  &          $5.75_{-0.59}^{+0.69}$  & $0.42_{-0.62}^{+0.36}$  &  $1.60_{-0.14}^{+0.16}$  \\
      &          &          &       &      & $0.00 \pm 0.00$  &  $0.00 \pm 0.00$  &  $0.00 \pm 0.00$  &  $0.00 \pm 0.00$  &          $5.81_{-0.59}^{+0.60}$  & $-0.36_{-0.38}^{+1.10}$  &  $1.84_{-0.48}^{+0.38}$  \\
 6530 &  283.324 &   1.3971 &  3.84 & 0.46 & $1.32 \pm 0.76$  &  $0.71 \pm 0.36$  &  $7.50 \pm 3.66$  &  $7.50 \pm 3.66$  &          $4.72_{-0.32}^{+0.04}$  & $0.01_{-0.31}^{+0.35}$  &  $0.02_{-0.61}^{+0.60}$  \\
      &          &          &       &      & $9.03 \pm 1.00$  &  $13.32 \pm 0.84$  &  $9.19 \pm 2.97$  &  $9.19 \pm 2.97$  &         $5.18_{-0.38}^{+0.34}$  & $0.22_{-0.24}^{+0.22}$  &  $1.41_{-0.28}^{+0.08}$  \\
      &          &          &       &      & $0.00 \pm 0.01$  &  $0.02 \pm 0.02$  &  $0.02 \pm 0.03$  &  $0.02 \pm 0.03$  &          $6.36_{-0.20}^{+0.21}$  & $0.62_{-0.06}^{+0.14}$  &  $1.37_{-0.09}^{+0.08}$  \\
 6572 & 284.2129 &   1.3027 &  2.96 & 0.47 & $2.54 \pm 0.80$  &  $1.86 \pm 0.70$  &  $3.82 \pm 1.52$  &  $3.82 \pm 1.52$  &          $5.54_{-1.89}^{+1.16}$  & $-0.42_{-0.36}^{+0.46}$  &  $0.94_{-0.75}^{+0.20}$  \\
      &          &          &       &      & $4.12 \pm 0.88$  &  $6.45 \pm 0.89$  &  $6.68 \pm 1.58$  &  $6.68 \pm 1.58$  &          $5.92_{-0.30}^{+0.34}$  & $0.44_{-0.34}^{+0.12}$  &  $1.20_{-0.13}^{+0.08}$  \\
      &          &          &       &      & $0.00 \pm 0.01$  &  $0.01 \pm 0.01$  &  $0.00 \pm 0.01$  &  $0.00 \pm 0.01$  &          $6.56_{-0.36}^{+0.14}$  & $0.56_{-0.06}^{+0.09}$  &  $1.44_{-0.04}^{+0.05}$  \\
      &          &          &       &      & $2.77 \pm 0.62$  &  $7.61 \pm 0.96$  &  $16.58 \pm 1.78$  &  $16.58 \pm 1.78$  &        $6.54_{-0.32}^{+0.26}$  & $-0.03_{-0.20}^{+0.14}$  &  $1.44_{-0.12}^{+0.14}$  \\
 6590 & 283.3846 &   1.7837 &  3.61 & 0.47 & $3.30 \pm 0.74$  &  $4.59 \pm 0.93$  &  $8.03 \pm 2.07$  &  $8.03 \pm 2.07$  &          $5.16_{-0.00}^{+0.00}$  & $0.75_{-0.00}^{+0.00}$  &  $1.63_{-0.01}^{+0.03}$  \\
      &          &          &       &      & $2.10 \pm 1.03$  &  $0.99 \pm 0.64$  &  $0.88 \pm 0.70$  &  $0.88 \pm 0.70$  &          $6.36_{-0.18}^{+0.26}$  & $0.57_{-0.00}^{+0.10}$  &  $1.55_{-0.06}^{+0.00}$  \\
 6747 & 284.1658 &   2.3252 &  3.72 & 0.48 & $17.85 \pm 1.42$  &  $15.93 \pm 1.10$  &  $20.28 \pm 1.58$  &  $20.28 \pm 1.58$  &      $5.64_{-0.39}^{+0.45}$  & $0.30_{-0.36}^{+0.25}$  &  $0.68_{-0.35}^{+0.22}$  \\
      &          &          &       &      & $0.08 \pm 0.11$  &  $0.04 \pm 0.07$  &  $0.04 \pm 0.08$  &  $0.04 \pm 0.08$  &          $6.39_{-0.25}^{+0.26}$  & $0.65_{-0.05}^{+0.07}$  &  $1.39_{-0.04}^{+0.04}$  \\
      &          &          &       &      & $12.87 \pm 2.08$  &  $9.70 \pm 1.36$  &  $6.47 \pm 1.89$  &  $6.47 \pm 1.89$  &         $6.04_{-0.87}^{+0.33}$  & $-0.25_{-0.31}^{+0.43}$  &  $0.73_{-0.77}^{+0.25}$  \\
 7137 & 285.2823 &   5.1651 &  1.88 & 0.46 & $22.53 \pm 1.95$  &  $26.09 \pm 1.60$  &  $32.33 \pm 2.08$  &  $32.33 \pm 2.08$  &      $5.48_{-0.37}^{+0.62}$  & $0.42_{-0.35}^{+0.15}$  &  $1.32_{-0.09}^{+0.06}$  \\
      &          &          &       &      & $7.37 \pm 1.84$  &  $7.73 \pm 1.81$  &  $9.67 \pm 2.35$  &  $9.67 \pm 2.35$  &          $6.55_{-0.36}^{+0.20}$  & $0.54_{-0.07}^{+0.07}$  &  $1.37_{-0.04}^{+0.04}$  \\
 7143 & 285.2769 &   5.1949 &  1.88 & 0.46 & $6.90 \pm 2.13$  &  $6.92 \pm 1.76$  &  $7.62 \pm 2.56$  &  $7.62 \pm 2.56$  &          $5.58_{-0.54}^{+0.93}$  & $0.49_{-0.30}^{+0.12}$  &  $1.45_{-0.10}^{+0.05}$  \\
      &          &          &       &      & $2.40 \pm 0.56$  &  $5.75 \pm 0.94$  &  $9.75 \pm 1.39$  &  $9.75 \pm 1.39$  &          $5.54_{-0.40}^{+0.51}$  & $0.32_{-0.34}^{+0.24}$  &  $1.51_{-0.10}^{+0.07}$  \\
 7183 & 285.9298 &   5.1744 &  2.48 & 0.47 & $0.15 \pm 0.25$  &  $0.12 \pm 0.33$  &  $0.21 \pm 0.99$  &  $0.21 \pm 0.99$  &          $6.05_{-1.29}^{+0.58}$  & $0.64_{-0.15}^{+0.17}$  &  $1.68_{-0.11}^{+0.07}$  \\
      &          &          &       &      & $5.46 \pm 1.31$  &  $6.71 \pm 0.99$  &  $7.79 \pm 1.53$  &  $7.79 \pm 1.53$  &          $5.79_{-0.71}^{+0.56}$  & $0.02_{-0.47}^{+0.39}$  &  $1.32_{-0.16}^{+0.13}$  \\
 7276 & 286.1944 &   6.0907 &  8.39 & 0.64 & $3.43 \pm 1.53$  &  $1.95 \pm 0.84$  &  $2.90 \pm 1.43$  &  $2.90 \pm 1.43$  &          $6.06_{-1.20}^{+0.47}$  & $0.75_{-0.09}^{+0.21}$  &  $1.28_{-0.14}^{+0.07}$  \\
      &          &          &       &      & $1.12 \pm 0.85$  &  $1.08 \pm 1.70$  &  $3.19 \pm 3.08$  &  $3.19 \pm 3.08$  &          $5.65_{-0.66}^{+0.91}$  & $0.62_{-0.15}^{+0.12}$  &  $0.56_{-0.42}^{+0.26}$  \\
 7492 &  287.454 &   8.0791 &  5.19 & 0.90 & $25.11 \pm 1.45$  &  $29.55 \pm 3.01$  &  $37.64 \pm 2.34$  &  $37.64 \pm 2.34$  &      $6.37_{-0.32}^{+0.72}$  & $0.57_{-0.18}^{+0.11}$  &  $1.27_{-0.21}^{+0.13}$  \\
      &          &          &       &      & $3.36 \pm 0.71$  &  $8.53 \pm 1.10$  &  $13.46 \pm 1.59$  &  $13.46 \pm 1.59$  &        $6.28_{-0.13}^{+0.26}$  & $0.80_{-0.05}^{+0.04}$  &  $1.46_{-0.04}^{+0.03}$  \\
 7516 & 287.5937 &   8.9819 & 11.20 & 0.50 & $0.64 \pm 0.50$  &  $1.83 \pm 0.91$  &  $2.14 \pm 1.28$  &  $2.14 \pm 1.28$  &          $5.73_{-0.81}^{+0.47}$  & $0.95_{-0.09}^{+0.25}$  &  $1.70_{-0.14}^{+0.07}$  \\
      &          &          &       &      & $2.93 \pm 1.01$  &  $6.34 \pm 1.88$  &  $8.21 \pm 2.25$  &  $8.21 \pm 2.25$  &          $5.51_{-0.87}^{+0.79}$  & $0.83_{-0.19}^{+0.29}$  &  $1.58_{-0.15}^{+0.14}$  \\
 7569 & 287.9451 &   9.7828 &  5.93 & 1.44 & $2.39 \pm 1.22$  &  $3.46 \pm 1.87$  &  $4.87 \pm 2.47$  &  $4.87 \pm 2.47$  &          $5.95_{-0.94}^{+0.56}$  & $0.71_{-0.11}^{+0.08}$  &  $1.46_{-0.13}^{+0.08}$  \\
      &          &          &       &      & $1.25 \pm 0.93$  &  $1.09 \pm 1.06$  &  $2.28 \pm 2.53$  &  $2.28 \pm 2.53$  &          $5.47_{-0.63}^{+0.95}$  & $0.66_{-0.21}^{+0.10}$  &  $1.40_{-0.18}^{+0.11}$  \\
 7580 & 287.9335 &   9.9845 &  5.91 & 2.10 & $12.03 \pm 1.30$  &  $28.54 \pm 1.78$  &  $45.91 \pm 2.74$  &  $45.91 \pm 2.74$  &      $5.70_{-0.51}^{+0.88}$  & $0.60_{-0.14}^{+0.09}$  &  $1.37_{-0.11}^{+0.07}$  \\
      &          &          &       &      & $5.16 \pm 1.27$  &  $9.93 \pm 1.41$  &  $13.60 \pm 1.93$  &  $13.60 \pm 1.93$  &        $3.62_{-0.20}^{+0.33}$  & $0.33_{-0.05}^{+0.49}$  &  $1.08_{-0.14}^{+0.11}$  \\
 7590 & 287.9285 &  10.1183 &  4.76 & 1.14 & $3.16 \pm 1.23$  &  $5.58 \pm 1.73$  &  $6.48 \pm 1.84$  &  $6.48 \pm 1.84$  &          $6.23_{-1.45}^{+0.40}$  & $0.62_{-0.23}^{+0.33}$  &  $1.13_{-0.30}^{+0.13}$  \\
      &          &          &       &      & $2.50 \pm 0.85$  &  $3.59 \pm 1.00$  &  $8.74 \pm 1.85$  &  $8.74 \pm 1.85$  &          $5.26_{-0.56}^{+0.96}$  & $0.54_{-0.24}^{+0.16}$  &  $1.23_{-0.10}^{+0.12}$  \\
 8304 & 292.2463 &  17.8265 &  9.76 & 0.56 & $10.20 \pm 1.06$  &  $16.89 \pm 1.19$  &  $24.67 \pm 1.95$  &  $24.67 \pm 1.95$  &      $6.36_{-0.84}^{+0.21}$  & $0.69_{-0.07}^{+0.12}$  &  $1.44_{-0.13}^{+0.07}$  \\
      &          &          &       &      & $0.02 \pm 0.05$  &  $0.00 \pm 0.01$  &  $0.05 \pm 0.13$  &  $0.05 \pm 0.13$  &          $6.23_{-0.08}^{+0.21}$  & $0.93_{-0.08}^{+0.11}$  &  $1.62_{-0.11}^{+0.06}$  \\
      &          &          &       &      & $0.00 \pm 0.00$  &  $0.00 \pm 0.00$  &  $0.00 \pm 0.00$  &  $0.00 \pm 0.00$  &          $5.78_{-0.37}^{+0.40}$  & $0.19_{-0.34}^{+0.20}$  &  $0.48_{-0.59}^{+0.31}$  \\
      &          &          &       &      & $12.67 \pm 1.03$  &  $10.86 \pm 0.87$  &  $9.77 \pm 1.57$  &  $9.77 \pm 1.57$  &        $6.29_{-0.23}^{+0.31}$  & $-0.07_{-0.27}^{+0.20}$  &  $0.33_{-0.50}^{+0.27}$  \\
 8376 & 292.5529 &  18.3298 &  1.92 & 0.64 & $0.73 \pm 0.52$  &  $0.20 \pm 0.33$  &  $0.78 \pm 1.23$  &  $0.78 \pm 1.23$  &          $5.40_{-0.63}^{+1.12}$  & $0.44_{-0.34}^{+0.15}$  &  $1.11_{-0.29}^{+0.17}$  \\
      &          &          &       &      & $1.74 \pm 0.64$  &  $4.20 \pm 1.65$  &  $5.43 \pm 2.00$  &  $5.43 \pm 2.00$  &          $5.79_{-0.73}^{+0.54}$  & $-0.10_{-0.39}^{+0.41}$  &  $1.07_{-0.25}^{+0.13}$  \\
 8745 & 294.8893 &  23.9764 &  2.16 & 0.10 & $3.81 \pm 0.76$  &  $8.09 \pm 1.22$  &  $11.45 \pm 1.66$  &  $11.45 \pm 1.66$  &        $6.10_{-0.82}^{+0.52}$  & $0.48_{-0.32}^{+0.14}$  &  $1.67_{-0.10}^{+0.08}$  \\
      &          &          &       &      & $8.62 \pm 0.96$  &  $6.52 \pm 0.74$  &  $8.31 \pm 1.00$  &  $8.31 \pm 1.00$  &          $6.44_{-0.42}^{+0.29}$  & $0.59_{-0.08}^{+0.10}$  &  $1.71_{-0.07}^{+0.04}$  \\
 8870 & 296.7318 &  25.2053 &   2.5 &      & $0.61 \pm 0.55$  &  $0.25 \pm 0.42$  &  $0.12 \pm 0.24$  &  $0.12 \pm 0.24$  &          $6.00_{-0.41}^{+0.60}$  & $0.49_{-0.15}^{+0.06}$  &  $1.08_{-0.13}^{+0.08}$  \\
      &          &          &       &      & $3.22 \pm 1.34$  &  $5.61 \pm 1.26$  &  $7.97 \pm 2.37$  &  $7.97 \pm 2.37$  &          $5.99_{-0.49}^{+0.43}$  & $-0.23_{-0.30}^{+0.40}$  &  $0.81_{-0.36}^{+0.19}$  \\
18398 & 265.9022 & -30.5484 & $<5.22$ &      & $0.02 \pm 0.02$  &  $0.02 \pm 0.02$  &  $0.01 \pm 0.01$  &  $0.01 \pm 0.01$  &          $5.82_{-1.18}^{+0.64}$  & $0.81_{-0.16}^{+0.27}$  &  $1.65_{-0.15}^{+0.11}$  \\
      &          &          &       &      & $2.04 \pm 1.46$  &  $3.81 \pm 1.54$  &  $6.33 \pm 2.51$  &  $6.33 \pm 2.51$  &          $6.07_{-0.81}^{+0.44}$  & $-0.10_{-0.43}^{+0.32}$  &  $0.40_{-0.45}^{+0.34}$  \\
      &          &          &       &      & $6.77 \pm 1.94$  &  $7.63 \pm 1.49$  &  $7.41 \pm 2.11$  &  $7.41 \pm 2.11$  &          $5.75_{-1.01}^{+0.70}$  & $0.75_{-0.19}^{+0.25}$  &  $1.63_{-0.15}^{+0.12}$  \\
18695 & 265.7287 & -29.3484 & $<6.01$ &      & $2.36 \pm 1.47$  &  $2.04 \pm 1.23$  &  $2.34 \pm 1.55$  &  $2.34 \pm 1.55$  &          $5.31_{-0.59}^{+1.21}$  & $0.64_{-0.16}^{+0.21}$  &  $0.95_{-0.38}^{+0.18}$  \\
      &          &          &       &      & $0.79 \pm 0.62$  &  $0.90 \pm 0.79$  &  $0.96 \pm 0.90$  &  $0.96 \pm 0.90$  &          $5.67_{-0.79}^{+0.81}$  & $0.59_{-0.22}^{+0.34}$  &  $0.91_{-0.35}^{+0.18}$  \\
      &          &          &       &      & $3.70 \pm 1.09$  &  $1.77 \pm 0.95$  &  $6.85 \pm 1.86$  &  $6.85 \pm 1.86$  &          $5.47_{-0.85}^{+0.91}$  & $0.60_{-0.30}^{+0.27}$  &  $1.40_{-0.16}^{+0.12}$  \\
18738 & 266.1215 &  -29.402 & $<7.00$ &      & $4.10 \pm 1.42$  &  $16.53 \pm 1.32$  &  $21.10 \pm 2.11$  &  $21.10 \pm 2.11$  &       $6.36_{-0.25}^{+0.25}$  & $0.75_{-0.09}^{+0.12}$  &  $1.64_{-0.06}^{+0.08}$  \\
      &          &          &       &      & $6.27_{-0.15}^{+0.22}$  & $0.89_{-0.07}^{+0.11}$  &  $1.71_{-0.05}^{+0.06}$ 
\enddata
\end{deluxetable}
\end{longrotatetable}
\clearpage
%\end{landscape}


\begin{thebibliography}{}
\expandafter\ifx\csname natexlab\endcsname\relax\def\natexlab#1{#1}\fi

\bibitem[{{Andr{\'e}} {et~al.}(2014){Andr{\'e}}, {Di Francesco},
  {Ward-Thompson}, {Inutsuka}, {Pudritz}, \& {Pineda}}]{Andre14}
{Andr{\'e}}, P., {Di Francesco}, J., {Ward-Thompson}, D., {et~al.} 2014,
  Protostars and Planets VI, 27

\bibitem[{{Azimlu} {et~al.}(2015){Azimlu}, {Mart{\'{\i}}nez-Galarza}, \&
  {Muench}}]{Azimlu15}
{Azimlu}, M., {Mart{\'{\i}}nez-Galarza}, J.~R., \& {Muench}, A.~A. 2015, \aj,
  150, 95

\bibitem[{{Bastian} {et~al.}(2010){Bastian}, {Covey}, \& {Meyer}}]{Bastian10}
{Bastian}, N., {Covey}, K.~R., \& {Meyer}, M.~R. 2010, \araa, 48, 339

\bibitem[{{Bate}(2012)}]{Bate12}
  {Bate}, M.~R. 2012, \mnras, 419, 3115

\bibitem[Benjamin et al.(2003)]{Benjamin03} Benjamin, R.~A.,
  Churchwell, E., Babler, B.~L., et al.\ 2003, \pasp, 115, 953

\bibitem[{{Bonnell} {et~al.}(2001){Bonnell}, {Bate}, {Clarke}, \&
  {Pringle}}]{Bonnell01}
{Bonnell}, I.~A., {Bate}, M.~R., {Clarke}, C.~J., \& {Pringle}, J.~E. 2001,
  \mnras, 323, 785

\bibitem[{{Bonnell} {et~al.}(2003){Bonnell}, {Bate}, \& {Vine}}]{Bonnell03}
{Bonnell}, I.~A., {Bate}, M.~R., \& {Vine}, S.~G. 2003, \mnras, 343, 413

\bibitem[{{Bonnell} {et~al.}(1998){Bonnell}, {Bate}, \&
  {Zinnecker}}]{Bonnell98}
{Bonnell}, I.~A., {Bate}, M.~R., \& {Zinnecker}, H. 1998, \mnras, 298, 93

\bibitem[{{Bonnell} {et~al.}(2004){Bonnell}, {Vine}, \& {Bate}}]{Bonnell04}
{Bonnell}, I.~A., {Vine}, S.~G., \& {Bate}, M.~R. 2004, \mnras, 349, 735

\bibitem[{{Bressert} {et~al.}(2010){Bressert}, {Bastian}, {Gutermuth},
  {Megeath}, {Allen}, {Evans}, {Rebull}, {Hatchell}, {Johnstone}, {Bourke},
  {Cieza}, {Harvey}, {Merin}, {Ray}, \& {Tothill}}]{Bressert10}
  {Bressert}, E., {Bastian}, N., {Gutermuth}, R., {et~al.} 2010, \mnras, 409, L54

\bibitem[Carey et al.(2005)]{Carey05} Carey, S.~J., Noriega-Crespo, A.,
  Price, S.~D., et al.\ 2005, Bulletin of the American Astronomical Society, 37, 63.33 

\bibitem[{{Carlson} {et~al.}(2012){Carlson}, {Sewi{\l}o}, {Meixner}, {Romita},
  \& {Lawton}}]{Carlson12}
{Carlson}, L.~R., {Sewi{\l}o}, M., {Meixner}, M., {Romita}, K.~A., \& {Lawton},
  B. 2012, \aap, 542, A66

\bibitem[{{Caswell} {et~al.}(2010){Caswell}, {Fuller}, {Green}, {Avison},
  {Breen}, {Brooks}, {Burton}, {Chrysostomou}, {Cox}, {Diamond}, {Ellingsen},
  {Gray}, {Hoare}, {Masheder}, {McClure-Griffiths}, {Pestalozzi}, {Phillips},
  {Quinn}, {Thompson}, {Voronkov}, {Walsh}, {Ward-Thompson}, {Wong-McSweeney},
  {Yates}, \& {Cohen}}]{Caswell10}
{Caswell}, J.~L., {Fuller}, G.~A., {Green}, J.~A., {et~al.} 2010, \mnras, 404,
  1029

\bibitem[{{Chabrier}(2003)}]{Chabrier03}
  {Chabrier}, G. 2003, \pasp, 115, 763

\bibitem[Churchwell et al.(2009)]{Churchwell09} Churchwell, E.,
  Babler, B.~L., Meade, M.~R., et al.\ 2009, \pasp, 121, 213 

\bibitem[{{Duch{\^e}ne} \& {Kraus}(2013)}]{Duchene13}
{Duch{\^e}ne}, G., \& {Kraus}, A. 2013, \araa, 51, 269

\bibitem[{{Elmegreen}(2000)}]{Elmegreen00}
{Elmegreen}, B.~G. 2000, \apj, 539, 342

\bibitem[{{Elmegreen} {et~al.}(2008){Elmegreen}, {Klessen}, \&
  {Wilson}}]{Elmegreen08}
  {Elmegreen}, B.~G., {Klessen}, R.~S., \& {Wilson}, C.~D. 2008, \apj, 681, 365

\bibitem[Fazio et al.(2004)]{Fazio04} Fazio, G.~G., Hora, J.~L.,
  Allen, L.~E., et al.\ 2004, \apjs, 154, 10

\bibitem[{{Forbrich} {et~al.}(2010){Forbrich}, {Tappe}, {Robitaille}, {Muench},
  {Teixeira}, {Lada}, {Stolte}, \& {Lada}}]{Forbrich10}
{Forbrich}, J., {Tappe}, A., {Robitaille}, T., {et~al.} 2010, \apj, 716, 1453

\bibitem[{{Green} {et~al.}(2015){Green}, {Schlafly}, {Finkbeiner}, {Rix},
  {Martin}, {Burgett}, {Draper}, {Flewelling}, {Hodapp}, {Kaiser}, {Kudritzki},
  {Magnier}, {Metcalfe}, {Price}, {Tonry}, \& {Wainscoat}}]{Green15}
{Green}, G.~M., {Schlafly}, E.~F., {Finkbeiner}, D.~P., {et~al.} 2015, \apj,
  810, 25

\bibitem[Hewett et al.(2006)]{Hewett06} Hewett, P.~C., Warren, S.~J., Leggett,
  S.~K., \& Hodgkin, S.~T.\ 2006, \mnras, 367, 454
  
\bibitem[{{Huff} \& {Stahler}(2006)}]{Huff06}
  {Huff}, E.~M., \& {Stahler}, S.~W. 2006, \apj, 644, 355

\bibitem[Indebetouw et al.(2005)]{Indebetouw05} Indebetouw, R., Mathis, J.~S.,
  Babler, B.~L., et al.\ 2005, \apj, 619, 931 

\bibitem[{{Indebetouw} {et~al.}(2007){Indebetouw}, {Robitaille}, {Whitney},
  {Churchwell}, {Babler}, {Meade}, {Watson}, \& {Wolfire}}]{Indebetouw07}
{Indebetouw}, R., {Robitaille}, T.~P., {Whitney}, B.~A., {et~al.} 2007, \apj,
  666, 321

\bibitem[{{Joncour} {et~al.}(2017){Joncour}, {Duch{\^e}ne}, \&
  {Moraux}}]{Joncour17}
{Joncour}, I., {Duch{\^e}ne}, G., \& {Moraux}, E. 2017, \aap, 599, A14

\bibitem[{{Kroupa}(2001)}]{Kroupa01}
{Kroupa}, P. 2001, \mnras, 322, 231

\bibitem[{{Kroupa} \& {Boily}(2002)}]{Kroupa02}
{Kroupa}, P., \& {Boily}, C.~M. 2002, \mnras, 336, 1188

\bibitem[{{Kroupa} {et~al.}(2013){Kroupa}, {Weidner}, {Pflamm-Altenburg},
  {Thies}, {Dabringhausen}, {Marks}, \& {Maschberger}}]{Kroupa13}
{Kroupa}, P., {Weidner}, C., {Pflamm-Altenburg}, J., {et~al.} 2013, {The
  Stellar and Sub-Stellar Initial Mass Function of Simple and Composite
  Populations}, ed. T.~D. {Oswalt} \& G.~{Gilmore}, 115

\bibitem[{{Lada} \& {Lada}(2003)}]{Lada03}
{Lada}, C.~J., \& {Lada}, E.~A. 2003, \araa, 41, 57

\bibitem[{{Larson}(1982)}]{Larson82}
{Larson}, R.~B. 1982, \mnras, 200, 159

\bibitem[{{Larson}(2003)}]{Larson03}
{Larson}, R.~B. 2003, in Astronomical Society of the Pacific Conference Series,
  Vol. 287, Galactic Star Formation Across the Stellar Mass Spectrum, ed. J.~M.
  {De Buizer} \& N.~S. {van der Bliek}, 65--80

\bibitem[{{Lomax} \& {Whitworth}(2018)}]{Lomax18}
{Lomax}, O., \& {Whitworth}, A.~P. 2018, \mnras, arXiv:1711.07385

\bibitem[{{Lomax} {et~al.}(2015){Lomax}, {Whitworth}, {Hubber}, {Stamatellos},
  \& {Walch}}]{Lomax15}
{Lomax}, O., {Whitworth}, A.~P., {Hubber}, D.~A., {Stamatellos}, D., \&
  {Walch}, S. 2015, \mnras, 447, 1550

\bibitem[{{Maschberger} \& {Clarke}(2008)}]{Maschberger08}
{Maschberger}, T., \& {Clarke}, C.~J. 2008, \mnras, 391, 711

\bibitem[{{McKee} \& {Tan}(2003)}]{McKee03}
{McKee}, C.~F., \& {Tan}, J.~C. 2003, \apj, 585, 850

\bibitem[{{Morales} \& {Robitaille}(2017)}]{Morales17}
{Morales}, E.~F.~E., \& {Robitaille}, T.~P. 2017, \aap, 598, A136

\bibitem[{{Morales} {et~al.}(2013){Morales}, {Wyrowski}, {Schuller}, \&
  {Menten}}]{Morales13}
{Morales}, E.~F.~E., {Wyrowski}, F., {Schuller}, F., \& {Menten}, K.~M. 2013,
  \aap, 560, A76

\bibitem[{{Mottram} {et~al.}(2011){Mottram}, {Hoare}, {Urquhart}, {Lumsden},
  {Oudmaijer}, {Robitaille}, {Moore}, {Davies}, \& {Stead}}]{Mottram11}
{Mottram}, J.~C., {Hoare}, M.~G., {Urquhart}, J.~S., {et~al.} 2011, \aap, 525,
  A149

\bibitem[{{Myers}(2011)}]{Myers11}
{Myers}, P.~C. 2011, \apj, 743, 98

\bibitem[Myers(2012)]{Myers12} Myers, P.~C.\ 2012, \apj, 752, 9

\bibitem[{{Oey} \& {Clarke}(2005)}]{Oey05}
{Oey}, M.~S., \& {Clarke}, C.~J. 2005, \apjl, 620, L43

\bibitem[{{Robitaille}(2008)}]{Robitaille08b}
{Robitaille}, T.~P. 2008, in Astronomical Society of the Pacific Conference
  Series, Vol. 387, Massive Star Formation: Observations Confront Theory, ed.
  H.~{Beuther}, H.~{Linz}, \& T.~{Henning}, 290

\bibitem[{{Robitaille} {et~al.}(2006){Robitaille}, {Whitney}, {Indebetouw},
  {Wood}, \& {Denzmore}}]{Robitaille06}
{Robitaille}, T.~P., {Whitney}, B.~A., {Indebetouw}, R., {Wood}, K., \&
  {Denzmore}, P. 2006, \apjs, 167, 256

\bibitem[{{Robitaille} {et~al.}(2008){Robitaille}, {Meade}, {Babler},
  {Whitney}, {Johnston}, {Indebetouw}, {Cohen}, {Povich}, {Sewilo}, {Benjamin},
  \& {Churchwell}}]{Robitaille08}
{Robitaille}, T.~P., {Meade}, M.~R., {Babler}, B.~L., {et~al.} 2008, \aj, 136,
  2413

\bibitem[{{Simon} {et~al.}(2007){Simon}, {Bolatto}, {Whitney}, {Robitaille},
  {Shah}, {Makovoz}, {Stanimirovi{\'c}}, {Barb{\'a}}, \& {Rubio}}]{Simon07}
{Simon}, J.~D., {Bolatto}, A.~D., {Whitney}, B.~A., {et~al.} 2007, \apj, 669,
  327

\bibitem[{{Stamatellos} \& {Whitworth}(2009)}]{Stamatellos09}
{Stamatellos}, D., \& {Whitworth}, A.~P. 2009, \mnras, 392, 413

\bibitem[{{Szymczak} {et~al.}(2005){Szymczak}, {Pillai}, \&
  {Menten}}]{Szymczak05}
{Szymczak}, M., {Pillai}, T., \& {Menten}, K.~M. 2005, \aap, 434, 613

\bibitem[{{Tobin} {et~al.}(2016){Tobin}, {Looney}, {Li}, {Chandler}, {Dunham},
  {Segura-Cox}, {Sadavoy}, {Melis}, {Harris}, {Kratter}, \& {Perez}}]{Tobin16}
{Tobin}, J.~J., {Looney}, L.~W., {Li}, Z.-Y., {et~al.} 2016, \apj, 818, 73

\bibitem[{{Urquhart} {et~al.}(2013){Urquhart}, {Moore}, {Schuller}, {Wyrowski},
  {Menten}, {Thompson}, {Csengeri}, {Walmsley}, {Bronfman}, \&
  {K{\"o}nig}}]{Urquhart13}
{Urquhart}, J.~S., {Moore}, T.~J.~T., {Schuller}, F., {et~al.} 2013, \mnras,
  431, 1752

\bibitem[{{Weidner} \& {Kroupa}(2004)}]{WeidnerKroupa04}
{Weidner}, C., \& {Kroupa}, P. 2004, \mnras, 348, 187

\bibitem[{{Weidner} {et~al.}(2010){Weidner}, {Kroupa}, \&
  {Bonnell}}]{Weidner10}
{Weidner}, C., {Kroupa}, P., \& {Bonnell}, I.~A.~D. 2010, \mnras, 401, 275

\bibitem[{{Weidner} {et~al.}(2013){Weidner}, {Kroupa}, \&
  {Pflamm-Altenburg}}]{Weidner13}
{Weidner}, C., {Kroupa}, P., \& {Pflamm-Altenburg}, J. 2013, \mnras, 434, 84

\bibitem[{{Wienen} {et~al.}(2015){Wienen}, {Wyrowski}, {Menten}, {Urquhart},
  {Csengeri}, {Walmsley}, {Bontemps}, {Russeil}, {Bronfman}, {Koribalski}, \&
  {Schuller}}]{Wienen15}
{Wienen}, M., {Wyrowski}, F., {Menten}, K.~M., {et~al.} 2015, \aap, 579, A91

\bibitem[Willis et al.(2013)]{Willis13} Willis, S., Marengo, M., Allen, L., et al.\ 2013, \apj, 778, 96 

\bibitem[{{Winston} {et~al.}(2009){Winston}, {Megeath}, {Wolk}, {Hernandez},
  {Gutermuth}, {Muzerolle}, {Hora}, {Covey}, {Allen}, {Spitzbart}, {Peterson},
  {Myers}, \& {Fazio}}]{Winston09}
{Winston}, E., {Megeath}, S.~T., {Wolk}, S.~J., {et~al.} 2009, \aj, 137, 4777

\end{thebibliography}
\end{document}